# Surface Engineering for Phase Change Heat Transfer: A Review

Daniel Attinger,[1,*] Christophe Frankiewicz,[1] Amy R. Betz,[2] Thomas M. Schutzius,[3] Ranjan Ganguly,[3,4] Arindam Das,[3] C.-J. Kim,[5] Constantine M. Megaridis[3,*]

[1] *Department of Mechanical Engineering, Iowa State University, Ames IA 50011, USA*
[2] *Department of Mechanical and Nuclear Engineering, Kansas State University, Manhattan, KS 66506, USA*
[3] *Department of Mechanical Engineering, University of Illinois at Chicago, Chicago IL 60607-7022, USA*
[4] *Department of Power Engineering, Jadavpur University, Kolkata 700098, India*
[5] *Department of Mechanical and Aerospace Engineering University of California, Los Angeles, Los Angeles, CA 90095-1597, USA*

**Abstract**

Among numerous challenges to meet the rising global energy demand in a sustainable manner, improving phase change heat transfer has been at the forefront of engineering research for decades. The high heat transfer rates associated with phase change heat transfer are essential to energy and industry applications; but phase change is also inherently associated with poor thermodynamic efficiencies at low heat flux, and violent instabilities at high heat flux. Engineers have tried since the 1930's to fabricate solid surfaces that improve phase change heat transfer. The development of micro and nanotechnologies has made feasible the high-resolution control of surface texture and chemistry over length scales ranging from molecular levels to centimeters. This paper reviews the fabrication techniques available for metallic and silicon-based surfaces, considering sintered and polymeric coatings. The influence of such surfaces in multiphase processes of high practical interest, e.g. boiling, condensation, freezing, and the associated physical phenomena are reviewed. The case is made that while engineers are in principle able to manufacture surfaces with optimum nucleation or thermofluid transport characteristics, more theoretical and experimental efforts are needed to guide the design and cost-effective fabrication of surfaces that not only satisfy the existing technological needs, but also catalyze new discoveries.

**Keywords:** Interface, wettability, boiling, condensation, ice/frost, phase change, heat transfer

* E-mail: attinger@iastate.edu, cmm@uic.edu; Tel: D. Attinger +1 (515) 294 1692, C. M. Megaridis +1 (312) 996-3436

**Contents**

## Introduction

This review discusses the increasingly important contributions of advanced materials and novel fabrication techniques in enhancing phase change heat transfer and the related energy applications. Technologies based on phase change heat transfer have been serving mankind for more than two millennia,[1] with some examples shown in Figure 1. Phase change is essential to energy applications, where it drastically enhances heat transfer because latent heat is typically much larger than sensible heat.[2] For instance, the latent heat needed to boil one gram of water ($h$=2260 J $g^{-1}$) is more than 100 times larger than the heat required to increase the temperature of the same amount of water by 5K ($c_p \Delta T$ = 21 J $g^{-1}$). In addition, phase change is frequently accompanied by fast and large changes in specific volume, which result in enhanced heat transfer due to convection.

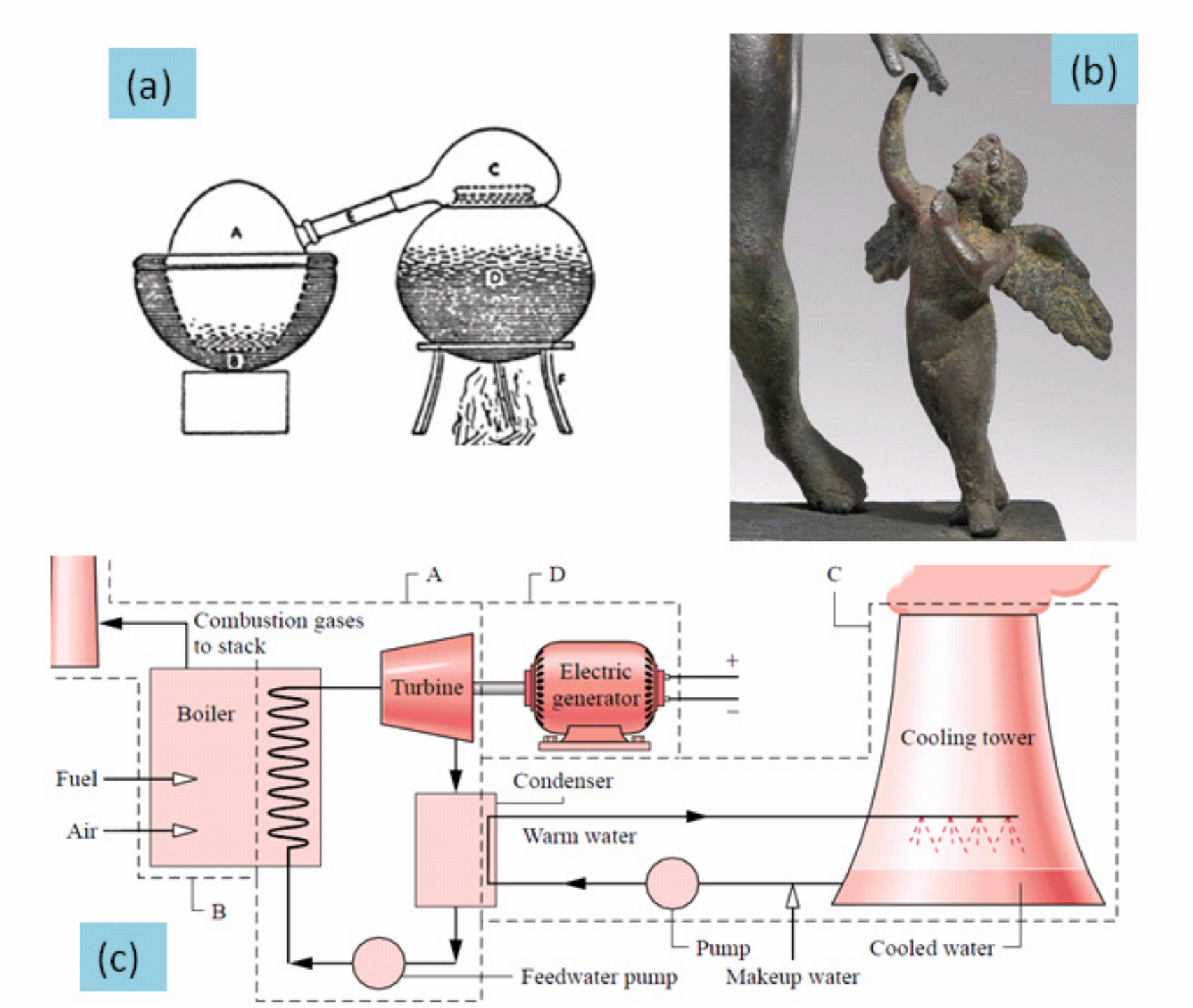

**Figure 1. Human beings have relied on phase change heat transfer from Antiquity to now, in order to satisfy needs such as (a) alcohol production by distillation[1] (ca. 600 BCE, India), (b) metal casting for military or artistic purposes (Eros with Aphrodite, bronze, ca. 100 BCE), and (c) boiling and condensation for mechanical power, during the industrial revolution of the 19th century [4]. Figures (a, b, c) reprinted with respective permissions of ASME, by courtesy of the Getty's Open Content Program, and with the permission of the Taylor and Francis Group.**

Three phase-change processes with industrial relevance are considered in a synthetic manner: boiling, condensation and ice/frost formation. After a presentation of how to characterize phase change, the properties of an optimum surface for each of these three modes of phase change is described; then the fabrication/manufacturing techniques available to modify surface texture and chemistry are described; next, the ways that these techniques have been used to date to enhance phase change heat transfer are reviewed. An emphasis is placed on the fast-developing area of wettability engineering, which relies on manipulating both surface texture and chemistry using special features, such as advanced materials. Then, an assessment is given on the prospective energetic, economic and environmental benefits of advanced surface engineering for phase change heat transfer. Finally, several research needs and priorities are identified and outlined, towards the goal of engineering optimum surfaces for phase change heat transfer. The review also aims at bridging the gap between the materials and heat transfer communities towards achieving the common goal of designing and deploying optimum material systems to enhance heat transfer performance and minimize thermodynamic inefficiencies, towards more sustainable energy transmission and production.

## 1. Characterization of phase change heat transfer

As mentioned in the introduction, the amount of heat exchanged during phase change heat transfer is very high. Therefore, phase change heat transfer is needed in many industrial applications [5], such as during thermal generation of electricity, desalination, metallurgy, electronics cooling and food processing. Industrial applications with phase change typically use water-based fluids or refrigerants, depending on considerations of the amount of heat to be transferred, operating conditions, and interactions of the fluid with electronic components[6]. Figure **2** describes and compares the three phase change processes considered in this review, according to their physical mechanisms, time scales and length scales. Boiling, condensation, and frost formation are each multiphase and quasi-steady processes, as shown with arrows arranged in circles. For instance, nucleate boiling involves the temporal sequence of nucleation, bubble growth and detachment, and a similar cycle is observed for condensation and freezing, as described in the caption of Figure **2**. As will be demonstrated later in section 1, these quasi-steady processes exchange heat optimally when the solid surface is in contact with both the discrete and continuous phase, e.g. the respective vapor and liquid in the case of nucleate boiling. However, phase change processes can experience instability towards a steady state of lower thermodynamic efficiency. After this transition, shown with an upward arrow in Figure **2**, the solid surface is in contact with only one fluid phase, e.g. with vapor in film boiling. In fact, the art of designing optimum surfaces for phase change heat transfer relies on keeping the solid surfaces in contact with both the continuous and discrete fluid phases, for the sake of optimum thermodynamic performance.

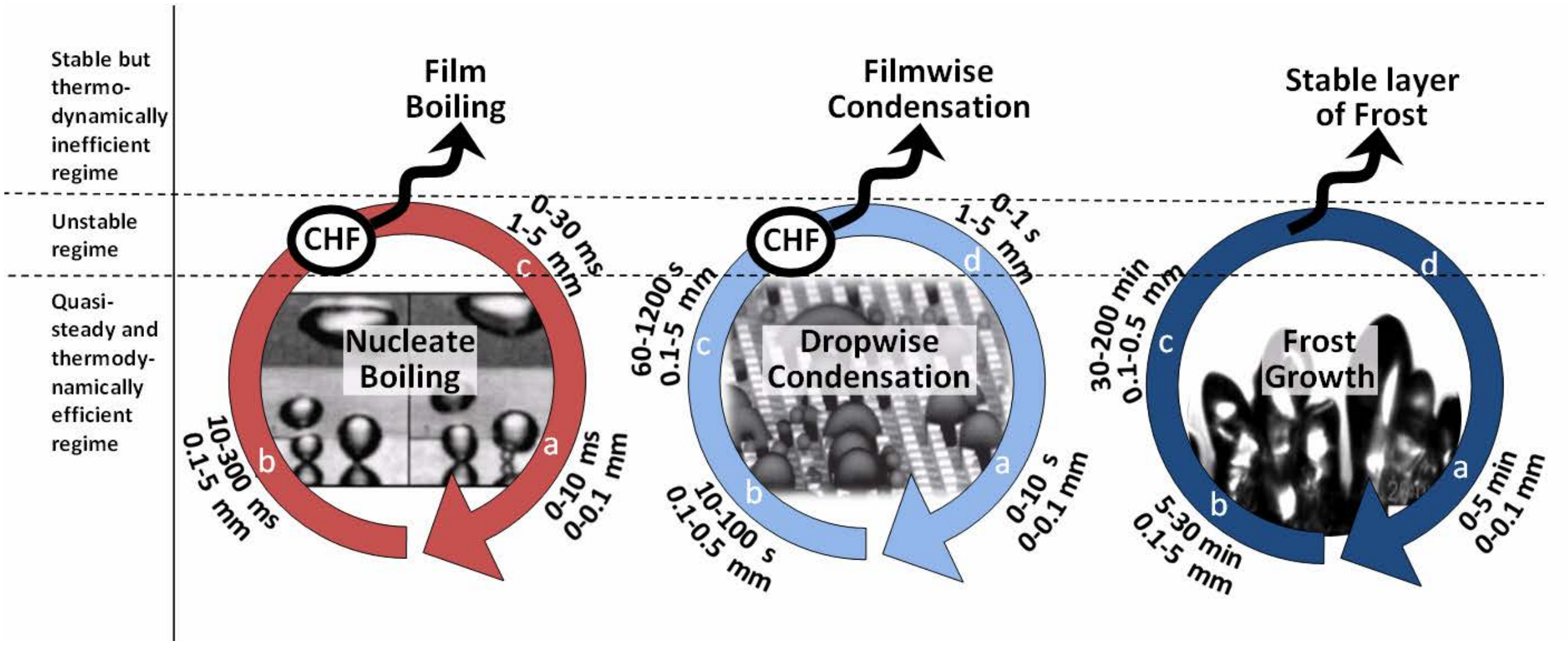

**Figure 2. Synthetic representation and universal features of phase change heat transfer during e.g. boiling [7-13], condensation [14-17] or frost formation [18, 19]. The quasi-steady mode is represented with circular arrows; it is most efficient thermodynamically and the time and length scales are mentioned. In boiling, it involves the succession of bubble nucleation (a), growth (b), and detachment (c). In condensation, it involves the succession of drop nucleation (a), growth (b), coalescence (c), and detachment (d). In desublimation or atmospheric frost growth, it involves crystal growth (a), layer growth (b), then dendrite or bulk growth (c), and removal by external forces, like scraping (d). When phase change occurs faster than the removal of the discrete phase, instability occurs and the process transitions towards a single-phase mode which is thermodynamically less efficient. Photograph of the boiling process is reprinted from [11], copyright 2012, with permission from Elsevier. Photograph of the condensation process is reprinted with permission from [16], copyright 2012 American Chemical Society. Photograph of the frost process is reprinted with permission from [20]. copyright 2010 American Chemical Society.**

A traditional description of these phase change processes is given with the phase diagram in Figure 3. The diagram also describes the terminology used in this review: Boiling (liquid to vapor phase change); condensation (vapor to liquid); sublimation (solid to vapor); desublimation (vapor to solid), also known as atmospheric deposition. The term frosting can also be used for the vapor to solid phase change. However, frosting may also describe the transition from vapor to solid via the liquid phase; melting (solid to liquid); solidification or freezing (liquid to solid); icing describes the transformation of water to ice.

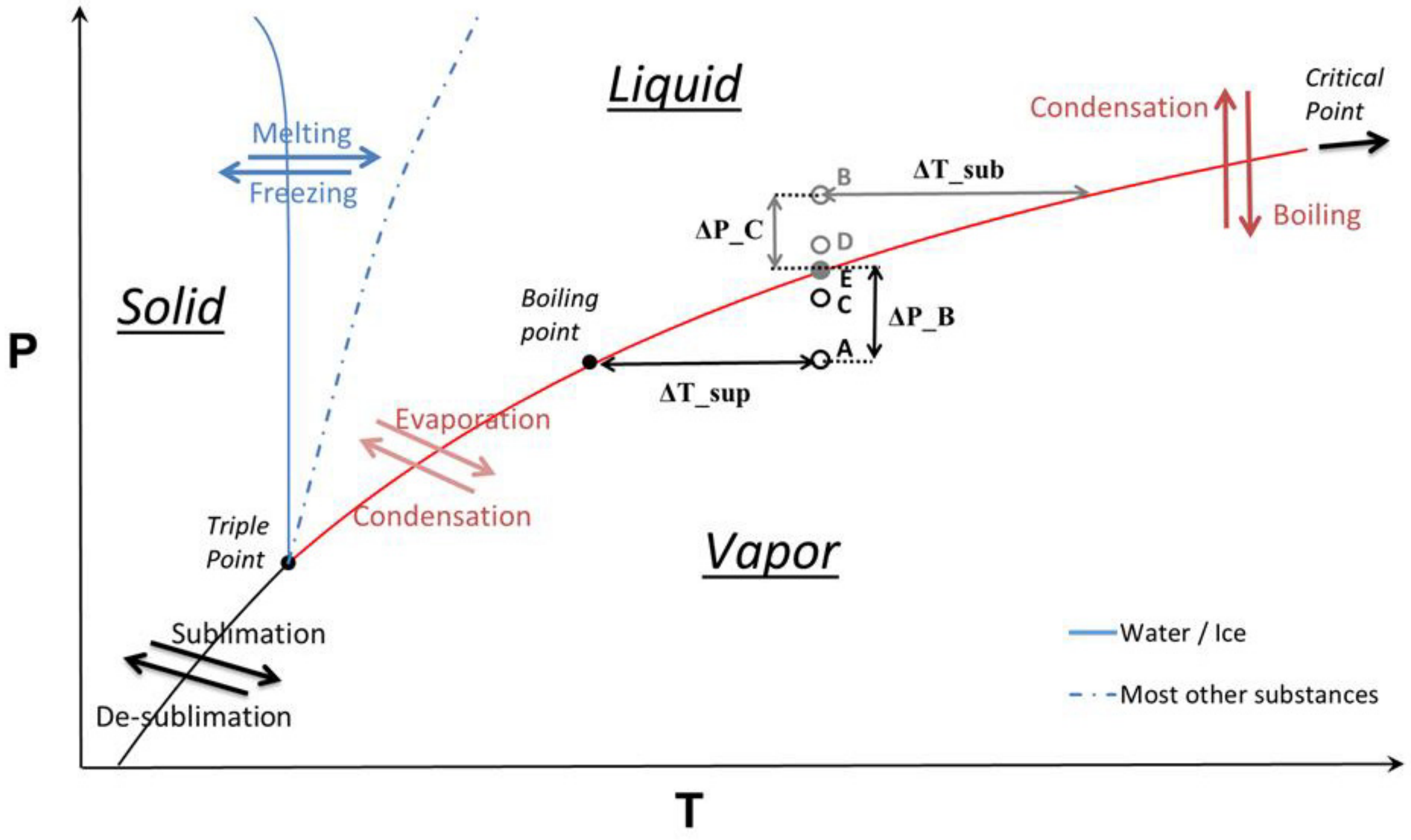

**Figure 3. The phase change processes considered in this review can all be described in a Pressure (log scale) - Temperature diagram. The points A, C and E illustrate the thermodynamic discussion in boiling, section 1.1, whereas the points B, D and E are relevant to the condensation section 1.2.**

Thermodynamically, a first measure of phase change heat transfer performance is the maximum heat flux exchanged between a solid surface and the fluid. For a boiling fluid, *the critical heat flux (CHF) is the highest heat flux that can be exchanged, before the individual vapor bubbles merge into a vapor layer that insulates the surface from the liquid.* Optimizing *CHF* can be seen as optimizing the heat transport, in the sense implied by the first principle of thermodynamics.

A second measure of performance of phase change heat transfer is the heat transfer coefficient HTC=$q''/\Delta T$, the ratio of the heat flux $q''$ to the difference $\Delta T$ between the solid surface temperature and the temperature of the fluid. HTCs are a measure of thermodynamic efficiency of heat transport, in the meaning of the second law. HTCs associated with phase change are typically one to two orders of magnitude larger than HTCs in single phase, because the latent heat minimizes $\Delta T$.

The characteristics of a solid surface that maximize HTC depend on the fluid and the $\Delta T$. The largest HTCs are typically achieved in regimes where both the continuous and discrete phases are in contact with the solid surface, such as the nucleate boiling regime or the analogous regime of dropwise condensation [21]. In these regimes, the solid surface interacts with multiple liquid and vapor volumes, maximizing interfacial and convective transport. For that reason, wettability engineering could play a critical role to improve phase change heat transfer.

If $q''$ is increased beyond CHF, the HTC is drastically reduced because heat is no longer transferred convectively but radiatively. This causes a significant and often destructive surface temperature increase [22-24], called dry-out and recently seen in Fukushima, Japan [25]. Similar considerations can be made for condensation, where the transition from dropwise condensation to film condensation drastically reduces HTC.

An optimum surface for phase change heat transfer is therefore a surface that optimizes heat transport according to the first and second principles of thermodynamics.

Phase change heat transfer is difficult to model or investigate experimentally, because it is a transient and multiscale phenomenon involving dynamic motion of multiple interfaces. The inherent complexity of phase change heat transfer appears in Table 1, which lists the forces and mechanisms governing the associated transport of heat and mass. Traditionally, the relative importance of pairs of forces or mechanisms is expressed with dimensionless numbers. Heat and mass transport is characterized with the Fourier, Biot, Nusselt and Jakob numbers. Fluid flow is characterized with the Reynolds, Bond, Weber, Froude, capillary, and Galilei numbers, as well as the drag coefficient. The relative importance of fluid flow transport -as compared with heat and mass transport- is expressed in terms of the Prandtl, Schmidt, Grashof and Eckert numbers. The influence of geometric/physical properties is expressed with numbers such as void fraction, wetting angle, density ratio, viscosity ratio and surface roughness. This extraordinary large quantity of *twenty* dimensionless numbers suggests that phase change heat transfer cannot be described with general theories, and that full similarity is impossible to obtain in experimental studies [26].

| | **Boiling** | **Condensation** | **Desublimation** | **Freezing** |
|---|---|---|---|---|
| **Continuous phase** | Liquid | Vapor | Vapor | Liquid |
| **Discrete phase** | Vapor | Liquid | Liquid or solid | Solid |
| **Wettability that promotes nucleation** | Hydrophobic | Hydrophilic | Hydrophilic | Debated |
| **Wettability that prevents transition from high to low heat flux** | Hydrophilic | Hydrophobic | Hydrophobic | Debated |
| **Forces that promote discrete phase detachment from substrate** | Buoyancy, shear, inertia | Gravity, surface tension, shear | Shear | Inertia, external stress |
| **Forces that promote discrete phase attachment to substrate** | surface tension, capillarity | Surface tension, gravity, inertia | Gravity, surface tension | Adhesion |
| **Dominant heat transfer mechanism at low q/ΔT** | Single phase convection | Single phase convection | Single phase convection | Single-phase (liquid) convection |
| **Dominant heat transfer mechanism at high q, below CHF** | Multiphase convection | Multiphase convection | Convection (primarily) and conduction | Conduction, convection |
| **Dominant mechanism for q above CHF** | Radiation | Single phase convection | Conduction | Conduction |

**Table 1: List of the fluid phases for typical phase change heat transfer processes, with the mechanisms and forces involved.**

Finally, let us consider how the development of micro- and nanotechnologies has impacted phase change heat transfer over the last 40 years. Drastic improvements have been made in implementing phase change heat transfer solutions closer to the heat source [6], as in electronics packaging. However, has the development of micro- and nanotechnologies improved the primary performance of surfaces for phase change heat transfer, such as the CHF and HTC? Figure 4 shows that, for the widely-studied process of water boiling on a copper surface, HTCs and CHFs have respectively increased by a factor 2 and 3 over the last 40 years. These improvements are significant; however several other technologies, such as computing power or energy density in batteries, have improved their performance by orders of magnitude over the same 40 years. Also, the theoretical limits of these improvements are either largely unknown or much higher, as exemplified by the "kinetic limit" obtained using the kinetic theory for the maximum heat flux. This limit can be reached if all molecules are emitted at the speed of sound from the liquid-vapor interface. As established by Schrage [27], for water at atmospheric pressure, the kinetic limit $q_{CHF,K} = 0.741 c \rho_v h_{fg} \approx 1.6 \times 10^4$ W cm$^{-2}$, with $c$ the speed of sound, is plotted in red in Figure 4. In that context, it seems worthy to review the fundamental mechanisms that govern phase change heat transfer, and new ways of engineering surfaces to enhance these mechanisms with the ultimate goal to raise the practically attained performance closer to theoretical limits.

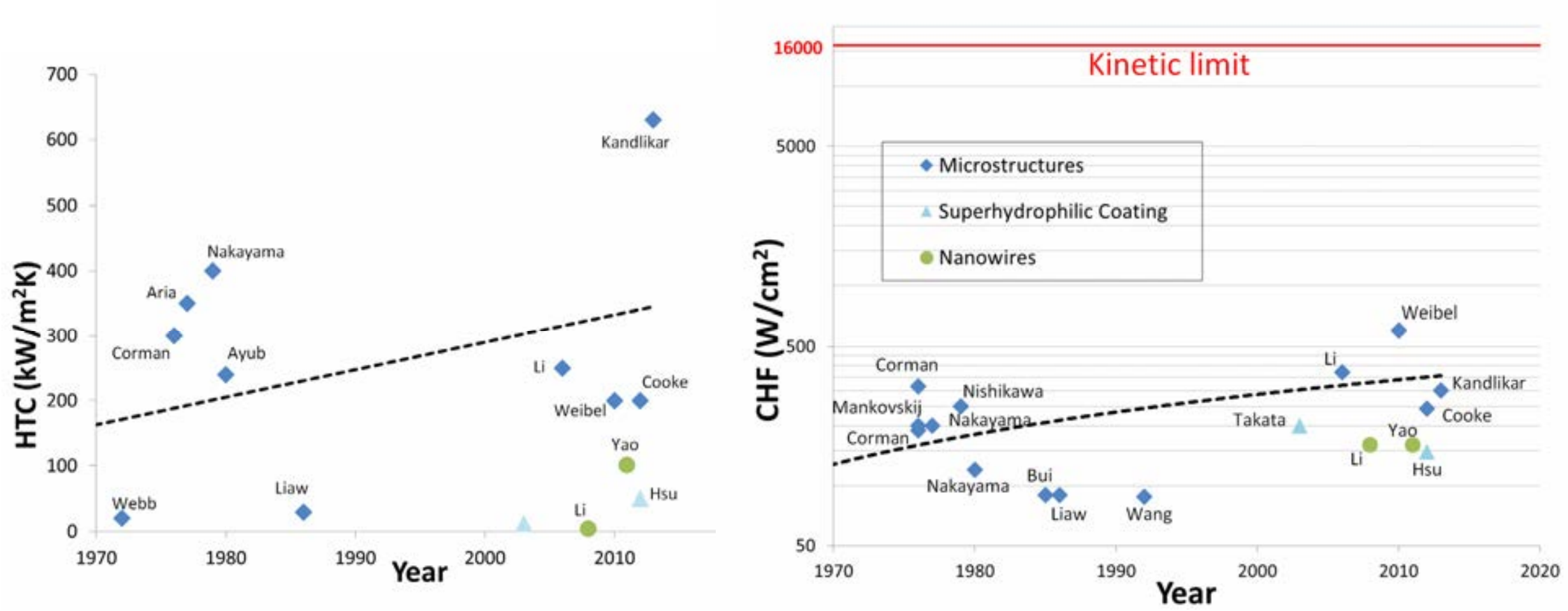

**Figure 4. In the representative process of water boiling on a heated copper surface, 40 years of micro- and nanoengineering of surfaces have improved the critical heat flux and heat transfer coefficients (at ΔT=5K). Studies [28-38] are referenced with the name of the first author. The dashed line shows an average trend of moderate improvement, with CHF values remaining well below the thermodynamic limit.**

## 1.1. Characteristics of optimum surfaces for boiling heat transfer

Boiling is a multiphase and multiscale energy transfer process from a heated surface to a liquid resulting in its vigorous vaporization. The first boiling curves, relating the heat flux to the superheat of a solid surface, were measured by Nukiyama [39]. Applications of boiling vary, but in general, surfaces can be engineered to facilitate nucleation or maximizing the critical heat flux by preventing film boiling [27, 40-51].

Let us describe here the physical aspects of the nucleation process, nucleate boiling and CHF, with emphasis on the contributions of the solid-fluid interface.

As a starting point, we summarize the thermodynamic processes of vapor formation in boiling, based on the extensive descriptions in [5, 52]. In Figure 3, the solid line corresponds to the co-existence or binodal curve, where liquid and vapor can both exist in a stable state. Metastable states can also exist in the vicinity of the binodal curve. For instance, point A is a metastable liquid state, B a metastable vapor state. Technically, boiling refers to the phase change of a metastable liquid (A) to a stable vapor phase (C).

***Nucleation***

Nucleation can occur at a solid surface, or within the bulk of the metastable liquid; the former process is called “heterogeneous nucleation,” while the latter is called “homogeneous nucleation.” Equilibrium thermodynamics can be used to predict the conditions for the incipience of boiling, i.e. the generation of an enough stable vapor nucleus.

For a closed system at thermodynamic equilibrium, constant pressure and temperature, boiling occurs under conditions of thermal, mechanical and chemical equilibrium [52]. The thermal equilibrium means that the temperature of the vapor phase and liquid phase are equal. The radius $r^*$ of a vapor nucleus in mechanical equilibrium with the surrounding liquid is given by the Young-Laplace equation:

$$p_V - p_L = \frac{2\sigma}{r^*}. \quad (1)$$

In engineering systems, the requirement of chemical equilibrium corresponds to $p_V \approx p_{sat}$, i.e. $p_C \approx p_E$ (in Figure 3 for a pure liquid). Thus, boiling involves the phase change of a metastable liquid at a pressure lower than the saturated pressure (point A at $T_L$ and $p_L$ in Figure 3) into a stable vapor phase (point C, usually approximated by E at $T_L$ and $p_{sat}>p_L$).

The superheat temperature $\Delta T=T_L-T_{sat}$ required for the nucleus to grow spontaneously is obtained from the pressure difference in eq. (1) , using a linear approximation of the P-T saturation curve [52] as

$$\Delta T = \frac{dT}{dP}\frac{2\sigma}{r^*}\,. \tag{2}$$

The incipience of boiling can be determined by considering the evolution of vapor nuclei. Nuclei occur naturally in the bulk liquid by density fluctuations, and may either collapse or grow depending on their Gibbs free energy [5]

$$G_{\mathrm{hom}} = 4\pi r^2\sigma\left(1-\frac{2r}{3r^*}\right). \tag{3}$$

The Gibbs free energy corresponds to the thermodynamic work required for the creation of a nucleus with radius $r$ in the bulk fluid. By the second principle of thermodynamics, the nuclei that will grow spontaneously are the ones with $\partial G_{\mathrm{hom}}/\partial r \le 0$. From Equation (3), $G_{hom}$ is maximum at $r^*$, and only the nuclei with $r>r^*$ will grow spontaneously [53]. In engineering systems, the incipience of boiling corresponds to the production of stable vapor nuclei at a rate $J$ of [$10^3$-$10^7$] $cm^{-3}s^{-1}$, with J depending [5] exponentially on the ratio of the free energy of the equilibrium vapor nucleus, over the Boltzmann constant $k_B$ times the system temperature:

$$J \sim C\exp\left(\frac{-G_{\mathrm{hom}}(r^*)}{k_B T_L}\right), \tag{4}$$

with $C$ being a function of the liquid properties. For water at atmospheric pressure, a nucleation rate J = $10^7$ $cm^{-3}s^{-1}$ corresponds to a superheat temperature equal to 204°C, a pressure difference larger than 8atm, and vapor nuclei with sizes O(30nm). These temperature and pressure values are much larger than experimental values on technical surfaces, where the onset of nucleate boiling typically occurs at a superheat of 6-10°C [54, 55]. Homogeneous nucleation appears therefore to play a weak role in technical boiling.

Indeed, it is commonly assumed that the solid-fluid interface facilitates the nucleation process. This *'heterogeneous nucleation'* process is described, for atomically smooth surfaces, by a reduction of the free energy when nuclei are created on a surface [56]:

$$G_{het} = \phi\, G_{\hom}\text{, with}\, \phi = \frac{2 + 2\cos\theta + \cos\theta \sin^2\theta}{4} \quad \text{and} \quad J \sim C^{2/3} \exp\left(\frac{-G_{het}(r^*)}{k_B T_L}\right) \qquad (5)$$

Here, $\theta$ is the wetting angle, as defined in section 2.5, and heterogeneous nucleation is observed for $J=O(10^{10}\ m^{-2}\ s^{-1})$ [52]. For superhydrophobic surfaces, $\phi$ tends exponentially toward 0 so that nucleation occurs at negligible superheat; for metallic surfaces with $\theta \approx 30°$, $\phi$ tends toward 1, negating the thermodynamic advantage of nucleating on a surface rather than in the bulk. In fact, the superheat estimated using the above two equations (ΔT≈188°C for smooth Teflon with $\theta$ =120°, or 211°C for smooth metal with $\theta = 30^o$) is again much higher than the 6-10°C superheat observed with water on real surfaces.

In fact, surfaces typically facilitate nucleation because of their natural pits and cavities. Gas or vapor entrapped in these geometries enhances the process of heterogeneous nucleation. Rigorous analysis of how a complex geometry enhances nucleation is possible and presented in section 3.1, it is however a challenging task given the inherent randomness in surface texture and chemistry. As a result, experiments are typically conducted to measure the rate of nucleation or the spatial density of active nucleation sites *n*. Qi and Klausner [57] describe the influence of the material and surface texture on the density of nucleation sites. Wang and Dhir [42] used optical microscopy to measure distributions of cavity sizes on surfaces with different wettabilities (θ=90°, 35° and 18°). They obtained a distribution of cavities of the form $n_s \propto D^{-m}$ with 2<*m*<5.4, depending on the diameter of the cavity mouth *D*. They also found the number of active nucleation sites *n* to be proportional to $(1-\cos\theta)D^{-6}$. Recently, Betz *et al.* [58] measured $n$ as a function of $\Delta T$ for a wide range of wetting angles (Figure 5a). They found that nucleation occurred on a superhydrophobic surface at superheats that were two orders of magnitude lower than those on a smooth hydrophilic surface.

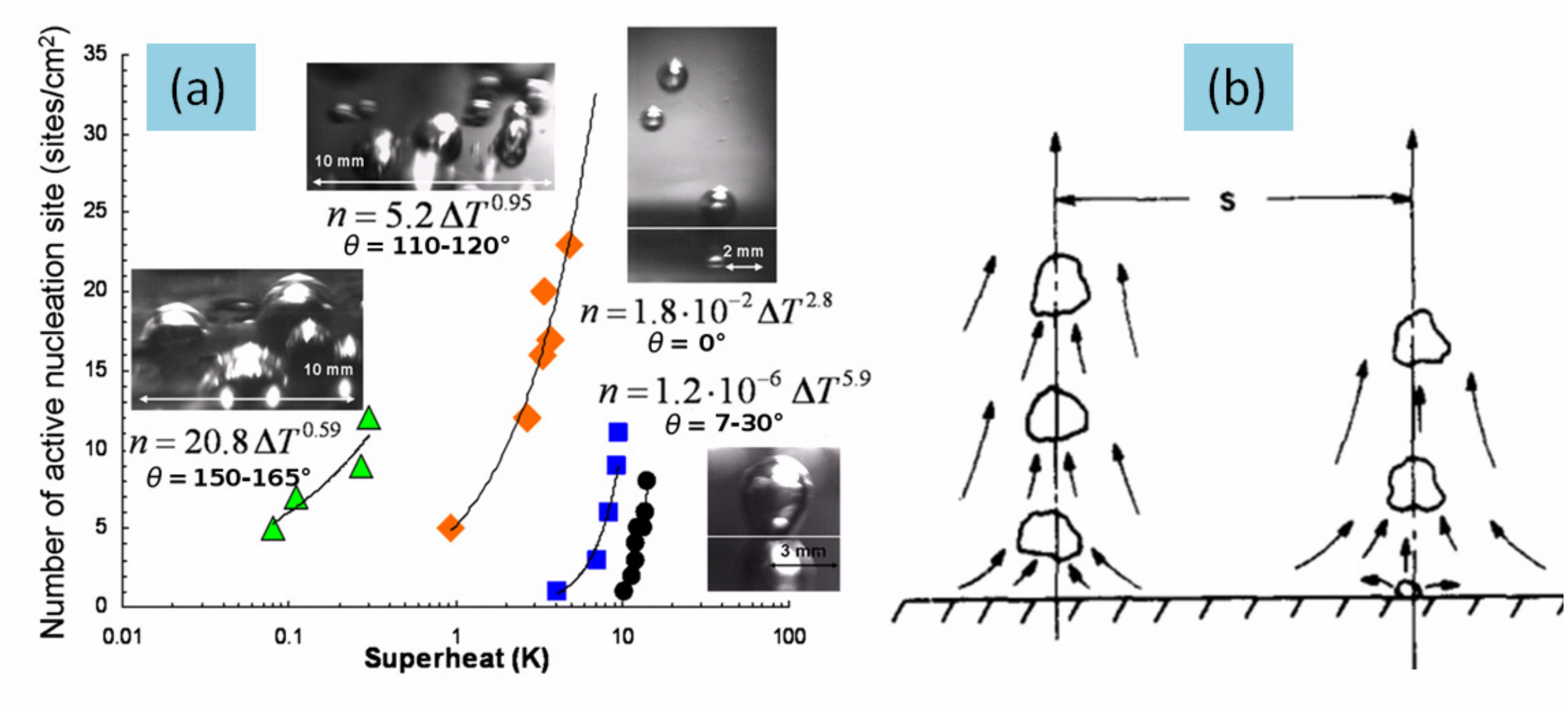

**Figure 5. Boiling occurs over a wide range of length scales. Plot (a) shows how nucleation is influenced by the surface microtexture and chemistry, that is the wettability[58], with $\theta$ the wetting angle as defined in section 2.5. In nucleate boiling (b), the length scales [59] range from μm (the thermal boundary layer), to mm (the bubble diameter), to cm (the pitch *s* between bubble columns). Figure (a) is reprinted from [58], Copyright 2013, and (b) is reprinted from [59], Copyright 1962, both with permission from Elsevier.**

For *real surfaces*, a general trend similar to the one obtained on *ideal surfaces* is illustrated in Figure 5a: *the lower the surface wettability, the lower superheat needed for nucleation.*

### *Nucleate boiling*

At higher superheat, sufficient to sustain the quasi-steady regime of nucleate boiling (see Figure **2**), the heat and mass transfer is complex and transient, involving (i) convective heat transport driven by buoyancy and moving bubbles [60, 61] [40], (ii) evaporation at the surface and wetting line of the growing bubbles [62, 63], and (iii) motion of the wetting line in the presence of fast evaporation [8, 64]. Numerous models have been proposed to describe nucleate boiling [65]. For instance, Figure 5b pictures the inverted stagnation flow model (b) of Tien[59], widely used to estimate the *HTC* in nucleate boiling. It assumes a similarity between the hydrodynamic flow of a bubble column rising from a heated surface and the stagnation flow against a wall, and illustrates the wide range of length scales present in boiling.

Most models of nucleate pool boiling describe the heat flux as a power law of the superheat $\Delta T$ and the density of active nucleation sites $n$ [52]:

$$q'' \propto \Delta T^{A}\, n^{B}, \tag{6}$$

with $1 \leq A \leq 1.8$ and $0.3 \leq B \leq 0.5$. A numerical study of Abarajith and Dhir [66] investigated the influence of the wetting angle on the growth and departure of a single bubble. The simulation method was based on the finite-difference method with interface reconstruction by the level set method. The contact angle was varied from 1° to 90° by changing the Hamaker constant, and HTC was found to increase with increasing contact angles for moderate levels of superheat.

When the heat flux increases, transition from nucleate to film boiling occurs, and radiative heat transport becomes the dominant transport mechanism. The *CHF* is the maximum heat flux reached at this transition, and its mechanism has been considerably investigated because of its industrial relevance [67]. For hydrophilic systems, the model by Zuber[46] considers that at high heat flux, single bubbles coalesce to form vapor columns. A velocity shear between the bulk liquid and these columns is induced by buoyancy, and CHF occurs from the resulting Helmholtz instability that merges columns into a vapor layer, insulating the heated surface from the liquid [52]. Lienhard and Dhir [68] refined this model, assuming that the pitch between the columns equals the wavelength of Rayleigh-Taylor instabilities. They obtained the following equation for the heat flux at CHF:

$$\mathrm{q''}_{\mathrm{CHF}} = C_L \rho_V h_{LV} \left[\sigma g(\rho_L - \rho_V)/\rho_V{}^{2}\right]^{1/4}, \tag{7}$$

where $C_L$ is a constant equal to 0.149 [68] or 0.131[46].

A second model by Haramura and Katto [69] assumed that the Helmholtz instability along large vapor bubbles results in the formation of a stable liquid microlayer between the heated surface and the bubbles.

Small Helmholtz-stable vapor jets pierce the liquid microlayer, connecting the nucleation sites with the bubbles. CHF occurs when the microlayer has totally evaporated. Experimental studies with water were found in very good agreement with this model [52].

The two hydrodynamic models above assume a very wettable surface. The effect of wettability on CHF have been studied experimentally in [50, 70], establishing that hydrophobic surfaces have lower CHF than hydrophilic surfaces.

The effect of wettability on CHF was modeled analytically by Kandlikar [71], who proposed that CHF occurs when the momentum flux caused by evaporation at the contact line overcomes gravity and surface tension forces, creating a vapor blanket. This can be modeled replacing the constant $C_L$ by a function of θ in Eq. (7). Dhir and Liaw [72] developed a computational model valid from nucleate to transition boiling, assuming that the energy from the superheated wall is transferred by conduction into the liquid through the microlayer that surrounds pre-existing stems. Their results estimated CHF for mostly non-wetting surface, with $\theta$ as a parameter. Finally, Theofanous [48] suggests that "CHF is controlled by the microhydrodynamics and rupture of an extended liquid microlayer, sitting and vaporizing autonomously on the heater surface," a process which is influenced by surface wettability.

From this body of experiments and theories, a dilemma appears given the above considerations on nucleation and critical heat flux. On the one hand, HTCs at low heat flux are enhanced on surfaces with low wettability. On the other hand, HTCs at high heat fluxes are enhanced on surfaces with high wettability, and so is the CHF.

## 1.2. Characteristics of optimum surfaces for condensation heat transfer

Like boiling, condensation is a multiscale and multiphase phenomenon. In technical applications, condensation refers to the phase change of a metastable vapor (point B in **Figure 3**) into a stable liquid (D).

Condensation has been studied on typical engineering geometries, along plates and channels, or inside channels as reviewed in [73]. For the sake of simplicity, this review is focused on external condensation, which may occur either as a liquid film or in the form of individual drops, see Table 1. *Dropwise condensation* (DwC) is facilitated for fluids with high surface tension or on low-energy surfaces [74], and occurs via the quasi-steady process described in Figure 2. For many engineering applications, the condensate is water which tends to wet metal surfaces, and the condensing droplets eventually end up forming a film on the surface, leading to *film condensation* (FC). By its geometry and multiphase structure, DwC is a process analog to convective boiling. DwC typically occurs in dehumidification or ammonia-based processes, while FC is prevalent in most other technical applications.

When the affinity between the condensing liquid and the solid surface is high, droplets of condensate *nucleate* on myriads of nucleation sites (nanoscale surface defects) on the cooled substrate. The droplets grow by sustained condensation caused by continuous cooling; droplet growth also occurs by impact and coalescence during the sliding of drops caused by gravity or by coalescence induced by droplet growth [75]. Heat transfer takes place from the gas phase to the substrate through the condensing droplets (except on areas that are not covered with droplets, where the heat transfer is direct). In DwC, progressive coalescence exposes new domains on the substrate for fresh nucleation and growth. DwC has received much attention from the heat transfer community due to its potential to produce heat transfer coefficients (HTC) that are an order of magnitude larger than in the film condensation mode [76]. Despite the initial promise for high performance, and extensive research over several decades -barring a hiatus in the '90s- achievement of sustained DwC has remained elusive till this date. The DwC process, its performance and technical applications have been extensively reviewed in [74], which also describes a theoretical prediction of the HTC. Several groups have made notable contributions to the theoretical and experimental characterization of DwC and the pertinent heat transfer correlations since its first report [77].

The mechanism responsible for DwC is subject to debate [52], be it either drops nucleating and growing on a dry surface [78], or drops resulting from the breakup of an unstable thin film [79]. The process of heterogeneous nucleation relies on reducing the Gibbs free energy associated with the creation of solid nuclei on the condensing surface. As in the case of boiling in section 1.1, the Gibbs free energy for heterogeneous nucleation of condensate also depends on the homogeneous free energy and the liquid contact angle (in its vapor), so that

$$\Delta G_{het} = \Delta G_{hom}\phi \text{, with } \phi = 1/4\left(2 - 2\cos\theta - \cos\theta \sin^2\theta\right). \quad (8)$$

The factor $\phi$ in Eq. (8)) is the analog to that in Eq. (5), and Eq. (8) implies the exact opposite trend as in boiling, that is, heterogeneous nucleation in condensation is easier on a hydrophilic surface rather than on a hydrophobic one.

Theoretical models [80-82] to predict DwC heat transfer rates typically consider the heat transfer across a single hemispherical condensing droplet as shown in Figure 6a. The temperature difference $\Delta T = \left(T_{\upsilon\infty} - T_{cool}\right)$ that drives the heat flux is partly offset by the subcooling required to condense on a convex liquid droplet of radius *r*. The subcooling can be approximated as $\Delta T \cong 2T_{sat}\sigma / \rho h_{fg} r$, with $h_{fg}$ the latent heat, combining Eq. (2) and the Clausius-Clapeyron equation, and the same equation can be used to estimate the minimum viable droplet radius for a given subcooling $\Delta T$, see Figure 6b.

Resistances to DwC heat transfer are due to thermal conduction ($R_{liq}$) through the liquid drop and mass transfer at the liquid-vapor interface ($R_{int}$)[52]. The substrate thermal resistance ($R_{sub}$) may also play a role due to droplet *recalescence*, i.e., the rejection of the latent heat of condensation [83]. On an actual condensing substrate, droplets of different sizes co-exist: the ratio of the largest drops (~1mm) to the smallest drops (1nm) being typically $10^6$. For inclined surfaces, the largest size $r_{max}$ at which the droplet slides by gravity is found to increase with increasing liquid surface tension and contact angle hysteresis, and decreases with increasing equilibrium contact angle and tilt angle [84].

The determination of the overall condensation heat transfer coefficient across the condenser surface requires integrating the heat flux at each droplet base (as obtained from the thermal resistance model described in **Figure 6**a over the entire spectrum of droplet sizes ($r_{min} < r < r_{max}$) and estimating the driving degree of subcooling $\Delta T = \left(T_{\upsilon\infty} - T_{cool}\right)$. A closed form solution of this model does not exist, and even numerical solutions are challenging due to the large range of droplet sizes involved, see Figure 6b. Measurements show that the drop-size distribution follows a statistically time-invariant pattern. For smaller droplets with $r_{min} < r < r_E$, where $r_E = (4n)^{-1/2}$ with the nucleation site density $n$ on the order of $10^{13}$ to $10^{15}$ $m^{-2}$ [85], the distribution of drop sizes can be obtained from a population balance [86]. For larger droplets with $r_E < r < r_{max}$, the fractional surface area covered by a droplet with base radius greater than $r$ follows the distribution in [75, 86, 87]. Substantial effort has been expended in numerical simulations of droplet nucleation [88] and drop growth and coalescence [87].

Besides the thermophysical properties of the fluid and the solid substrate, the HTC in DwC also depend on other key parameters like degree of subcooling, heat flux, maximum droplet radius, presence of non-condensable gases, and gas phase pressure. The HTC increases with the degree of subcooling or the surface heat flux due to enhanced creation of active nucleation sites [89]. The droplet size also has an impact on HTC. Bonner [90] showed that for droplet radii less than a few hundred nanometers, the overall heat transfer is limited by $R_{int}$, while for larger droplets, the influence of $R_{liq}$ dominates (see **Figure 6**b). Effective DwC technologies, therefore, should typically aim at removing the droplets from the surface before they grow too large, preferably at an order of magnitude lower than the capillary length of the liquid [14]. Interestingly, the HTC is severely reduced in the presence of even minute amounts of a non-condensable gas [74, 91]: HTCs are reduced by up to 30% for DwC [92] and 50% for FC [93] with a concentration of noncondensable gas of 0.5% by volume. This illustrates that noncondensable gases influence the nucleation process in condensation, as they also do in boiling. Besides, the presence of air adds a diffusion resistance ($R_{bl}$) to steam-air boundary layers [93] (see **Figure 6**a) and alters the relative contribution of latent and sensible heat transfer [92]. Most importantly, several industrial applications involve condensation of binary fluids, where Marangoni stresses promote similar structures to dropwise condensation [94, 95]. Vapor-side pressure influences the HTC primarily through the $R_{int}$. Tanasawa [96] reported that for water vapor DwC, the interface heat transfer coefficient varied from 0.383 $MWm^{-2}K^{-1}$ at 0.01 atm to 15.7 MW $m^{-2}$ $K^{-1}$ at 1 atm.

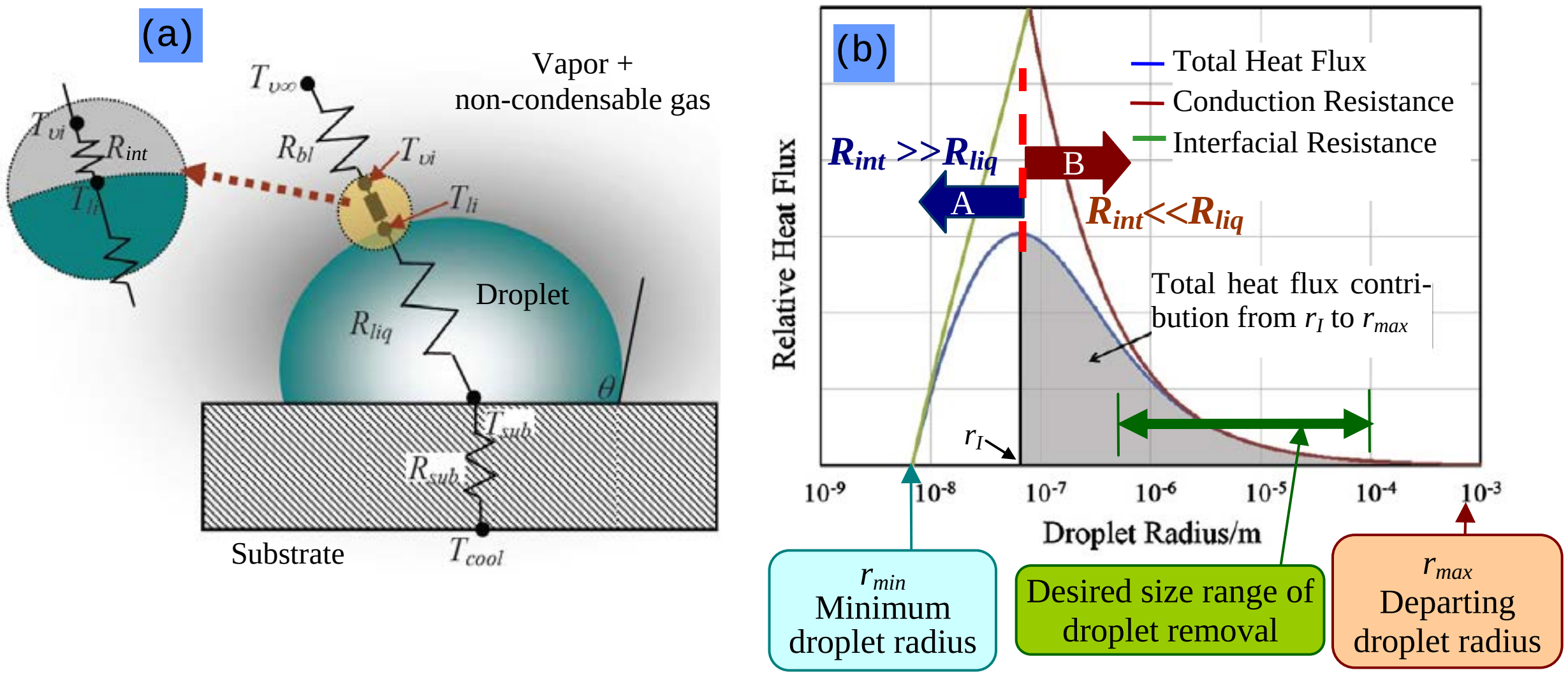

**Figure 6: Thermal aspects and length scales in dropwise condensation. (a) Schematic representation of thermal resistances associated with dropwise condensation (DwC) [$R_{bl}$: Vapor boundary layer thermal resistance (present if the vapor phase has noncondensable gases), $R_{int}$: Vapor-liquid interface thermal resistance, $R_{liq}$: Liquid conductive thermal resistance, $R_{sub}$: Substrate thermal resistance. Temperature legends $T_{\upsilon\infty}$: far stream vapor; $T_{\upsilon i}$: interface (vapor side); $T_{li}$: interface (liquid side); $T_{sub}$: substrate (droplet base); $T_{cool}$: subcooled surface]. (b) Influence of the drop size on the relative heat flux contributions. Reprinted from [90], Copyright 2013, with permission from Elsevier.**

Forces responsible for the removal of the condensate are different in nature and weaker in condensation than in boiling, see Table 1. This removal typically occurs by shear or gravity forces; rather than shear and buoyancy in boiling.

When the solid surface removes less condensate than that condensing on the surface, a thick film forms on the surface, and film condensation (FC) occurs. FC typically features one order of magnitude lower

HTC's than DwC [52], because (i) much of the condensing surface that is covered with the liquid film cannot offer heterogeneous nucleation sites, and (ii) the liquid film offers a considerable thermal resistance.

The transition from drop- to film-wise condensation, schematically shown in **Figure 2**, is a process similar to the boiling transition from nucleate to film boiling, and similarly, it is poorly understood. A typical value of the peak heat flux corresponding to that transition for steam is 10MW $m^{-2}$, as reported in [74]. At least two transition mechanisms have been proposed for the transition from DwC to FC [74]: (1) under high heat flux, the droplet growth rate exceeds the condensate removal rate, leading to increased coverage of condenser surface area by the condensate and merging; or (2) the density of active nucleation sites increases with increasing $\Delta T$, leading to a situation where condensates from neighboring active nucleation sites merge into a wetted patch.

Nusselt provided a simple model for HTC in film condensation, assuming laminar flow, pure vapor and negligible convection [97]

$$HTC = 0.943\left[\frac{\rho_L(\rho_L-\rho_V)gh_{fg}k_L{}^3}{\mu_L L\left(T_{\upsilon\infty}-T_{cool}\right)}\right]^{1/4}, \qquad (9)$$

with $L$, $\mu_L$ and $k_L$ standing, respectively, for the length of a vertical plate, viscosity of the liquid, and thermal conductivity of the liquid. Later, Rohsenow [98] included thermal advection effects with a latent heat term that depends on the Jakob number.

To summarize, the salient issues specific to condensation on solid substrates are (i) the design of substrates that promote heterogeneous nucleation, (ii) the efficient drainage of the film or the evacuation (removal) of the drops from the condensing surface. The first factor bears a direct influence of the surface wettability, while the second depends both on the condensing surface and the acting physical forces.

## 1.3. Characteristics of optimum surfaces for freezing and desublimation

As per the definitions related to Figure 3 and **Table 1**, there are at least eight terms describing phase change from or towards the solid phase. Solidification (freezing) commonly occurs in technological processes, such as free-form manufacturing, rapid prototyping, metals processing, microelectronic fabrication, etc. Despite the large number of technologically relevant fluids that can undergo solidification, for the sake of simplicity, this section concentrates only on aqueous systems.

During the densification and bulk-growth phase, the frost thickness and density increase, resulting in significantly lower thermal conductivity and reduced heat transfer, see Figure 7a. Eventually, the frost thickness reaches a maximum, as shown in Figure 2; the temperature of the frost surface is no longer cold enough to promote further frost growth, and an equilibrium heat flux is reached [99]. Measuring HTCs in freezing configurations poses formidable challenges, as roughness changes continuously on ice-covered surfaces. The roughness controls the boundary layer structure in the vicinity of the solid, in turn affecting the convective flow structure around the body, the resulting heat transfer and the droplet collection efficiency, which in turn, influence the ice shape [100]. The high dependence of HTC on the specific conditions of each problem prevents generalizations, and makes the development of HTC correlations difficult. The complexity of the problem is depicted schematically in which demonstrates the relevant processes involved in ice formation and accretion [100]. HTCs up to 2kW $m^{-2}$ $K^{-1}$ have been measured for ice accreted on airfoils in wet (vapor rich) and dry (vapor poor) regimes [100]. For water frosts deposited on a cooled plate from a humid air stream, HTCs in the range 20-60 W $m^{-2}$ $K^{-1}$ were reported [99]. Various correlations for the HTC under frosting conditions as related to refrigeration applications are given in the review by Iragorry et al. [101]

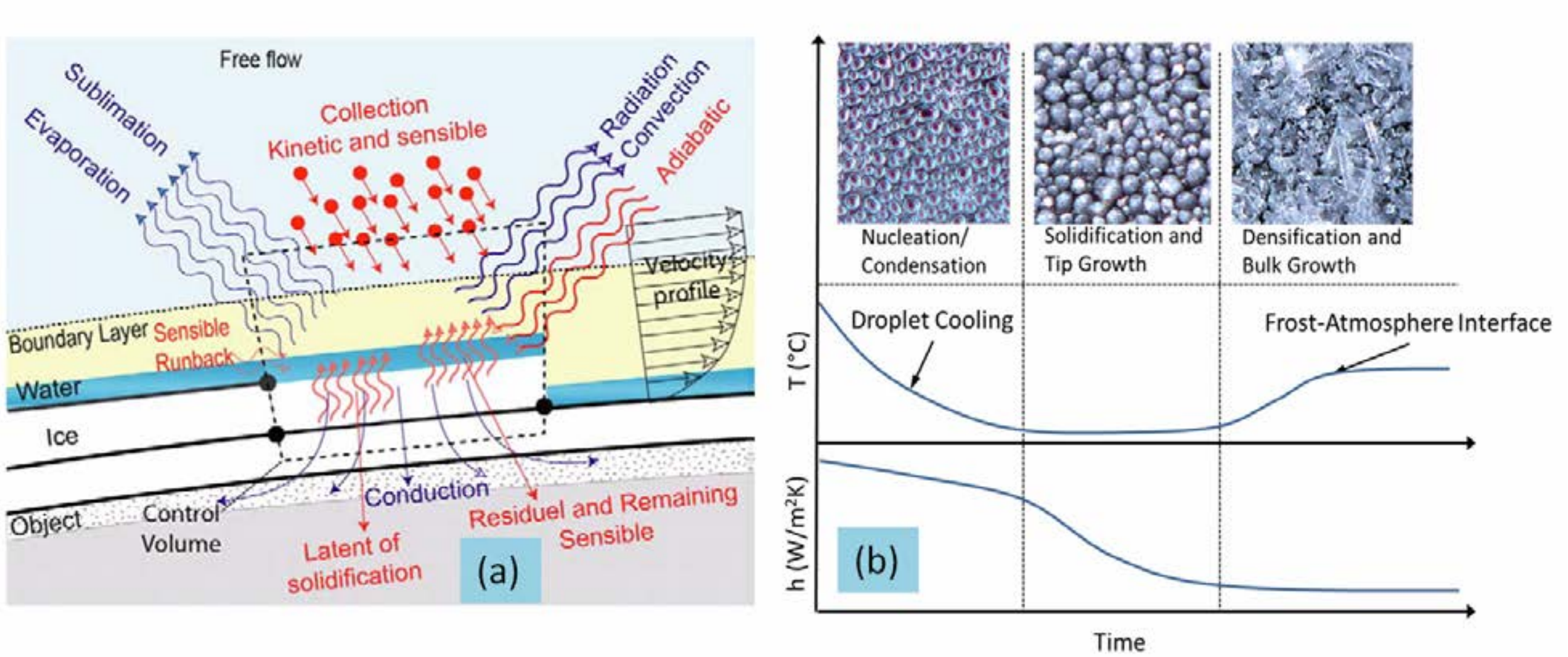

**Figure 7: Thermal aspects and spatial features in icing and frosting. (a) Schematic depiction of processes involved in ice accretion on top of a solid object exposed to a free flow in conditions conducive to freezing. Reprinted from [100], copyright 2005, with permission from Elsevier. (b) Photographs (top) of the three stages of condensation frosting on a hydrophilic surface shown with the corresponding temporal evolution of temperature and heat transfer coefficient (bottom). Pictured areas are 1mm × 1mm.**

For most technical applications, ice and frost have negative consequences and surfaces that mitigate their buildup are desired. For example, in refrigeration systems, frost can form an insulating layer which impedes heat transfer [101, 102] or increase pressure drop in ducts. Ice formation on planes can prevent the movement of ailerons and decrease aircraft safety. Ice can also adversely impact the safety and operation of wind turbines, marine vessels, and offshore oil platforms [103]. Moreover, frost and ice buildup can make systems brittle or heavy, leading to catastrophic failure. For applications where frost avoidance is desired, an ideal surface must prevent nucleation.

## Freezing/Icing

The formation of ice from a metastable supercooled liquid occurs via ice nuclei, i.e. tiny transient crystalline clusters existing in various sizes within the water phase. Continuous size fluctuations of these clusters occur due to the simultaneous addition of molecules and detachment of the same. Based upon the various thermodynamic parameters, the nucleus size may grow large enough to become stable, i.e. for growth to become more probable than decay. This designates the critical size for stability, and hence further nucleation of the solid phase. Stability of ice nuclei mainly depends upon their surface-to-volume ratio. As the size of the ice phase increases, this ratio decreases, thus making the number of molecules returning to the liquid from the ice surface far less significant compared to the number of interior molecules, hence establishing stability of the nucleus and facilitating growth. Nucleation of this type is known as *homogeneous* nucleation. There exists another type of nucleation where the nucleus forms on some pre-existing solid structure, thus making the stability of the nucleus more probable. This type of nucleation is known as *heterogeneous* nucleation and is far more common than homogeneous nucleation due to its smaller free energy barrier, as described in section 1.1. Hence, heterogeneous nucleation is sought for technical applications; it is also similar to the nucleation process discussed in the condensation section. As seen from Eq. (8), which also applies here, increasing the contact angle delays condensation, and thus ice formation [104-110]. In fact, since high contact angles decrease the contact area between the water and the cold surface, if the pressure is high enough to condense first and then freeze water, the decreased contact area will slow heat transfer and delay freezing time [105]. Additionally, low energy surfaces usually have low adhesion to water and ice [105]; coupled with engineered forced convection, these surfaces efficiently prevent ice buildup.

In addition to ice formation within the liquid water phase, ice can also form on a solid surface directly from the vapor state; this phenomenon is known as desublimation, atmospheric deposition. The term frosting is ambiguous (see section 1) and will not be used here.

**Desublimation/Frosting**

Frost is formed when humid air comes in contact with a solid surface at a temperature below the freezing temperature of water. Frost can form either by condensing then freezing or directly through solid deposition [104]. In the former, frost formation has two distinct phases, similarly to the condensation process: nucleation and growth. Before frost forms, the heat transfer is dominated by single phase convection. At the onset of nucleation, heat transfer is dominated by multiphase convection. In frosting, nucleation depends on surface structure and the availability of nucleation sites.

Several studies have been conducted on the ice/frost repellency of superhydrophobic surfaces [20, 105, 110]. However, due to the increased surface area and roughness, some superhydrophobic surfaces have shown increased frost nucleation compared to smooth hydrophobic surfaces [105, 108]. Other research has also shown that a low contact angle hysteresis prevents frost nucleation [111]. Kim et al. [111] studied liquid-infused porous surfaces, which provide a defect-free molecularly flat liquid interface, while maintaining a high contact angle. The lowest rates of frosting were observed on surfaces that had both a high contact angle and low contact angle hysteresis.

After nucleation and cooling, frost goes through two distinct growth phases, see Figure 7b. First, the frost goes through the solidification and tip growth stage, with conduction and multiphase convection dominating the heat transfer. Then, during the densification and bulk-growth stage, the temperature increases at the interface between the frost and the atmosphere and the frost layer eventually reaches an equilibrium thickness and temperature with the environment. During this period, the heat transfer is dominated by conduction. The final frost thickness depends on the surface temperature and relative humidity. The thermal conductivity of the surface has been shown to affect frost growth [112]. On low thermal conductivity surfaces, a supercooled mm-sized droplet freezes slowly, with evaporation triggered by the latent heat released upon recalescent freezing. Eventually, the gas volume around the droplet reaches saturation and vapor microdroplets coalesce to form smaller droplets that deposit around the droplet, eventually crystallizing and forming a frost halo. Experiments in [112] under controlled humidity conditions establish how a balance between heat diffusion and vapor transport determines the final expansion of the frozen condensate halo, which, in turn, controls frost formation and propagation.

The surface energy may also influence the frost growth and density [109]. Once a thin layer of frost is deposited, the surface energy of the base surface should no longer affect the growth of subsequent frost [107, 113]. However, the surface energy influences the initial shape of the crystal, therefore modifying the kinetics and changing the dendritic frost growth patterns [109, 114, 115]. Experimental evidence suggests that hydrophilic surfaces produce a higher density frost [114], while hydrophobic surfaces form more dendrites creating a looser frost structure [114].

In summary, surfaces with low wettability prevent, delay or slow down the formation of ice from a metastable, supercooled liquid. The role of roughness on frost formation is not clear at present. Also, the more wettable the surface, the denser the frost layer.

## 2. State of the art methods for micro- and nanofabrication of surfaces

Several approaches and numerous techniques are available to fabricate the surfaces of interest for phase change heat transfer. The phase change on a solid surface is affected by the energy and texture of the surface. The surface energy is of chemical nature but manifests itself physically as a contact angle. Aside from some practical issues, such as changes over time and temperature, the surface energy is a relatively simple parameter as it is inherent to each material. The surface texture, on the other hand, is geometric in nature and can have many variables, such as pitch (periodicity), spacing, depth, shape, and randomness. Another important variation for phase change heat transfer is the local control of the surface energy and texture, i.e., a surface can juxtapose regions with different chemistries and textures. These infinite possibilities of multiscale spatial arrangements leave ample room for heat transfer enhancement. Here we discuss the fabrication methods used to control the energy and texture of surfaces for phase change heat transfer.

### 2.1. Surface machining and roughening

Compared with a smooth surface, a roughened surface enhances the phase change heat transfer by increasing the net contact area for heat transfer and the number of nucleation or precipitation sites [116]. To roughen the surface, various mechanical methods have been used, such as sandpapering and abrasive blasting [117]. The surface structures obtained are relatively simple, i.e., consisting essentially of low peaks and valleys, and randomly distributed. Conventional machining has also been used to obtain more elaborate surface structures aimed specifically to enhance nucleate boiling, e.g., Figure 8a. These surface structures show a regular pattern due to the sequential nature of the sharp-tip machining processes. Chemical and electrochemical methods can produce more pronounced and complex structures, such as porous surfaces, e.g., Figure 8b. The surface structures are randomly distributed, but the roughness parameters (e.g., average pitch, porosity) can be controlled by the processing conditions, such as etchant chemistry, etching time, and current density. The roughening processes are suitable for large area and mass manufacturing, while the machining processes are less economical. Because of their high thermal conductivity, metallic surfaces are preferable for phase change heat transfer.

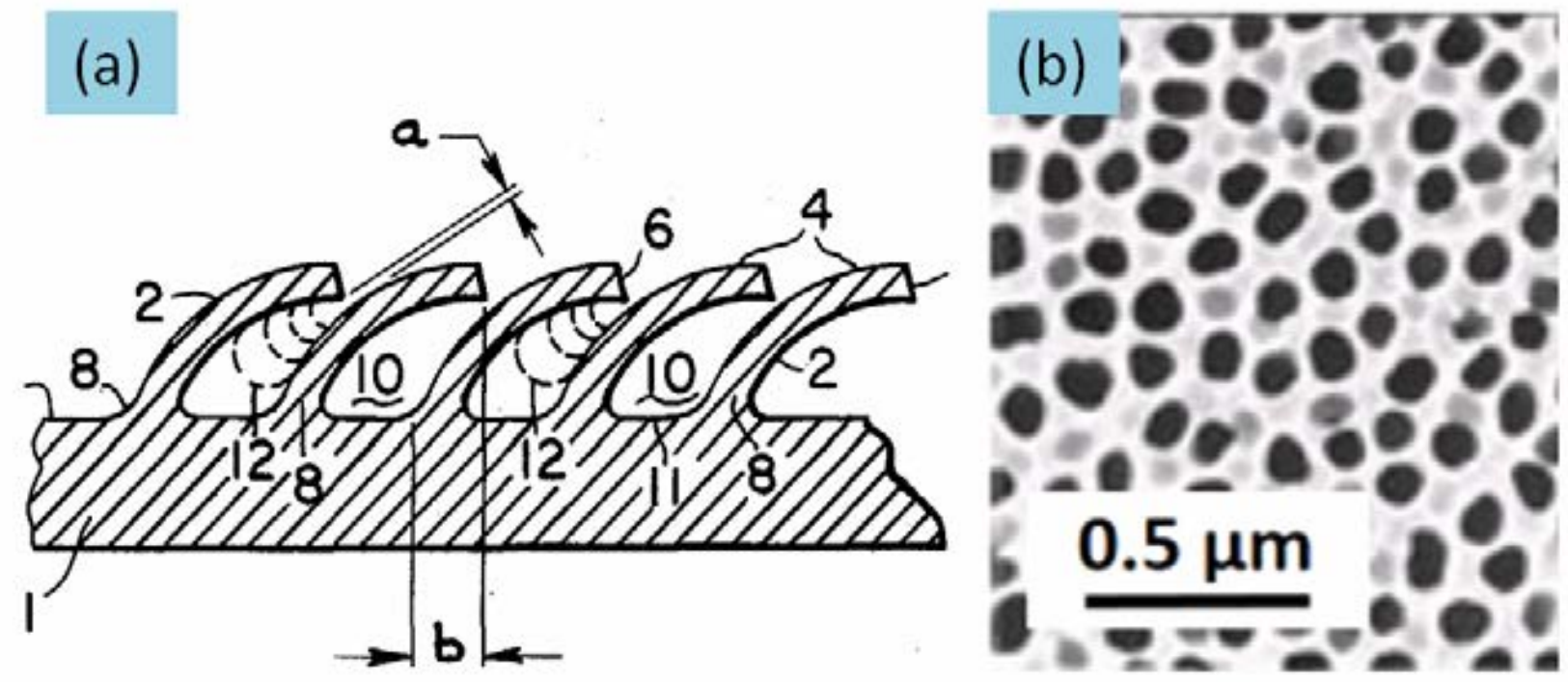

**Figure 8. Examples of surface machining and roughening: (a) a machined surface with bent fins to enhance nucleate boiling, from patent[118]; gap *a* is in the 25-125 μm range; (b) electrochemically processed porous alumina (AnodiscTM).**

## 2.2. Surface coating

The surface textures can be added on the base surface by coating. Compared with the subtractive methods of the previous section, the coated textures can be more complex in geometry and more versatile in material selection. The texture can be as complex as a three-dimensional porous network, and the material can be a metal, ceramic, polymer, or a combination of them. The coating methods are numerous, including sintering, spray coating, dip coating, plasma coating, electroplating, and gluing, just to name a few. For metal oxides, Zhou et al. [119] electrodeposited manganese oxide from an ionic liquid to grow a number of different morphologies, including nanoparticles and nanofibers, on glass substrates. Jiang et al. [120] used a hydrothermal process to grow titania nanowire arrays and porous frameworks on titanium substrate. For metals, Dhal and Erb [121] flame-sprayed metal-oxide particles on a metal substrate to form porously interconnected open-cell nucleation sites. Jiang and Malshe [122] electrosprayed cubic boron nitride (cBN) particulates with sizes smaller than 2 μm, as shown in Figure 9a, on a tungsten carbide cobalt (WC–Co) substrate. While most methods produced random structures over large areas, Kim and Bergles [123] formed a controlled porous surface by sintering a single layer of large copper particles on a layer of smaller copper particles, as shown in Figure 9b. The large pores formed between the large particles on top were intended to promote boiling nucleation, while the small doubly re-entrant cavities formed between the small particles below were to help preserve the trapped vapor during subcooling and assist the incipient nucleation when reheated. Not limited by the material, You and O'Connor [124] developed a method to glue dispersed particles into highly porous structures, as depicted in Figure 9c. Because a polymeric material glues the particles together, a wide range of materials can be used as the constituent particles. The resulting porous surface was used to enhance boiling heat transfer for cooling electronics by immersion.

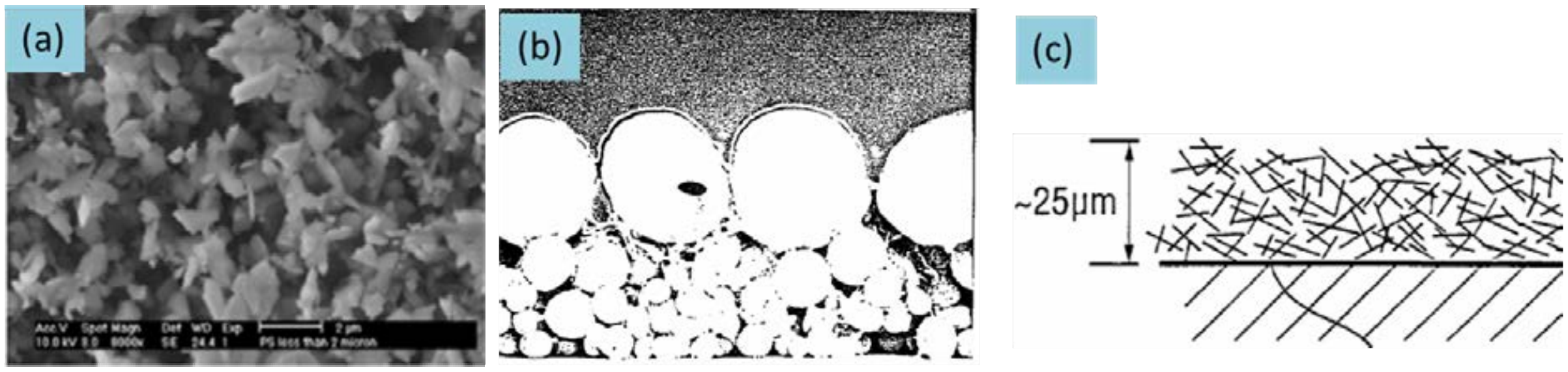

**Figure 9: Examples of surface coatings: (a) SEM image of an electrospray boron nitride coating, reprinted from [122], Copyright 2011, with permission from Elsevier; (b) Cross-sectional SEM picture of a controlled porous surface fabricated by sintering copper particles of increasing diameters on a copper substrate to study incipient boiling, reprinted from [123], permission requested from Hemisphere Publishing. The large particles on top are ~250 μm in diameter; (c) Schematic of a porous surface obtained by gluing particles for electronics cooling, from US patent [124].**

The surface structures made with coating techniques tend to be complex and randomly distributed in size and shape. The size of the structures and the thickness of the coating can vary from a few nanometers to hundreds of micrometers.

## 2.3. Lithographic fabrication

In contrast to the surface roughening and coating processes, which produce mostly a random distribution of certain surface structures and sizes, lithographic processes can produce surface textures of exact shapes and sizes. The lithographic patterns can be at the micrometer scale using standard photolithography or at the nanometer scale using electron-beam lithography. There are other related fabrication methods, such as interference lithography, but they usually have additional limitations. The patterns may be formed by a lithographic process of photoresist coated on the surface or by direct printing using the so-called soft lithography. Using the well-defined patterns as an etching mask, the substrate is carved out by various sophisticated etching processes to create surface textures with depths or heights up to hundreds of micrometer. The etching is usually not performed after soft lithography, which is most convenient when the surface texture is merely a pattern of a thin layer including a chemical coating. Requiring a clean room and expensive equipment, lithographic fabrication is not suitable for large-area applications. However, it provides a unique capability to study the phase change on textured surfaces with sizes and shapes that correspond exactly to the desired designs. Available are a variety of powerful micromachining techniques developed in the field of micro electro mechanical systems (MEMS) as well as conventional microfabrication techniques for integrated circuit (IC) manufacturing. The two main shortcomings of lithography are the cost of fabrication and the limited area (i.e., sample size). If acceptable, soft lithography [125] can reduce the cost significantly and interference lithography [126] can cover a larger area (possible even several meter [127]).While silicon is the material of choice for lithographic fabrication, other materials such as glass (Figure 10a), polymers and metals (Figure 10b) can also be processed with reduced capabilities (e.g., limited etching depth for glass) or with additional steps (e.g., molding step for metals).

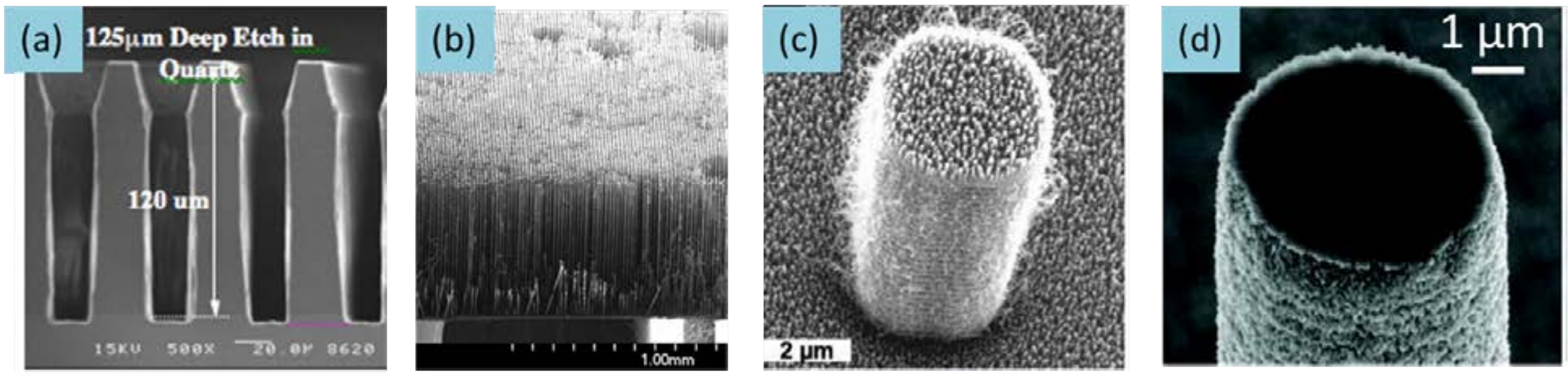

**Figure 10. Examples of photolithographic processes to create large aspect ratio geometries and multiscale features on solid surfaces. (a) Cross-sectional SEM picture of glass surface textures fabricated by photolithography and deep reactive ion etching on a glass (quartz) substrate; (b) SEM picture of a dense array of 600 µm-tall freestanding Ni posts (diameter = 5 μm; pitch = 14 μm; height-to-diameter aspect ratio = 120) fabricated by electroplating into deep pores formed by photoelectrochemical etching on silicon wafer Copyright 2011, IEEE. Reprinted with permission from [128]. (c-d) Microscale posts covered with carbon nanotubes. Figures are reprinted from [129] with the permission of the American Chemical Society.**

## 2.4. Multiscale surface textures

Surface textures featuring multiple length scales are called multiscale or hierarchical textures [130]; several such textures are reviewed by Lu and Kandlikar [131]. For example, a surface of interest may have millimeter scale features, whose surfaces are covered with microstructures, whose surfaces are covered with nanostructures, as shown in Figure 10c,d. After fabricating the post structure by silicon deep reactive ion etching, nanoscale roughness was formed by depositing carbon nanotubes to study dropwise condensation [15]. A similar result has been obtained by a galvanic displacement process [129] but with an additional level of sophistication of nanostructuring either all surfaces or all except the top surface of the microstructures. If the nanoscale roughness was desired only on the side surfaces of the posts, a masking step was added to protect the top surfaces of the posts from the roughening step. Recently, biotemplates have been used to fabricate multiscale surfaces [132].

As a summary, Table 2 reviews all the processing methods by category and mentions the distribution, size and shape of the patterns as well as the suitable material and scale of application.

| Category | Processing type | Fabrication method | Distribution of the patterns | Shapes (and size) | Suitable scale of the application | Material |
|---|---|---|---|---|---|---|
| Roughening or machining | Subtractive | Sandpapering, sandblasting | Random | Random (μm) | Large (mass production) | Copper [117] |
| | | Machining | Regular | Controlled (μm to mm) | Medium | Metal [118] |
| | | Chemical, electrochemical | Random | Controlled (10-200 nm) | Medium | Alumina (Anodisc™) |
| Coating | Additive | Sintering | Random | Complex (μm to mm) | Large | Oxide particles [121]; Copper particles on copper[123] |
| | | Electrospraying | Random | Random & complex | Medium | cBN particles on XC-Co [122] |
| | | Gluing | Random | Controlled (1 μm to ~100 μm) | Large | Metal or ceramic particles [124] |
| Lithographic fabrication | Additive or subtractive | Photolithography, e-beam lithography | Regular | Controlled (nm to ~100 μm) | Small | Si pillars on Si [133]; metal pillars on Si [128] |
| | | Interference lithography | Regular | Controlled (100 nm to ~1 μm) | Medium | Si pillars on Si[134]; Ti, Al and Au pillars on glass [135] |

**Table 2: Microfabrication techniques available for phase change heat transfer applications.**

## 2.5. Wettability engineering

Wettability describes the spreading of a liquid on a surface. Wettability engineering is important in e.g. the automotive industry for repelling water from windshields or paints [136], preventing fouling [137] and reducing friction in moving parts [138]; in the marine industry to prevent fouling [139]; in the electric power industry to prevent fouling on solar cells [140]; in biology to prevent bacterial contamination [141, 142], and in the electronic [143] and chemical industries [144].

The wettability of a surface can be modified by engineering its texture and (or) its chemistry. A first characterization of wettability is the static wetting of liquid droplets on flat substrates [145, 146 147]. Figure 11 shows the possible wetting regimes of sessile droplets on a substrate, where $\theta_E$ is Young's contact angle, and the spreading parameter $S$ is defined as

$$S = \sigma_{\mathrm{SV}} - \left(\sigma_{\mathrm{SL}} - \sigma_{\mathrm{LV}}\right), \qquad (10)$$

where $\sigma_{SV}$, $\sigma_{SL}$, and $\sigma_{LV}$ are the surface tension at the solid/vapor, solid/liquid, and liquid/vapor interfaces, respectively [146]. When $S > 0$, a liquid droplet will completely (or perfectly) wet the substrate (*i.e.*, $\theta_{\mathrm{E}}$=0°) to form a liquid film with nanoscale thickness. This property is typically obtained on surfaces that have high values of $\sigma_{\mathrm{SV}}$ (≈500-5,000 mN $m^{-1}$), *e.g.*, metals [146]. When $S < 0$, a liquid droplet will partially wet the substrate forming a non-zero contact angle (*i.e.*, $\theta_{\mathrm{E}}$> 0°). When $\theta_{\mathrm{E}} \leq 90°$ a substrate is called *wetting* or—in the special case of water — hydrophilic. When $\theta_{\mathrm{E}} > 90°$ a substrate is called *non-wetting* or — in the special case of water — hydrophobic. This hydrophobic regime is usually observed on substrates which have lower values of $\sigma_{\mathrm{SV}}$ ($\sigma_{\mathrm{SV}}$≈10-50 mN $m^{-1}$), *e.g.*, plastics [146]. By balancing the horizontal projections of capillary forces at the triple contact line, Young [148] and Dupré thereafter [149] obtained

$$\cos\theta_{\mathrm{E}} = \left[\sigma_{\mathrm{SV}} - \sigma_{\mathrm{SL}}\right] / \sigma_{\mathrm{LV}} \qquad (11)$$

From Eq.(11), $\theta_{\mathrm{E}}$ is a function of the liquid, solid and vapor properties, so that a surface which is *non-wetting* to a liquid may be *wetting* to another.

Thereafter, Wenzel [150] and Cassie and Baxter [151] described how surface texture influences wettability, see Figure 11. Wenzel assumed that the droplet completely wets the rough surface whereas Cassie and Baxter stated that air was trapped in the rough solid-liquid interface. Later [152], it was shown that both states could be obtained, depending on the scale of the roughness. Wetting states intermediate between the Cassie-Baxter and Wenzel states have also been reported [153-155]. The apparent contact angle ($\theta$*) of a liquid droplet in a Wenzel wetting state can be defined as

$$\cos\theta^* = r\cos\theta_{\mathrm{E}}, \qquad (12)$$

with $r$ being the roughness, defined as the ratio of the true surface area to the projected surface area. For the hydrophobic case ($\theta_{\mathrm{E}} > 90°$), the Cassie-Baxter case is preferred to the Wenzel state if $\theta_E$ is smaller than a critical angle $\theta_C$, with $\cos\theta_C = (\Phi_S + 1)/(r - \Phi_S)$, where $\Phi_S$ is the solid-liquid fractional surface area, and the value of the apparent Cassie-Baxter contact angle is [146]

$$\cos\theta^* = -1 + \Phi_S(\cos\theta_E + 1) \ . \tag{13}$$

For the hydrophilic case ($\theta_E \leq 90°$), the Cassie-Baxter case is preferred to the Wenzel state if $\theta_E > \theta_C$, with $\cos\theta_C = (1 - \Phi_S)/(r - \Phi_S)$, and the value of the apparent Cassie-Baxter contact angle is

$$\cos\theta^* = 1 + \Phi_S(\cos\theta_E - 1). \tag{14}$$

For hydrophobic surfaces, the more enhanced the surface texture is (*i.e.*, high $r$, low $\Phi$), the lower $\theta_C$; therefore, the Cassie-Baxter wetting state becomes more probable. When a droplet transitions to a Cassie-Baxter state, and $\theta^*>150°$, the solid surface is said to be *superhydrophobic*. In general, roughness tends to increase $\theta^*$ on a hydrophobic surface and the likelihood of a superhydrophobic wetting state ($\theta^* \to 180°$). Reciprocally, the superhydrophilic wetting state ($\theta^* \to 0°$) is also observed by roughening an originally hydrophilic substrate [146]. As it relates to phase change heat transfer, the Cassie-Baxter wetting state plays an important role in enhancing dropwise condensation behavior [156] and stabilizing Leidenfrost vapor layers [157].

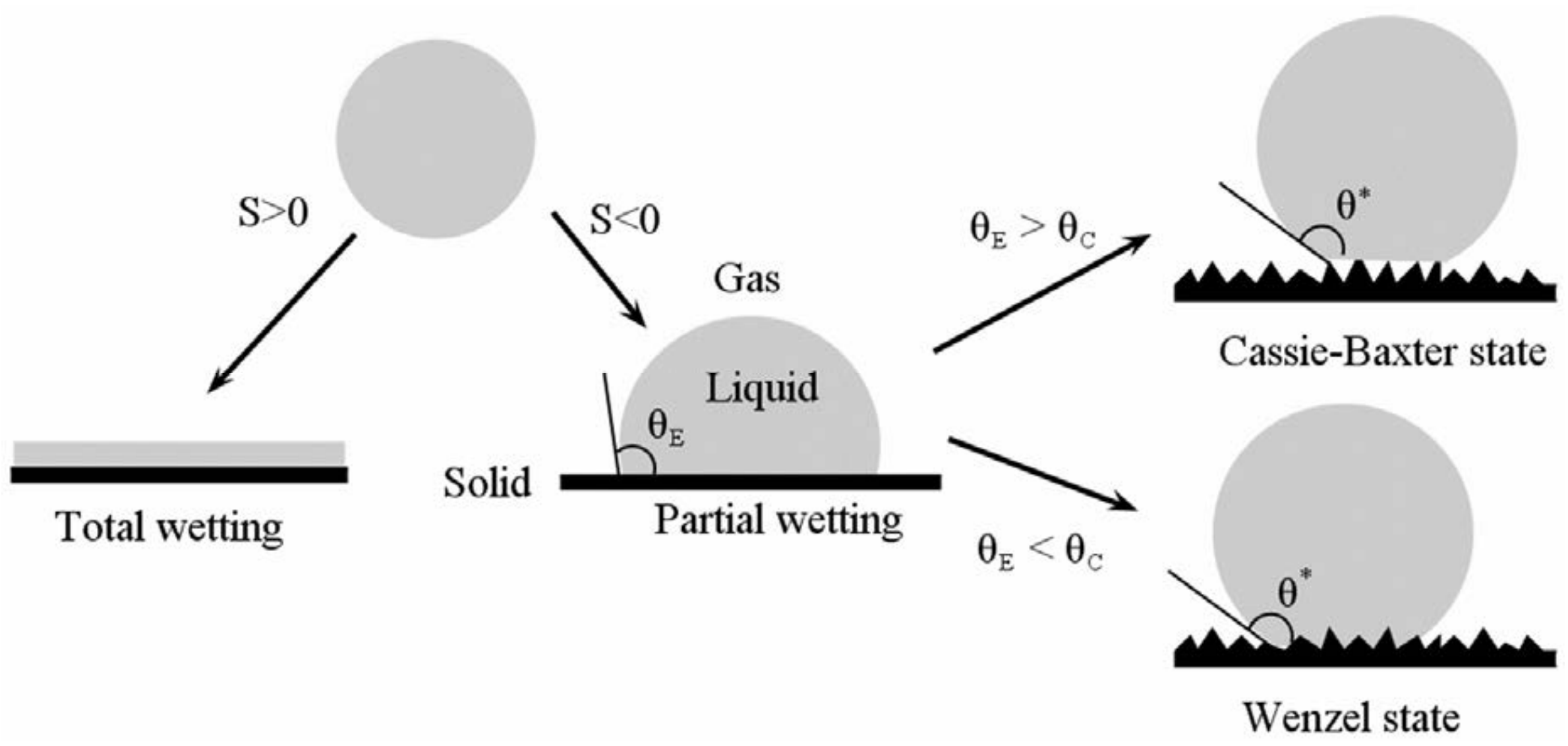

**Figure 11. Possible wetting states of a sessile liquid droplet on a flat substrate: total wetting (S > 0) and partial wetting (S<0) on an originally smooth hydrophobic surface. On the roughened hydrophobic states, two wetting states can be observed: Wenzel state, the droplet wet the substrate, whereas in Cassie-Baxter state, small gas pockets are located under the droplet, which is thus suspended in the so-called 'fakir' state.**

Additionally to these considerations, it is important to add that roughness influences both the static contact angle and the so-called contact angle hysteresis (difference between the advancing and receding values). In the classic experiment of Johnson and Dettre [158], advancing and receding contact angles were measured as a function of the roughness of a wax sample with constant chemical properties. The main outcome of this experiment was that an increase in the roughness caused an increase of the advancing contact angle and a decrease of the receding angle (increase of the hysteresis angle) until a saturation value. At this point, the hysteresis drastically decreased and the receding angle became closer to the advancing contact angle. Texturing the surface appears consequently as an efficient strategy to change the wetting behavior [146]. It is interesting to note that droplets on superhydrophobic surfaces in the Cassie-Baxter state have a low contact angle hysteresis. In many cases, though not all (see *rose petal effect*), superhydrophobic surfaces characteristically have low contact angle hysteresis; this leads to drops having low sliding angles and ultimately self-cleaning behavior (sliding angle < 10°; *lotus effect*). A small hysteresis has been proposed to be a necessary condition to the superhydrophobicity of a surface [147]. Tilting the surface will result in a rolling motion of the droplet, as shown experimentally [159, 160]. The sliding or roll-off angle has significance in condensation processes as it plays a role in removing (also called shedding) droplets of condensate [154], see section 3.2.

Table 3 and the related discussion describe the methods utilized to alter the intrinsic wettability ($\theta_e$) of a surface in a spatially uniform manner (*i.e.*, homogeneous wettability). Surface chemistry (or surface energy) can be modified e.g. by forming oxides on the surface, if hydrophilicity is desired, or by depositing low-surface energy materials (*e.g.*, fluoropolymers, fluorosilanes) if hydrophobicity is desired. Table 3 presents a representative series of coating techniques for modifying the chemistry of a surface.[i] For hydrophilic oxides (*e.g.*, copper oxide), the chemistry and resulting wettability can be tuned by varying the degree of oxidation of the surface [161] or by UV irradiation [162, 163]. For hydrophobic surfaces, the chemistry can be controlled by choosing low-surface energy solid materials. The chemical bonds in low-surface energy solid materials are in order of decreasing surface energy: $–CH_2 > –CH_3 > –CF_2 > –CF_2H > –CF_3$ (*i.e.*, $CF_3$ is the least wettable) [164]. As an example of an original synthesis method for forming a hydrophobic surface, Hummel spray deposited poly(tetrafluoroethylene) on a metallic surface [51]. Phan et al. [165, 166] describe specific methods capable of tuning static contact angle values from 22° to 112° as well as their performance under nucleate boiling conditions. An issue with hydrophobic materials is their low thermal conductivity, typical of organic materials. Other approaches to producing hydrophobic surfaces are promoters (*i.e.*, waxes), silanes, fluorinated acids, and self-assembled monolayers (SAMs).

[i] Table 3 is by no means a comprehensive list of coating techniques. References were chosen based upon applicability to phase change heat transfer.

| Material | Processing Technique | Type | Surface Chemistry | Coating Thickness | Substrates | References |
|---|---|---|---|---|---|---|
| Metallic | MOCVD | hydrophilic | Pt | 20 nm | Metal | 165, 166 |
| Metal-oxides | Oxidation | hydrophilic | Copper oxide etc. | N/A | Copper | 161, 167 |
| Metal/Metalloid-oxides | NNBD, PECVD, MOCVD | hydrophilic | $TiO_2$, $SiO_x$, $Fe_2O_3$ | 20-100 nm | Stainless steel | 165, 166 |
| Metal/ Metalloid-organic | PECVD, AR (E)-MSIP | hydrophobic | SiOC, CrCF | 20 nm, 1-5µm | Stainless stell, copper | 165, 166, 168, 169 |
| Lanthanide-oxides | Sintering, l spray, sputtering | hydrophobic | -- | 200-350 nm | Silicon | 170 |
| Organic | Many processes | hydrophobic | $-CH_3$, $-CF_3$, etc. | 0.02-8 µm | Unrestricted | 165, 166, 168, 171 |
| n-alkanethiols | Self-assembled monolayers | hydrophobic, hydrophilic | $-CF_3$, $-CH_3$, $CH=CH_2$, -CN, -OH, $-CO_2H$ | monolayer | Gold | 172-175 |
| Wax, etc. | Promoters | hydrophobic | $-CH_3$, $-C_6H_5$ | 0.5-10 monolayers | Metal | 74, 176, 177 |
| Fluorinated acids, etc. | Langmuir-Blodgett | hydrophobic | $-CF_3$ | 1-3 monolayers | Unrestricted | 178, 179 |
| Ionic polymers/$SiO_2$ | Layer-by-Layer | hydrophobic, hydrophilic | -- | 10-40 bilayers | Nickel, Stainless steel | 180 |
| $TiO_2$ | UV Irradiation | hydrophilic | | 0.25-4.0 µm | Unrestricted | 162, 163 |
| Silanes | Chemical vapor deposition | hydrophobic | $-CF_3$ | molecular | Glass, metals | 180, 181 |
| Fluoro-POSS | Spin coating | hydrate-phobic | $-CF_3$ | -- | Steel | 182 |
| p(PFDA-co-DVB) | iCVD | hydrophobic | $-CF_3$ | 40 nm | Al, Cu | 183 |

**Table 3. Select coating techniques to modify the chemistry of a surface —in a spatially uniform manner— for heat transfer applications. List of acronyms: iCVD (initiated CVD); MOCVD (Metal-Organic Chemical Vapor Deposition); NNBD (Nanofluids Nucleate Boiling Deposition); PECVD (Plasma Enhanced Chemical Vapor Deposition); AR-MSIP (Actived Reactive-Magnetron Sputtering Ion Plating); Fluoro-POSS (fluorodecyl polyhedral oligomeric silsesquioxane); p(PFDA-co-DVB) poly-(1 H ,1 H ,2 H ,2 H -perfluorodecyl acrylate)-co-divinyl benzene**

Self-assembled monolayers (SAMs) are an assembly of molecules which form spontaneously on surfaces through adsorption into ordered structures. They are a straightforward means to achieving homogeneous wettability modification (15°-115°) without requiring a relatively thick coating on thermally conductive metallic substrates, and have been characterized for boiling conditions [173-175]. Modification of $\theta_E$ can also be achieved by utilizing additives (*i.e.*, surfactants [184] or nanoparticles) in the boiling liquid; however, this approach makes it difficult to interpret the effect of $\theta_E$ on boiling as the surface tension is altered too, and the surface typically keeps modifying itself by deposition during boiling. Moreover, monolayers are not as durable as thicker coatings.

Besides surfaces with spatially homogeneous chemistry and wettability, there are surfaces with wettability contrasts, juxtaposing hydrophobic regions with hydrophilic regions [185]. Such surfaces have an affinity towards both the liquid and vapor phases, a quality that we name *biphilic*. In nature, the biphilic wings of the Namib desert beetle (Figure 12 optimize its water intake [186]; while hydrophilic regions of the wing help collect water drops from fog-laden winds, the hydrophobic regions help the drops detach and roll down to the beetle's mouth [187]. In 1965, the first biphilic surfaces by Hummel [51] (see Figure 13), who sprayed hydrophobic polymer drops onto a steel surface, showed a heat transfer coefficient about 2 to 7 times higher than the bare steel surface. The various methods to fabricate surfaces with non-spatially homogeneous wettability, either biphilic or with wettability gradients, are described schematically in Figure 14 and listed in detail in Table 4.

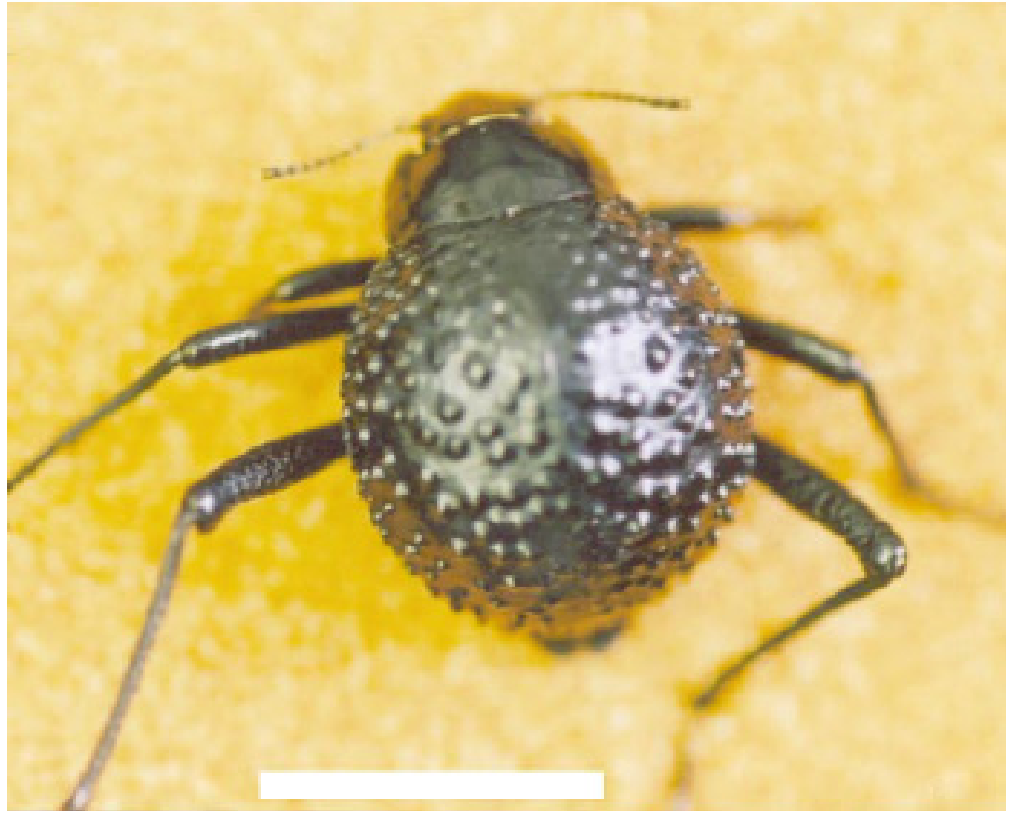

**Figure 12. The Namib desert beetle features a network of hydrophobic and hydrophilic spots on its back, facilitating collection and transport of water in scarce humidity conditions. Reproduced by permission from Macmillan Publishers Ltd: [187], Copyright 2001.**

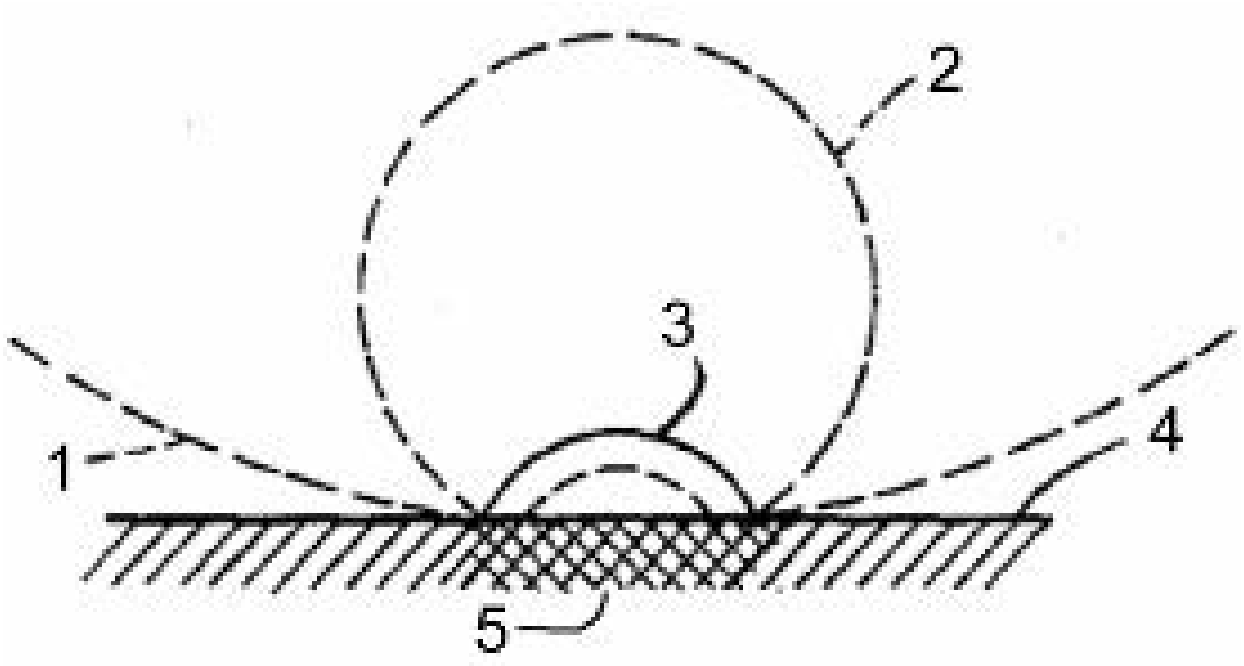

**Figure 13. Schematic showing the rapid growth of a vapor bubble from a non-wettable island (hydrophobic) onto a wettable background (hydrophilic). 1-3 represent liquid-vapor interfaces at the different growth stages; 4 indicates a hydrophilic region; 5 indicates a hydrophobic region. Reproduced with slight modifications from [51].**

Essentially, most methods for biphilic surfaces can be classified into one (or a combination) of the following three categories (see Figure 14): 1) hydrophobic or hydrophilic domains on a coating with initially homogeneous wettability are achieved by an additive process, such as depositing a polymer coating (*additive*); 2) hydrophobic or hydrophilic domains are revealed by a subtractive process on a coating with initially homogeneous wettability, *e.g.*, a pristine PTFE coating on a metal substrate is removed at specified areas, by say mechanical or etching methods, to reveal the underlying hydrophilic metallic substrate (*subtractive*); 3) surface chemistry of an initially homogeneous coating is either chemically degraded (*e.g.*, plasma) or reacts to transform its wettability (*e.g.*, hydrophobic to hydrophilic) usually through exposure to radiation (*reactive*).

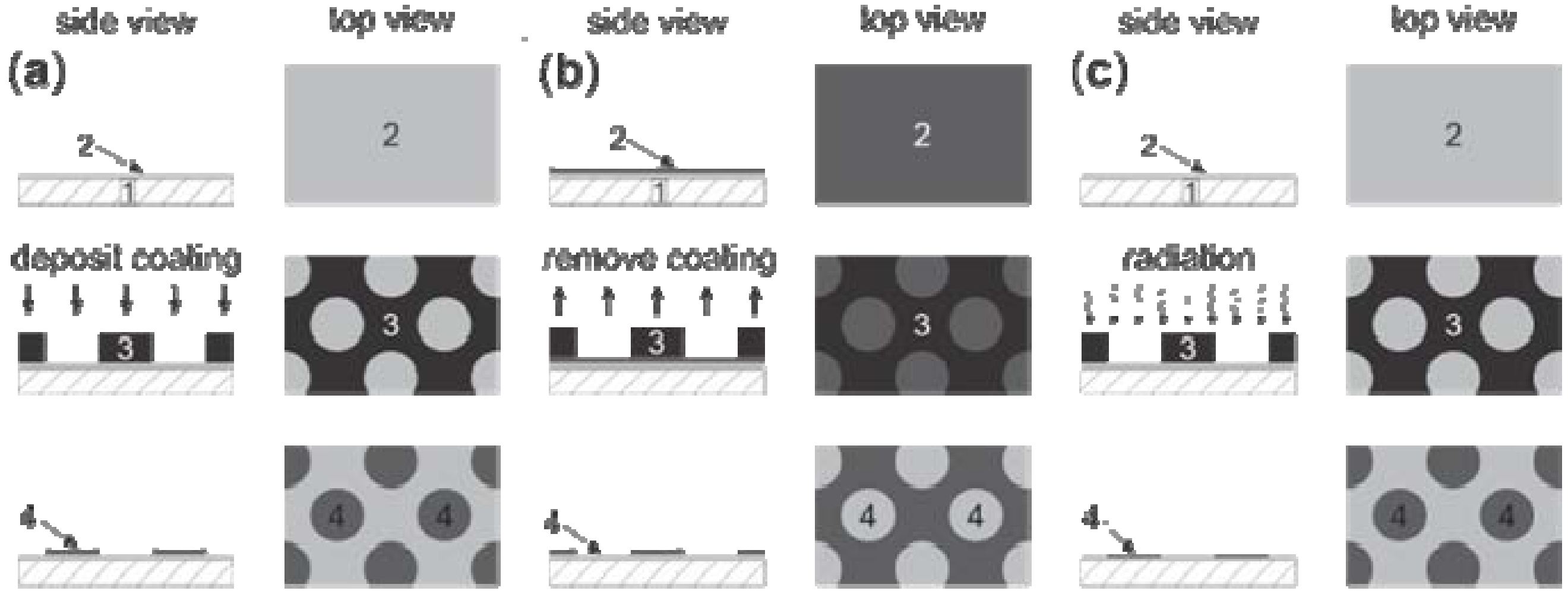

**Figure 14. Schematics demonstrating three techniques for achieving biphilic surfaces. List of identification numbers: (1) substrate; (2) coating with homogeneous chemistry; (3) mask (*e.g.*, lithography); (4) coating region with differential chemistry as compared with coating (2). In (a), biphilic surfaces are achieved by depositing (4) onto (2); *additive*. In (b), biphilic surfaces are achieved by removing (2) to reveal (4); *subtractive*. In (c), radiation or other reactive mechanisms are utilized to transform the surface chemistry of (2) into (4); *reactive*.**

| Material | Processing Techniques | Processing Type | Surface Type | Surface Chemistry | Coating Thickness | Substrates | References |
|---|---|---|---|---|---|---|---|
| silanes | vapor diffusion, others | additive | chemical gradient | $-CH_3$, others | 0.6 nm | many | 185, 188 |
| hydrophobic polymers: PTFE, epoxy resin, etc. | "pitting"; dip coating; abrading | subtractive | biphilic | -- | 0.25-12.7 μm | many | 51, 116, 189 |
| alkanethiolates | micromachining; molecular self-assembly | additive; subtractive | biphilic | $-CH_3$, -OH, -CN | mono-layer | gold, glass | 190, 191 |
| polystyrene, poly(4-vinylpyridine) | spin coating; annealing | additive | biphilic | $-C_5H_5N$, $-C_6H_5$ | 0.1-2.5 μm | many | 192 |
| silicon, fluorosilane | photolithography; UV assisted surface modification | subtractive; reactive | biphilic | -- | -- | silicon | 193 |
| silicon, fluorosilane, poly-vinylalcohol, AgI | reactive ion etching; oxygen plasma; silane treatment; polymer/particle solution deposition | additive | biphilic | -- | -- | silicon | 194 |
| silicon, fluoropolymer | fluoropolymer coating; photolithography; oxygen plasma | subtractive | biphilic | $-CF_3$, others | ~0.1 µm | silicon | 43, 58 |
| silicon fluoropolymer | deep reactive ion etching; plasma coating; photolithography; oxygen plasma | subtractive | biphilic | $-CF_3$, others | -- | silicon | 195 |
| silicon, fluoropolymer | oxidation; spin coating; photolithography | subtractive | biphilic | $-CF_3$, others | ~0.5 µm | silicon | 44 |
| silicon, hexamethyldisiloxane, | etching; photolithography, air plasma | subtractive | biphilic | -- | -- | silicon | 196 |
| alumina, $TiO_2$, FAS | photolithography, photocatalytic | reactive | biphilic | -CF3, others | ~200 nm | glass | 197 |
| silica, fluoroalkyl compound | photolithography | reactive | biphilic | $-CF_3$, others | 2 µm | plastic, glass, silicon, metals | 198 |
| polyelectrolyte, silica, fluorosilane | layer-by-layer | additive | biphilic | many | ~80 layers | -- | 186 |
| PTFE, polybutadiene, hydrophilic monomers | plasma chemical | additive | biphilic | $-CF_3$, many polymers | -- | silicon, PTFE | 199 |
| perfluoroazides | surface initiated polymerization | reactive | biphilic | $-CF_3$, -OH | -- | carbon nanotube forest | 200 |
| Fe-oxide, others | plasma etching, hydrophobic coating, oxygen plasma | subtractive | biphilic | many | ~1 µm | metals | 201 |
| Fluoro-POSS, PMMA | electrospinning, oxygen plasma | reactive | biphilic | $-CF_3$, oxygen containing groups | -- | many | 202 |
| methylsilsesquioxane, hydrophobic silica | $CO_2$ laser | subtractive | biphilic | $-CH_3$, others | ~10 µm | metal, glass, etc. | 203 |

**Table 4. Engineering techniques for producing wettability gradients and biphilic coatings.**

In the open literature, alternating patterns of superhydrophobic and superhydrophilic areas with well-defined repeatable features—while initially (ca. 2000) not well reported—has now been demonstrated by an ever increasing number of techniques. See the recent review article by Ueda and Levkin[204] for a general description of the state-of-the-art for superhydrophobic/superhydrophilic surfaces.

Tadanaga et al. [197] were one of the first groups to report on the subject of *superbiphilic* surfaces, *i.e.*, the juxtaposition of superhydrophobic and superhydrophilic regions in a patterned manner, by creating superhydrophobic alumina surfaces and then utilizing a photocatalytic reaction induced by UV-light to produce superhydrophilicity. Zhai et al. [186] generated hydrophilic patterns on superhydrophobic surfaces by selective deposition of polyelectrolyte/water/2-propanol mixtures, mimicking the fog collection behavior of the Namib Desert beetle (see Figure 12). Garrod et al. [199] utilized a two-step, plasma chemical approach to generate biphilic surfaces for applications in fog harvesting. While most coatings for heat transfer applications are metallic (due to their high thermal conductivity), polymeric and ceramic materials [205, 206] can also be coated by numerous different methods. Polymeric coatings do not provide high conductivity, but allow complex textures of a wide choice of materials. The fabrication methods used are usually simple and economic, such as dip coating, spin coating, plasma spray, and electrospray. Thickett et al. [192] used sequential spin coating to prepare polymer bilayers of hydrophobic polystyrene (acting as underlayer) and hydrophilic poly(4-vinylpyridine) (a porous network of beads and strands) on clean, smooth silicon substrates. Both polymer layers were smooth, with RMS roughness values below 0.5 nm. These surfaces were cooled and exposed to humid air to study dynamic condensation rates and their dependence on pattern features during dropwise condensation of water. Her et al. [201] created wettability contrast on alloy steels by applying nano-flake or needle patterns of multiscale micro/nanostructured iron-oxides on steel surface and subsequently tuning the surface energy. These surfaces were exposed to water mist, which readily covered the entire hydrophilic patterns, and created droplets with spherical shape on the surrounding superhydrophobic domains. The main advantage of this approach was that it utilized an industrially relevant, thermally conductive substrate, with scalable processing techniques. Betz *et al.*[58] fabricated biphilic and superbiphilic surfaces by coating silicon wafers with nanometer-thick layers of Teflon, after an optional nm-scale roughening using the black silicon technique. Zimmermann et al. [207] deposited silicone nanofilament coatings that could be locally activated with plasma to yield well-defined superhydrophobic, superhydrophilic, superoleophobic or superoleophilic domains on a single substrate. Kobaku et al. [202] fabricated liquid-repellent surfaces by electrospinning solutions of fluorodecyl POSS and poly(methyl methacrylate) (PMMA). The highly porous, re-entrant, morphology of the electrospun surfaces led to super-repellency (*i.e.*, high advancing contact angle and low contact angle hysteresis for water and various low-surface tension liquids, such as heptane). When these surfaces were exposed to oxygen plasma, they turned super-wettable (*i.e.*, advancing and receding contact angles near zero for both water and heptane). These surfaces displayed preferential nucleation behavior under boiling and condensation conditions of low-surface tension liquids.

## 3. Review of engineered surfaces for multiphase flow applications

### 3.1. Engineered surfaces for boiling heat transfer

The concept of engineering surfaces to achieve higher performance in boiling dates as far back as 1931 [208]. Fundamental studies began to appear in the 1950s [117, 209, 210]. In the 1960s, fabrication techniques headways allowed a more precise manufacturing of these enhanced geometries, followed by an increasing number of patents [116]. Webb reviews the historical evolution until 2004 of enhanced surfaces for boiling in great details [49, 116], and recent developments have been reviewed by Bergles and Manglik [211]. This section describes the main engineering attempts to produce surfaces with high boiling performance, i.e. high HTC and CHF. Historically, engineers have first designed surfaces with enhanced nucleation rates and high HTC. A second direction was later taken to increase CHF. More recently, engineers have designed multifunctional surfaces which increase both HTC and CHF. The reliability and repeatability of boiling experiments, crucial for industrial applications, will also be discussed.

*Controlling the surface texture to enhance thermodynamic efficiency:* An early attempt with enhanced surfaces to improve the HTC is by Jakob and Fritz [208] who roughened copper plates with machined groves of 16μm depths and ½ mm pitch; they observed 3 times improvement in HTC in nucleate boiling. The improvement did not last longer than 3 days, because of a slow degassing of the cavities, a phenomenon explained analytically in [212]. Berenson[50] increased HTC by a factor 6 by roughening metallic surfaces.

Kurihara and Myers [210] showed -on roughened surfaces- that HTC ~ $n^{1/3}$, with *n* the spatial density of active nucleation sites. Boiling improvements due to surface roughening were seen as of little industrial relevance because of the fast aging of the surfaces. Thus, to better understand and control the influence of surface texture on the nucleation site density in boiling, researchers have focused their efforts on the nucleation process on surfaces with engineered or well characterized cavities.

As shown in section 1.1, heterogeneous nucleation reduces the superheat needed to nucleate, by decreasing the free energy and by providing natural pits and cavities that entrap gas[ii] or vapor [5, 213]. To study the influence of pits and cavities on real surfaces, researchers have engineered surfaces with controlled geometries and cavities [49, 116]. Bankoff [212] observed three different states of these nucleation sites: the *active cavities* are nucleating, the *dormant cavities* are not nucleating but contain vapor that may nucleate, and the *extinct (or flooded) cavities do not contain vapor* and will only nucleate once the superheat reaches the value estimated for heterogeneous nucleation in section 1.1.

As shown in Figure 15, cavities are geometric means to manipulate the curvature of the liquid-vapor interface during the evolution of the volume of the vapor pockets, for the purpose of helping nucleation or preventing extinction or flooding.  Geometrically, a cavity is formed when a portion of the horizontal surface enters into the substrate. For a typical cavity, its inner surface enters downward with a slope at an angle between 0° and 90° from the horizontal surface. A conical cavity is a common example. If the slope increases beyond 90°, the inner surface is considered to re-enter outwards from a vertical surface, and the cavity is called a re-entrant cavity. If the slope further increases beyond 180°, the inner surface is considered to re-re-enter upwards from a horizontal surface, and the cavity is called a doubly re-entrant cavity. Note that this definition assumes an axisymmetric or 2D cavity. Physically, a re-entrant cavity is a

---

[ii] The rest of this discussion neglects the role of gas in cavities, because it is mostly transient and unpredictable, in contrast to the role of vapor

cavity for which the curvature of the liquid–vapor interface changes its sign once it penetrates the cavity mouth. During penetration in a doubly re-entrant cavity, the interface curvature reverses twice.

The state of cavities can be predicted from their geometry, the superheat, wettability and surface energies. Section 1.1 shows that the phase change in boiling occurs when a metastable liquid (point A, Figure 3) becomes stable vapor (point E). According to equation (2), the superheat needed to nucleate from a cavity with mouth radius R is proportional to the maximum curvature $1/r_{min}$ of the vapor pocket during its growth. Figure 15a shows that the maximum curvature is found at the mouth of the cavity [214] for typical conical cavities with $\theta > 2\gamma$, where $\gamma$ is the cone angle. This explains why the curvature *1/R*, with *R* the mouth radius, is used to estimate the superheat needed to activate nucleation in the cavity [214].

Flooding of cavities can also be described from their geometry using the same equation (2). If the fluid-solid system is brought to a temperature inferior to the fluid saturation temperature, or if the solid surface becomes colder than the fluid, the pressure of the vapor phase in the cavity also decreases below the pressure of the liquid, as shown on Figure 3. The condition of mechanical equilibrium, equation (1), implies a change in the sign of the curvature of the vapor-liquid interface. If the geometry or chemistry of the cavity does not allow this change in curvature, the cavity becomes flooded. Thus, conical cavities with $\theta < 90°$ cannot sustain subcooling, contrary to conical cavities with $\theta > 90°$, i.e. hydrophobic cavities [180, 215].

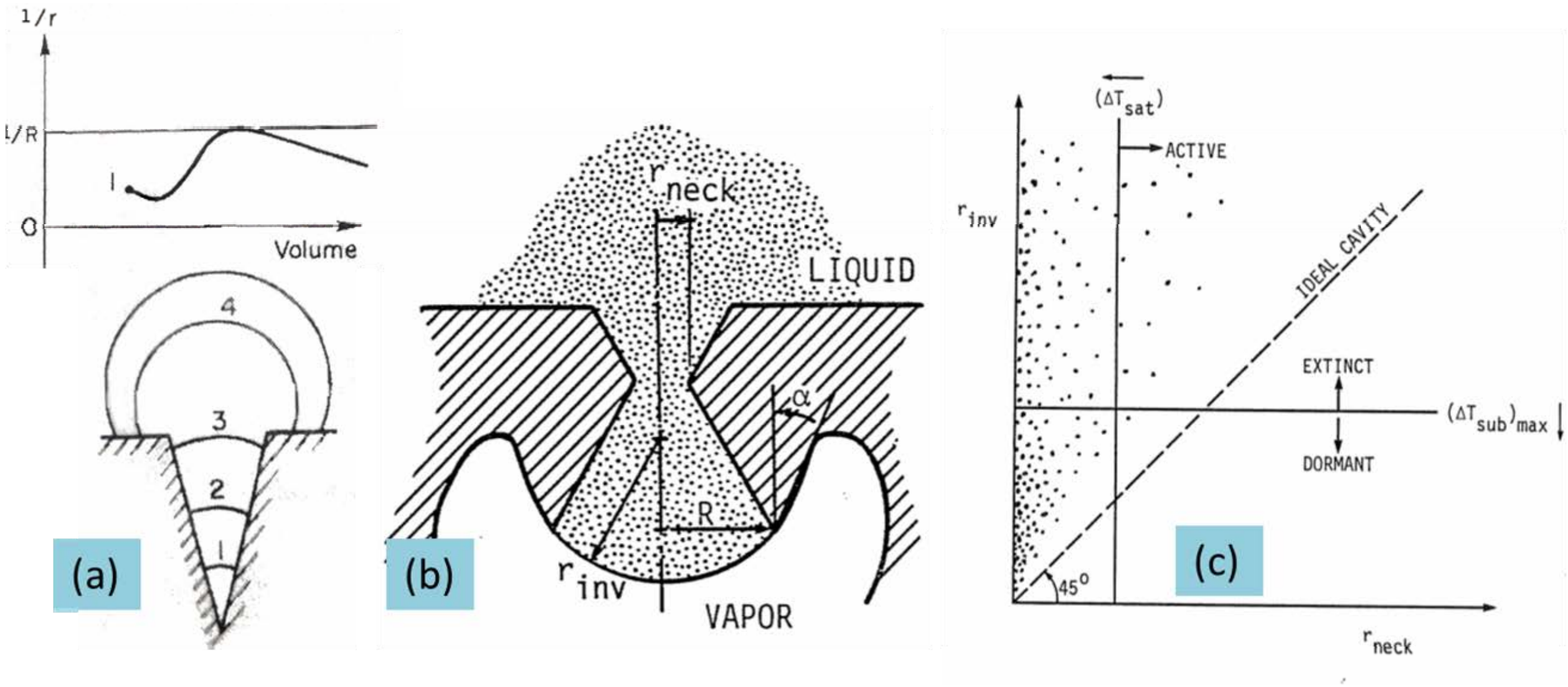

**Figure 15. Cavities on solid surfaces control the curvature of the vapor-liquid interface to facilitate nucleation and prevent flooding during subcooling. (a) Schematic of a typical conical cavity with variation of the curvature 1/r during nucleation; R is the radius of the cavity mouth. Figures are reproduced from [52] with minor modifications with the permission of Taylor and Francis Group. (b,c) the geometry of a doubly-reentrant cavity in a fully wetting situation and the superheat/subcooling determine the state (active, extinct, dormant) of the cavity. Reproduced with modifications from [216] and permission of the author of the thesis.**

Hsu [41] improved earlier models for nucleation in conical cavities by adding an upper limit on the size of active cavities, considering the variation of temperature in a thermal boundary layer with height δ above the cavity. This model evaluated the size range of active cavities with the relation

$$\{r_{\max}, r_{\min}\} = \frac{\delta}{2C_1}\left[1+\frac{T_{SAT}}{T_W} \pm \sqrt{(1-\frac{T_{SAT}}{T_W})^2 - \frac{4AC_3}{\delta T_W}}\right], \quad (15)$$

where $C_1$ and $C_3$ are functions of the contact angle and the constant $A$ is obtained from the Clausius-Clapeyron equation. This model has been widely used in boiling studies [54, 217, 218].

With the improvement of micromachining techniques, other types of cavities have been studied to further improve the performance of engineered surfaces in boiling and increase the stability during subcooling in comparison to conical cavities. One of the first analytical and experimental studies on machine-designed cavities was by Griffith and Wallis [219]. From their study, re-entrant cavities were shown to act as stable, easily activated and long-lasting nucleation sites. Thereafter, many studies on re-entrant cavities have been carried to enhance the entrapment of vapor, as reviewed by Shoji [220]. For instance, Marto and Rohsenow [221] manufactured doubly reentrant cavities (100-600μm) by drilling and tapering. They found a seven fold enhancement in HTC compared to mirror surfaces. One key issue is to find a practical and economical method to make these cavities at the suitable length scales. Later, Kim [216] investigated analytically and experimentally the relation between geometry and state of re-entrant cavities. He showed that for fully wetting systems, such as the ones involving refrigerants, three parameters determine the cavity state: the superheat, the radius of the neck $r_{neck}$ and the inverted radius of the cavity $r_{inv}$ (see Figure 15b). For a doubly re-entrant cavity, $r_{neck} \leq r_{inv}$, and an ideal doubly re-entrant cavity has $r_{neck} = r_{inv}$. Clusters of points in Figure 15c show the distribution of cavities on a real surface. The ΔT corresponding to the respective activation and extinction of the cavity can be obtained using equation (2), with r* respectively equal to $r_{neck}$ or $r_{inv}$. It appears from Figure 15c that for fully wetting systems, small cavities are more resistant to flooding than large cavities in case of subcooling, but also require larger superheat to nucleate.

To manufacture re-entrant cavities, two industrially relevant approaches have been developed. The first two types of cavities (Gewa-T, Thermoexcel-E) are fabricated by flattening or bending fin-like protrusions, while the latter (High Flux, [222-224]) are fabricated by *sintering metallic particles*. These complex surface geometries enhance not only nucleation, but also fluid transport. For instance Chien and Webb studied the interactions of cavities interconnected via pores and channels [49, 225]. Their suction evaporation model predicts optimum pore diameter that balance CHF and HTC [49, 226]. More recently, Ujereh et al.[227] reduced the onset of nucleate boiling and enhanced HTC in pool boiling of refrigerant FC 72 (up to 400%) by coating Silicon surfaces with carbon nanotubes.

*Control of surface chemistry to enhance thermodynamic efficiency*: Surface texture (or topography) is not the only property that has been modified to enhance multiphase heat transport. Engineering the chemistry of surfaces has also been shown to enhance multiphase heat transfer. As shown in section 1.1, nucleation is facilitated on low wettability surfaces, while high wettability enhances performance at high heat flux.

Gaertner [228] was among the first to modify the wettability of a metallic surface with a thin fluorocarbon film coating and showed that the maximum heat flux decreased as the surface became non wettable. Liaw and Dhir [167] modified the wettability by controlling the degree of oxidation of a surface or by coating a thin coating of fluoro-silicone sealant. They obtained a maximum heat flux lowered by 50% on the hydrophobic surface ($\theta$=107°) compared to the hydrophilic surface ($\theta$=38°). More recent studies also show that hydrophobic surfaces enhance HTC by increasing the number of nucleation sites [42], by reducing the waiting time between bubbles [166] and by reducing the superheat at the onset of nucleate boiling [174]. In [229], comparisons of several nanometrically smooth surfaces showed how HTC increases with the increasing surface hydrophobicity. The exhaustive study of Jo et al. [44] of boiling on smooth silicon samples, showed that a hydrophobic coating ($\theta$=123°) enhance the HTC by a factor up to 3 compared to a hydrophilic coating ($\theta$=54°). Hydrophobic surfaces had an onset of nucleate boiling at superheats five times lower than hydrophilic ones; however the CHF was reached at 250 kW $m^{-2}$ on the hydrophobic coatings, whereas the CHF on hydrophilic coatings occurred at fluxes close to four times higher. Finally, the surface energy of nickel wires was modified with layer-by-layer coatings of silica nanoparticles to obtain hydrophobic, hydrophilic or superhydrophobic surfaces [180]. The coating caused drastic changes in wettability but barely changed the roughness. The superhydrophobic coating was shown to increase the HTC and the CHF by a factor 2 compared to the bare metallic wire. The higher HTC was attributed to a higher number of available active nucleation sites on hydrophobic coatings. The higher CHF was attributed to a low value of the receding angle (20°), similar to angles on hydrophilic coatings. In Takata et al. [230], superhydrophobic surfaces were found to induce stable film boiling at very small values of superheat, and no nucleate boiling region was observed.

*Control of surface texture to enhance CHF:* Several attempts have been made to use micro and nanostructures, to delay the occurrence of dry-out, or minimize its effects, as reviewed by Lu and Kandlikar [131]. The mechanism by which these nanostructures enhance heat flux is still a matter of debate [231], be it enhanced capillary transport [232], enhanced surface area or superwettability [162]. Li and Peterson [34] used micro meshes to increase capillary transport, resulting in higher CHF. Other examples of materials used to enhance CHF are copper nanowires [35, 232-234], silicon nanowires [231, 232], carbon nanotubes[235], zinc oxide nanoparticles [236], nanoporous copper [237, 238], nanoporous zirconium [239], nanoporous silicon [240], and nanoporous aluminum oxide [241]. Figure 16 presents a micrograph of superhydrophilic silicon nanowires which achieved a 100% increase in the CHF value for water, as compared with smooth silicon [232]. Also, Ahn and co-workers [242] investigated the effect of carbon nanotube coatings on pool boiling. The surface was coated with a carpet of vertical multiwalled carbon nanotubes. Interestingly, the heat flux in the film boiling regime was increased by 175%, possibly by disruption of the microlayer or by increase of the contact line length. Wicking tracks along the surface have been obtained by deposition of nanoparticles after boiling nanofluids on metallic microwires. These coatings have been shown to enhance wettability and wicking, thereby increasing CHF [243]. Diamond particles have been coated to obtain a microporous layer [244] that resulted in a significant increase of CHF. Additionally, this microporous coating also decreased the required superheat temperature for nucleation from 30 to 5K. A microporous coating obtained by fusing copper microparticles has improved the boiling performance of refrigerant by respectively 300 and 120% in terms of HTC and CHF, compared to a bare copper surface [245]. Other coating materials, such as nano-silica particles [181], have also been deposited and led to similar results and conclusions. Takata et al. [162] showed that a $TiO_2$ coating could greatly enhance the critical heat flux in pool boiling and measured a 3-fold enhancement of the CHF values compared to a bare surface. In [246], over 100% enhancement of critical heat flux in pool boiling heat transfer has been demonstrated using atomic layer deposition of $Al_2O_3$.

.

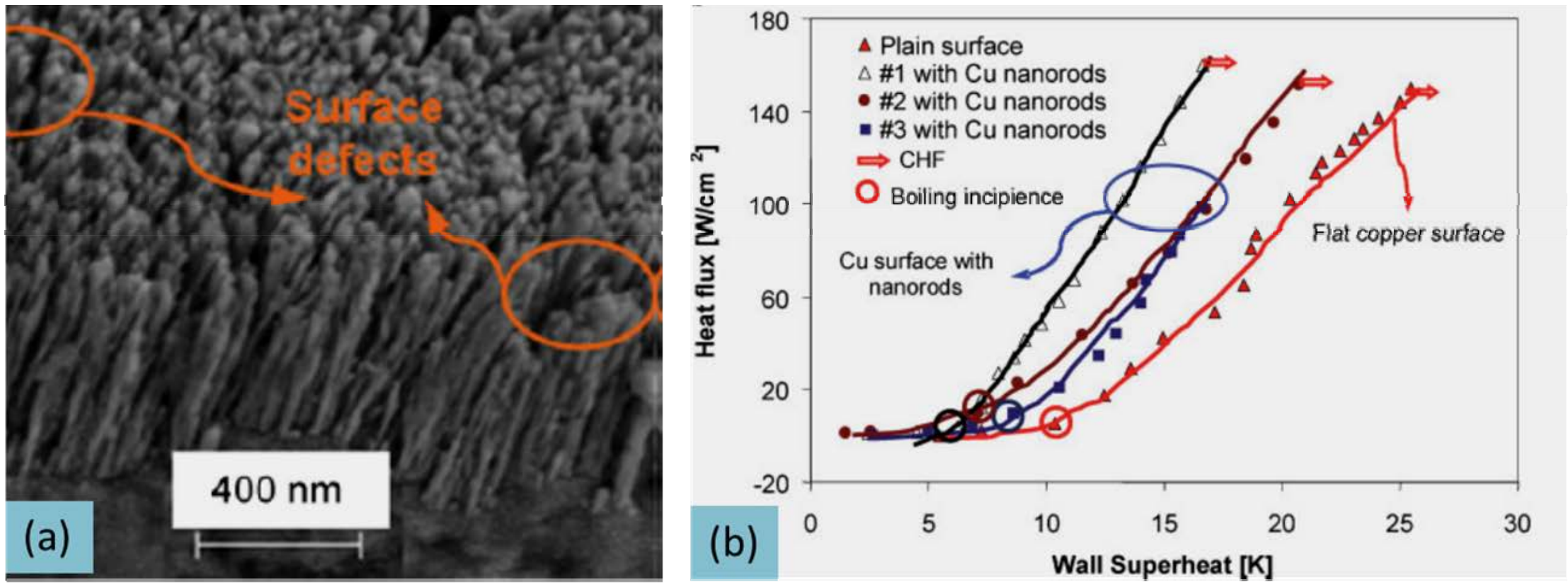

**Figure 16. Use of surface texturing for pool boiling enhancement: (a) SEM image of copper nano-rods used to enhance pool-boiling performance (b). The surface defects act as artificial cavities and enhance the nucleation. Reprinted from [35] with permission of John Wiley and Sons.**

*Chemistry and texture control to enhance CHF and HTC:* Superwettable coatings have been used to enhance critical heat flux, by maintaining a high degree of solid-liquid contact area, while at the same time, delaying dry-out by reducing the spreading of vapor bubbles [65, 71, 232, 239]. Indeed the study of the modification of the chemistry and texture to enhance the CHF did not receive much attention until recently with the pool boiling experiments of nanofluids. As shown by Kim and Kim [243], boiling nanofluids enhanced both the texture and chemistry by depositing a coating of nanoparticles contained in the fluid on the surface during boiling. Using nanoparticles of $SiO_2$, $TiO_2$ and $Al_2O_3$, they obtained hydrophilic surfaces with a range of wettabilities. They concluded that the CHF increases for improved wettability and by capillary pumping resulting from the presence of nanoparticles. Recently, Ahn et al. [239] controlled the degree of anodic oxidation of zirconium alloy plates to develop micro and nano structures on the surface, reducing wetting angles from 50° down to ≈0°. They concluded that the combined modification of chemistry and texture enhance the CHF by improving the wettability and the liquid spreading (from 177 kW $m^{-2}$ on a nanotextured surface, to 560 kW $m^{-2}$ on the micro/nanotextured surface).

Surfaces with *multiscale textures* or geometries offer the possibility to address the multiple length scales associated with boiling, from wetting and nucleation (O(nm)), to bubble dynamics (O(cm)). Surfaces that show distinct multiscale features, like those in Figure 17 or the pyramids of microparticles [232, 247-250], have been shown to improve both thermodynamic efficiency and CHF in pool boiling. Liter and Kaviany [249] designed a porous multi-scale texture by stacking of copper microsphere (200μm diameter) as in Figure 17b to form an array of cones (≈1.5 mm height). They found that the critical heat flux is reached when either of the following phenomena limits the flow of liquid towards the surface: (1) the vapor columns merge into a vapor layer on the surface that keeps the liquid away from the solid surface; or (2) the viscous drag in the wicking geometry exceeds the capillary suction. Figure 17 represents a multiscale texture created by the Microreactor Assisted Nanomaterial Deposition (MAND) process. This multiscale texture enhances both the CHF (4 times higher than a bare Al surface) and the HTC (10 times higher). Launay et al. [247] combined micro and nanoscale patterns for pool boiling applications: several surfaces including bare Silicon, fully coated CNT silicon, 3D microstructures with micropillars and nanotubes were characterized with water and a dielectric fluid. Also, by controlling the synthesis of CNT on porous capillary wicking surfaces [130], the temperature of superheat required for nucleation was reduced by 72% compared to baseline tests; importantly, the “temperature overshoot” observed on non-coated surfaces was eliminated. Recently, Kandlikar designed surfaces with low profile (0.5-1mm) embossed fins (Figure 17) featuring microscale cavities at their base; measured HTCs and CHFs were respectively, 8 and 2.5 times higher than on plain copper surfaces [38]. The microcavities were thought to enhance nucleation, while the fins induced lateral rather than vertical bubble trajectories. Kim et al. [250] designed hydrophilic surfaces with octagonal micro-posts combined with ZnO nanorods, with CHFs about twice as high as bare surfaces.

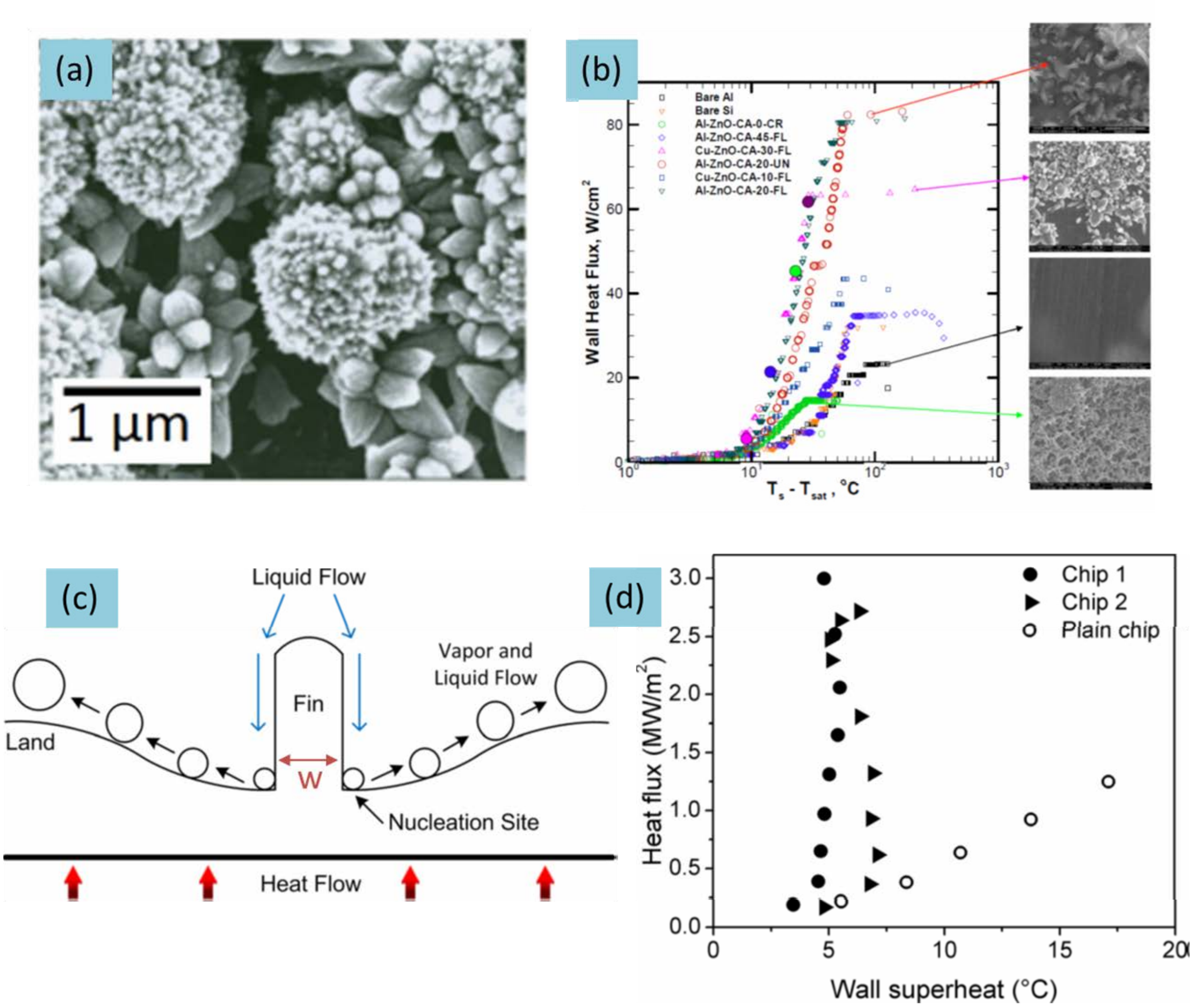

**Figure 17. Multiscale surfaces increase HTC and CHF in pool boiling. (a) ZnO nanostructures deposited on Al substrates by the MAND (Microreactor Assisted Nanomaterial Deposition) process, and corresponding (b) boiling curve on two different substrates (Silicon and Aluminum) with different MAND process parameters. Reprinted from [236], Copyright 2010, with permission from Elsevier.; (c) Embossed mm-scale fins with μm-scale cavities at their base drastically enhance HTC and CHF in pool boiling (d), by inducing lateral rather than vertical bubble trajectories and by increasing nucleation. Reprinted with permission from [38], Copyright 2013, AIP Publishing LLC.**

*Spatial patterning of wettability to improve both HTC and CHF:* The ability to spatially pattern wettability is important to the design of ideal boiling surfaces. As seen in the conclusions of section 1.1, an *ideal* boiling surface has complex requirements on its wettability: it requires hydrophobicity to promote nucleation and HTC at low heat flux, and hydrophilicity to maintain water transport to the hot surface for preventing early CHF [42]. Therefore, biphilic surfaces, which juxtapose regions of high and low wettability, offer an elegant solution to this dilemma. Measurements have shown that biphilic surfaces *significantly enhance the heat transfer performances (HTC and CHF) in pool boiling* [43, 51]. The enhancement was explained by the efficient management of both the vapor and liquid transport, maximizing nucleation rates and delaying critical heat flux [43].

The first reports (given in the form of US Patents) of biphilic surfaces come in the late 1960's by Hummel [51] and by Gaertner [189], see Figure 12. The emphasis is on stabilizing gas and vapor cavities to prevent cavity flooding, that resulted, as an example, in an increase of a factor 5 of the HTC for the biphilic surface compared to clean surface [51]. Betz et al. have shown that biphilic and superbiphilic surfaces (see Figure 18) also act to delay the formation of the vapor film that insulates the surface from the liquid and therefore increase the CHF by 65% as compared with a uniform non-wettable surface [43]. Their superbiphilic surfaces also showed significant HTC enhancement for low superheat values, up to 10 times the HTC of a simple biphilic surface at 5K of superheat [58]. Other biphilic and superbiphilic surfaces have been synthesized and have demonstrated site-selective nucleation [203, 251] and enhanced boiling heat transfer [44, 235, 251-253] for water. For design considerations of these surfaces, it has also been shown that the size and pitch of the hydrophobic dots are the dominant parameters at low heat fluxes and that, at a high heat flux, a high number of hydrophobic dots could result in decreasing the value of the CHF [44, 235, 251-253]. Finally, biphilic surfaces have also been manufactured for lower surface tension liquids like methanol [202].

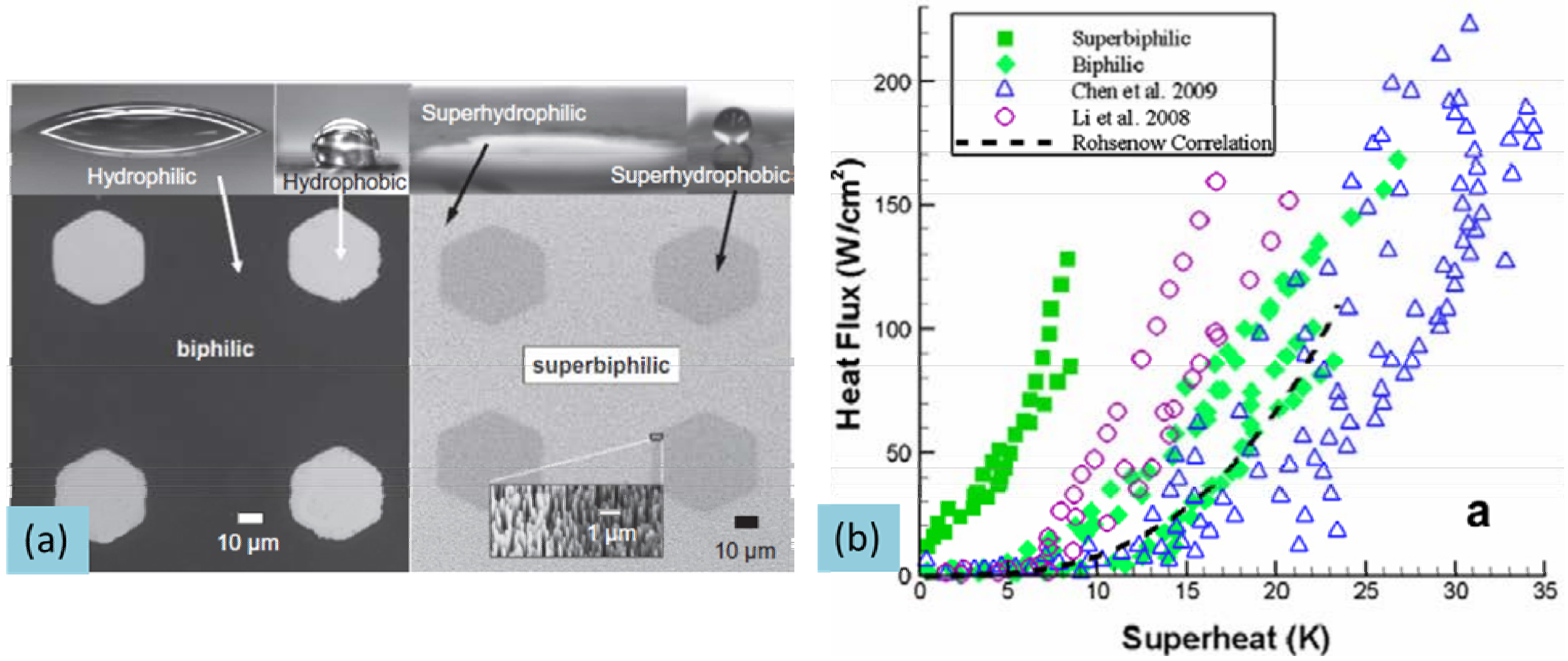

**Figure 18. Biphilic and superbiphilic patterns (a) on silicon surfaces drastically enhance pool boiling heat transfer (b) Reprinted from.[58], Copyright 2013, with permission from Elsevier.**

### 3.2. Engineered surfaces for condensation heat transfer

Theoretical insights of condensation phenomena have been used as an effective tool to engineer condenser surfaces. Early reviews of methods to enhance heat transfer in condensing equipment and technical ways to favor dropwise condensation have been provided by Bergles and Morton [254] and Williams et al. [255]. These involve (a) changes in geometry to either increase surface area and/or help remove the condensate, such as [75] surface roughening (mostly effective at high Reynolds numbers), fins, fluted or grooved tube, wavy surfaces such as the "*Gregorig surfaces*" which create gradients of capillary pressure to destabilize the film of condensate [256, 257], and wired coiled around tubes; (b) modifications of surface energy or chemistry, and (c) the use of force fields –which is outside the scope of this review.The use of micro- and nanostructured to enhance DwC on surfaces has recently been reviewed by Enright et al.[258]

Engineering the wettability of surfaces to improve condensation heat transfer seeks three purposes: (i) maximizing the DwC HTC by ensuring adequate nucleation and early removal of droplets, (ii) ensuring adequate drainage of condensate (larger drops or puddles caused by the merging of droplets) to avoid the transition to film condensation, and (iii) sustaining DwC over long times by minimizing surface degradation. While the first purpose is particularly important at low heat flux, the other two are typically investigated for high heat flux conditions. The design requirements for surfaces addressing these three purposes are distinct. Enhancing DwC HTC requires that the surface should be hydrophobic enough to shed the condensate droplets at a small size (thus minimizing the thermal resistance across a liquid droplet), and at the same time allow better surface rejuvenation (to have a higher nucleation rate). Enhanced droplet mobility on a hydrophobic condenser surface also facilitates condensate drainage, thus delaying a transition from the DwC to the FC mode. Multiscale wettability patterning is investigated to enhance the removal of condensate and prevent formation of liquid film on the surface. The third purpose however, to sustain DwC over the operational life of the condensing surface, is more of a manufacturing challenge. The following sub-section describes how these wettability engineering strategies have been implemented through homogeneous (monophilic) and heterogeneous (biphilic) chemical coatings, homogeneous and heterogeneous physical texturing, or a combination of chemical coating and physical texturing.

Patankar [21] performed a detailed theoretical analysis of superwettable surfaces to optimize nucleation in condensation (and boiling) applications. The analysis combined the Clausius-Clapeyron equation, chemical and mechanical equilibrium conditions, and assumed that the geometry consisted of arrays of cylindrical pillar or cavities. Geometries thermodynamically favorable to nucleation were identified in terms of height and radius. The study concluded that long slender pillars with hydrophobic sides and hydrophilic tops might optimally enhance nucleation associated with condensation, and facilitate roll-off. Other studies concluded that some penetration of the fluid into the porous layer would induce better thermal contact than a suspended Cassie State [259, 260]. It is important to note that such theoretical analysis focus on a specific aspect of the condensation cycle presented in Figure **2**, for example the nucleation or the conductive heat transfer. Nevertheless, some continuum assumptions need to be revisited at the nanoscale: for instance can one still apply the Young-Laplace law with radii of curvature on the order of the thickness of the liquid-vapor interface?

*Controlling the surface chemistry to induce hydrophobicity:* Tuning the chemistry of the condensing surface has been attempted, since the importance of early droplet roll-off in dropwise condensation has been acknowledged [81]. Organic polymers have been proposed since the early sixties [177] for promoting DwC of steam, and active research continued until the eighties [221]. Polymer films on metal surfaces have also been viewed as a potential solution for condensing organic vapors [168]. Although these coatings achieved enhanced steam DwC HTC and prevented transition to FC even under high heat flux conditions (~ 4000 $kWm^{-2}$) [261], their durability (~100 – 1000 h) still needs to be improved to prevent observed peeling or cracks after the heat transfer tests. Polymer coatings with higher durability (~ 20,000 h) have been evaluated [262], but their much larger thickness implied larger thermal resistance of the substrate, for which the pertinent HTC values were not reported. Recently, new methods have been introduced to achieve durable hydrophobic coatings on condensing metal surfaces. Modification of substrate surface through coating of self-assembled organic monolayers (SAM) [263-265] and Polytetrafluoroethylene (PTFE) [168] have been attempted. For a similar wettability, SAM has been found advantageous over PTFE coatings, because of the low thickness (~30 Å) and associated lower thermal resistance between the condensate and the substrate [266]. Typically, low-energy coatings of fluoroalkylsilanes have been used as a popular coating for investigating the improvement in dropwise condensation heat transfer [267]. Zhao et al. [179] reported a 30 fold enhancement in DwC heat transfer of steam (maximum heat flux ~ 1500 $kWm^{-2}$) on Langmuir–Blodgett treated copper surfaces (compared to the FC on the bare metal). Paxson et al. [183] have shown that the use of iCVD copolymer thin films grafted to a metal substrate shows more robust (no ageing up to ~ 50 h) hydrophobicity than fluorosilane SAMs and nearly seven times larger heat transfer (HTC ~ 40 $kWm^{-2}K^{-1}$) for steam condensation than FC. Superhydrophobic coatings typically enhance droplet mobility and induce faster droplet sliding due to the low contact angle hysteresis. This, in turn, augments convective heat transfer between the rolling droplet and the substrate [268]. However, condensate removal on all these monophilic coatings necessarily relies on an external force (e.g., gravity), and therefore has limited effectiveness for smaller droplets or on horizontal static surfaces. Also, homogeneous hydrophobic surfaces have a higher thermodynamic barrier to nucleation (see Eq (8)), compared to hydrophilic surfaces. More advanced features in terms of chemistry and texture of condensing surfaces have therefore been developed in the recent years to overcome these shortcomings.

*Effect of surface texture on droplet mobility:* Manufacturing surface textures for improving DwC heat transfer has been attempted with mm-size grooves [269], but with limited efficacy, as the largest fraction of heat transfer improvement can only be realized by controlling the growth and removal of condensates only at a smaller length scales. This has led to a number of studies on DwC on surfaces engineered with nearly homogeneous microscale patterns of triangular spikes [270], pillars of square and circular [156, 271-273] or other [274] cross sections. For all these cases, surface roughness imposed through the texturing led to increased apparent contact angle. A liquid drop on a roughened surface can exist either in a Cassie or Wenzel state, depending upon whether the equilibrium contact angle (on the smooth surface) $\theta_E$ is, respectively, larger or smaller than a critical angle $\theta_C$, which is a function of the surface geometry (see Eq. (13) and (14)) [151,275]. However, this theory only holds if the drop size is larger than the texturing length scales: water droplets ensuing from condensation on a rough superhydrophobic substrate have been found to behave differently from deposited or impinging droplets [270, 274, 276, 277]. If the minimum droplet size ($r_{min}$) is smaller than the length scale of the surface roughness (as in the case of micropatterned surfaces), the initial phase of droplet nucleation and growth does not differ from that on a smooth surface [277]. The nucleation and the early growth of the droplet occurs with the same probability at the base and the tip of the structure. For micro-textured surfaces having $\theta_E < \theta_C < \theta^*$, the coalescence of droplets through the formation of liquid bridges gives rise to a metastable Cassie-like state, which finally gives way to a more stable droplet in a Wenzel state that keeps growing in size [277] (the temporal growth of droplet diameter follows $D{\sim}t^{1/3}$ for the initial phase, until $D$ is smaller than the microstructure length scale, and then a self-similar trend of $D{\sim}t$). This apparent loss of hydrophobicity and affinity towards Wenzel wetting has been reported in other studies too [271, 274, 278]. Although the Wenzel mode of wetting on textured surfaces minimizes the droplet-to-substrate contact thermal resistance, it is detrimental to droplet mobility. Efforts have been devoted to achieve a transition from Wenzel to a Cassie state during droplet growth to enhance droplet mobility. Dorrer and Rühe [156] reported such transitions during condensation on a silicon surface textured with silicon micro-posts and coated with a hydrophobic fluoropolymer ($\theta_E$ ~ 118$^o$ on a smooth surface, which was greater than the $\theta_C$ for the textured surface). They reported that droplets nucleate and grow at the valleys and the tips of the structures and form liquid bridges in the microcavities within the posts during the initial phase. This was akin to the observation of Narhe and Beysens [277], except that the microdroplets for the case of Dorrer and Rühe [156] grew tall enough to touch the overlying liquid bridges (originating from the coalesced droplets at the tips of the micro-posts) and were drawn up to a more stable Cassie state. It has been demonstrated that for textured condensing surfaces to retain superhydrophobic behavior, topological features of aspect ratio (the height to width ratio of the micro-structures) must be greater than one, and the depth of the separation between the structures must be less than 500 nm [273]. With the advent of nanofabrication, a few recent works have attempted condensation studies on homogeneously textured surfaces having feature size ranging from 100 nm (electrodeposited nanopillars) to 10 µm (microfabricated on Si) [272], or carpeting with carbon nanotubes [279, 280], which offer promise of improved droplet mobility.

*Effect of chemical coating and surface texture on HTC:* While the combination of superhydrophobic and nanostructured surfaces can induce spontaneous drop evacuation [14] and are a topic of current research [281], the process of coating surfaces with superhydrophobic materials may add additional thermal resistance which offset the overall gain in heat transfer [282]. Prior studies [83, 283, 284] indicated the influence of the substrate thermal resistance on the overall heat transfer. More recent investigations on substrates with hydrophobic coatings [285] have considered the influence of the substrate thermal conductivity on overall heat transfer rate. Thermal resistance of vapor layer trapped between the droplet and the substrate also offer significant thermal resistance while the droplets reside on nano-pillars or a textured superhydrophobic surface in a Cassie Baxter mode [286]. Thus, an effective DwC strategy needs to resolve the dilemma over the appropriate combination of surface wettability and liquid mobility to maximize heat transfer. Recent studies have shown that a partially wetting droplet behavior during condensation offers the optimum configuration for improved heat transfer. Miljkovic et al. [259] showed that a nanostructured Si surface had 56% heat flux enhancement for partially wetting droplet morphologies, as opposed to a 71% heat flux degradation for suspended morphologies in comparison to a flat hydrophobic surfaces. In another study [287], nearly 25% increase in overall heat flux was demonstrated using CuO nanostructured surfaces with roughness height ~1000 nm, wetted fraction ~ 0.023 and roughness ratio ~10. Scalable metal oxide nanostructures have been identified as promising candidates for condensation heat transfer applications, as they produce partially wetting droplets, a compromise between good substrate-droplet thermal contact and droplet mobility [287]. As far as research on promoting dropwise condensation evolves, a clear trend towards using nanoscale features or a combined approach of microscale features with spatially patterned surface chemistry is observed.

*Effect of multiscale surface texture on condensate mobility and HTC:* A more recent type of condensing surface has used multiscale roughness to achieve better control of nucleation and droplet growth and removal. For example, biomimetic surfaces having both micro- and nano-scale roughness features (often called two-tier structure) produced by carbon nanotubes deposited on Si micro-pillars, and coated with hexadecanethiol, exhibited superhydrophobic behavior during condensation [15]. Chen et al. [288] investigated dropwise condensation on a microscale pyramidal Si microstructure having nanotextured surfaces and showed a 65% increase in drop number density and 450% increase in the droplet self-removal volume, as compared to a superhydrophobic surface with nanostructures alone. These multiscale structures essentially provide the substrate both locally wettable nucleation sites as well as global superhydrophobicity (see Figure 19 (a)). Boreyko and Chen [14] also supported the view that both nanoscale and microscale topological features are necessary to promote spontaneous drop motion. Cheng et al. [289] observed adaptive purging of condensing surfaces via self-propelled or coalescence-triggered condensate on surfaces having two-tier roughness where they attributed prompt removal as the dominant factor for sustaining continuous dropwise condensation. Liu et al. [290, 291] provided thermodynamic analyses explaining the effect of the multiscale architecture of a superhydrophobic surface during dropwise condensation. Rykazweski et al. [292] proposed the existence of drops in nanoscale as well as microscale Wenzel and Cassie-Baxter wetting states in a micro- and nano-textured surface and identified the optimal feature spacing of the multiscale superhydrophobic structures. The theory behind the enhanced droplet performance on such structure is still under development. Although these surfaces look promising from the point of view of maximizing DwC heat transfer in low heat flux applications (e.g., for condensation under humid air conditions), further investigations are warranted under more intensely condensing (e.g., in a 100% vapor mass fraction) environments regarding how these multiscale surfaces fare in preventing FC or sustaining prolonged DwC.

*Gradient and biphilic wettability patterning for enhanced condensate mobility:* To address the issues of improved condensate transport for preventing the onset of film formation on a condensing surface, a significant body of work has been reported on patterning of surface wettability (i.e., inducing biphilicity). The technique uses a combination of chemically heterogeneous deposition and surface texture to facilitate selective nucleation, growth and condensate mobilization on the substrate. In fact, condensation patterns have been used to monitor the fabrication of biphilic surfaces [293] by measuring diffraction intensity as the temperature of the surface was lowered. An important contribution of biphilic surfaces to condensation is the ability of transporting the condensate with wettability gradients to prevent the transition from dropwise to filmwise condensation, e.g., by inducing self-ejection of the condensing droplets from the substrate [14]. Daniel et al. [294] utilized wettability gradients under dropwise condensation conditions to passively transport water droplets from hydrophobic to hydrophilic regions, enhancing condensate transport, and droplet rejuvenation. Derby et al. [295] reported 3.2–13.4 times enhancement in heat transfer on hydrophobic and hydrophobic/hydrophilic patterned surfaces (HTC ~ 280–425 $kWm^{-2}K^{-1}$ under a highly advective flow) over a hydrophilic surface. Varanasi et al. [193] performed water condensation experiments in the ESEM on hybrid surfaces that consisted of alternating hydrophobic and hydrophilic segments on a silicon wafer. The intrinsic contact angles for these surfaces were ~110° and ~25°, respectively. ESEM images taken over a span of 30s (see Figure 19(b)) clearly demonstrate a large difference in the intrinsic wettability of the hydrophilic and the hydrophobic segments. This wettability difference resulted in preferential nucleation on the hydrophilic segments of the hybrid surface and subsequent droplet growth. The condensation pattern progressed from a few non-uniform droplets on some of the hydrophilic regions, to uniform droplets that covered all of the hydrophilic regions. These droplets ultimately merged over the hydrophobic terrain, thus covering the entire substrate with a liquid film.

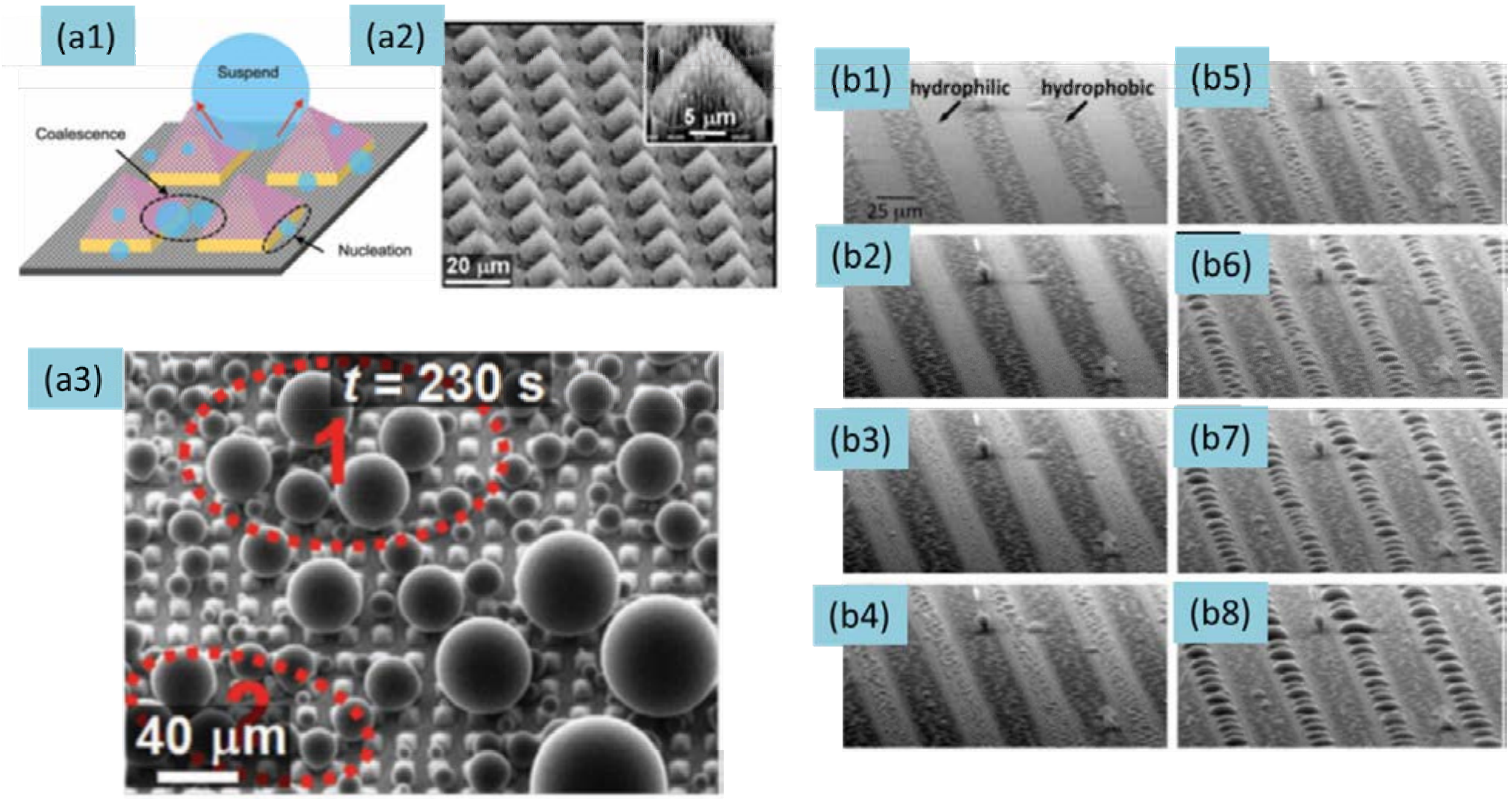

**Figure 19. Control and enhancement of condensation using surfaces with spatial variations of wettability. (a) DwC on multiscale pyramidal Si microstructure with nano-textured surface: (a1) schematic, (a2) SEM of the structures, (a3) ESEM image of dropwise condensation on such surface. The synergy between the micro and nanoscale roughness features results in a stable Cassie state and upwards surface tension force, which assist in drop departure. Reprinted from [288], reproduced with the permission of John Wiley and Sons, copyright 2011. (b) ESEM images of condensation of water vapor on a surface with alternating hydrophobic and hydrophilic segments: (b1) dry surface, (b2–8) Snapshot images of condensation on a biphilic surface. The width of each segment is about 25µm. The intrinsic contact angles of the hydrophilic and hydrophobic regions are ~25° and ~110°, respectively. The sample was maintained at a temperature of 274 K by means of a cold stage accessory of the ESEM. In order to ensure a dry surface at the beginning of the experiment, the chamber pressure started at ~400 Pa, well below the saturation pressure at that temperature. The vapor pressure in the chamber was then slowly increased until droplet nucleation was observed. Droplets were observed to preferentially nucleate and grow on the hydrophilic regions. Reprinted with permission from [193], Copyright 2009, AIP Publishing LLC.**

In the case of dropwise condensation, one of the earliest studies utilizing biphilic surfaces was performed by López et al. [190] The sample surfaces juxtaposed hydrophobic (water contact angle 120°) and hydrophilic (CA 30°) domains, and were fabricated by patterning SAMs of different alkanethiolates on gold and of alkyl siloxanes on glass. Water vapor condensation on such cold surfaces was used to visualize their spatial heterogeneity. Biomimetic micropatterning has been attempted by several groups to achieve enhanced DwC on superbiphilic surfaces. These patterns primarily used discrete superhydrophilic spots that exhibit highly enhanced nucleation rate on a superhydrophobic background [201, 203]. Garrod et al. [199] investigated the effect of varying the size and pitch of the superhydrophilic spots on condensation rate. They found that water collection rate and microcondensation efficiency peaked for superhydrophilic spot width and pitch of 500μm and 1000μm, respectively. The biphilic structures of Thickett et al. [192], consisting of a micropatterned P4VP layer (hydrophilic) on a PS background (hydrophobic), were used for atmospheric water. Water droplets detached from the low hysteresis (~ 23°) micropatterned PS/P4VP substrates at lower volumes than for flat P4VP films (hysteresis ~23°), thus accelerating water collection rates. Depending on the flow velocity of the incident humid air, they observed condensate collection rate in the range of 0.7 – 3.3 $Lm^{-2}h^{-1}$ on the patterned substrate. Flat PS films had the lowest hysteresis of the three films considered, however they condensed the least amount of water (~ 0.6 – 1.8 $Lm^{-2}h^{-1}$). Some other recent works [16, 296] have demonstrated that oil-infused micro- and nano-textured surfaces exhibit superior performance in terms of droplet mobility and condensation rate. Combined influence of surface heterogeneity and low adhesion of water in an optimally oil-infused textured surface results in enhanced nucleation density and greater mobility of condensate. Dropwise condensation in such surfaces has been observed at a moderate condition (heat flux ~160 $kWm^{-2}$ [16] and HTC ~ 10 $kWm^{-2}K^{-1}$ [296]), although longevity of these surface can be affected by lubricant cloaking, drainage and mixing (with condensate) in case of suboptimal surface design.

By using perforated masks with the desired geometry during $O_2$ plasma treatment, Kobaku et al. [202] fabricated superomniphilic domains on superomniphobic surfaces for condensation experiments of heptane vapors. The experiments revealed preferential condensation of heptane vapors on the superomniphilic domains. More recent studies have shown the efficacy of attaining spatial control over droplet nucleation and growth by using hybrid coatings, e.g., by selectively coating the tips of the hydrophobic microstructure posts with hydrophilic materials [193]. Droplets preferentially nucleate and grow on these hydrophilic post tops of the hybrid surface more uniformly than they would on a similarly textured superhydrophobic surface, thereby promoting Cassie growth and more uniform droplet shedding [297]. Recently, Mischenko et al. [194] developed a more robust hybrid surface of various geometrical microstructures and demonstrated controlled nucleation and growth of droplets during condensation.

However, like in the case of multi-scale textured surfaces, the HTC in DwC of these hybrid surfaces, and their ability to maintain sustained DwC under high heat flux conditions still needs to be quantified.

### 3.3. Engineered surfaces for applications with ice or frost

The technological challenges faced when attempting to avert frosting and icing are both multifaceted and numerous. Thus, engineers have long resorted to combining external means (mechanical, thermal, or chemical) to avoid ice formation/deposition even in the late 20$^{th}$ century. A passive strategy to achieve this goal includes delaying ice nucleation and the onset of mass solidification. This can be achieved by modifying the chemistry of the surface or its micro- and nanoscale texture. For example, surfaces modified with fluorinated and polysiloxane coatings have low water wettability and surface energy, resulting in icephobic behavior under subzero temperatures [298, 299]. However, this approach is limited under many technical circumstances: for example, when a cold surface encounters supercooled water droplets in a humid atmosphere, icing is ultimately inevitable. In this situation, ice adhesion hinders the effort to rid a surface of the ice. As a result, several engineering attempts to prevent ice/frost formation or prolong machinery operation in ice-promoting environments have concentrated on ice/frost delay, ice adhesion reduction or both. An even more severe technical challenge is the design of surfaces that maintain machinery performance in the presence of ice or frost. The density of the ice/frost is a function of temperature, humidity, thermal conductivity and wettability of the surface. Therefore, while it is desirable that a surface delay ice/frost nucleation, if ice/frost cannot be prevented, a surface that produces a dense phase is preferred in refrigeration systems for mitigating the heat transfer reduction due to icing. The present section describes engineering attempts to produce surfaces that delay ice/frost nucleation and reduce ice adhesion. Studies that focus on ice removal are not within the scope of this review.

*Surface chemistry and texture affect ice nucleation:* Ice nucleation rates were found to be one order of magnitude higher on smooth hydrophobic surfaces compared to smooth hydrophilic surfaces [300], where water microdroplets formed on flat, solid surfaces by an evaporation and condensation process. The first study of freezing on textured superhydrophobic surfaces was conducted by Saito and co-workers, and showed less snow accumulation on nanostructured superhydrophobic surfaces than on smooth surfaces [301]. Hydrophobic textured stainless steel surfaces functionalized with polyelectrolyte polymer brushes showed lower freezing temperatures than polished untreated steel for supercooled liquid droplets under slow cooling conditions [302]. Increased freezing delays and lower ice adhesion strength with decreasing contact angle hysteresis were observed by Arianpour et al. [303], who studied small water droplets cooled on nanostructured composite coatings consisting of vulcanized silicone rubber and various amounts of carbon-black, titania or ceria nanopowders as fillers. A recent condensation-frosting study made on superhydrophobic surfaces with enhanced self-propelled jumping of condensed water microdroplets, showed significant delay in frost initiation even at very low (subzero) temperatures and high degree of supersaturation [304, 305]. Another recent study [306] on superhydrophobic surfaces made of zinc oxide nanowires showed significant retardation in frosting as well as considerable delay in freezing of condensed liquid droplets on nanowires with smaller diameters. Condensed droplets on these surfaces were formed at temperatures well below 0°C and showed very high contact angle, even at smaller length scales (~50 μm). The mode of initial condensation also determined the freezing delay. When initial condensation occurred in filmwise mode, the freezing temperature was higher and freezing duration was an order of magnitude lower compared to dropwise condensation in [307], where nanostructured surfaces consisting of ZnO nanorod arrays were employed in an evaporative closed cell. Sessile water droplets freezing on surfaces with different wettabilities –ranging from superhydrophilic to superhydrophobic- were also studied by Yin et al. [308] under continuous substrate cooling. Ice accretion was tested by spraying supercooled water to samples at different horizontal inclination angles. Their results did not show any correlation of ice formation with surface wettability and revealed the greater role of surface texture in anti-icing performance. On micro/nanotextured superhydrophobic surfaces, the time delay in ice nucleation of sessile droplets was found to be extended, as was the total time of freezing [307, 309].

*Controlling ice adhesion with surface hydrophobicity and texture:* Several studies on shear force of ice adhesion on multiscale-texture hydrophobic/superhydrophobic surfaces have shown ice adhesion force reduction between a factor of 2 to 6 compared to the respective control surfaces [310-314]. Under axial stresses, the reduction factor was even higher (5-10).[310] Saleema et al. [315] reported almost zero ice adhesion on a superhydrophobic (with low contact angle hysteresis) textured surface with a PTFE coated layer. Ice adhesion strength on superhydrophobic surfaces has been shown to directly correlate with contact angle hysteresis [303, 312, 316] rather than contact angle itself, as found in earlier studies [301]. Ice adhesion decreases with increasing contact angle on surfaces of similar roughness and increases with decreasing icing/freezing temperature [310]. Cassie wetting on superhydrophobic surfaces creates voids between the solid surface and ice, which decrease the shear strength. These voids act as microcracks, which along with contact angle hysteresis play a major role in reducing ice adhesion force on superhydrophobic surfaces. Hence superhydrophobic surfaces can have strong ice adhesion only if they do not create sufficiently large voids at the interface [317]. Meuler et al. [318] extensively studied ice adhesion on many smooth surfaces of representative wettabilities and hysteresis. Their results showed strong relationship between the average ice adhesion strength and the practical work of adhesion, which is a function of receding contact angle on those smooth surfaces. Humid conditions led to higher ice adhesion due to the anchoring effect [319, 320]. Studies have been carried out in an attempt to understand the role of surface roughness on ice adhesion using superhydrophobic surfaces. Since the contact area between the ice and the solid on superhydrophobic surfaces is very low (typically less than 5% of the entire surface), such surfaces are expected to feature significantly lower ice adhesion force. Fibrous superhydrophobic surfaces with their increased robustness and resistance to condensation-induced wetting would also seem to be ideal for resisting ice adhesion, assuming they can be made robust enough not to be damaged [320, 321]. Wind turbine blades coated with highly porous superhydrophobic coating made from PTFE showed significant reduction in ice accretion rate [322]. Highly re-entrant surface structures reduce solid liquid contact area, and thus the low heat transfer rate may be responsible for their observed superior anti-frost performance [320].

*Hydrophobicity affects frosting/defrosting:* Jing et al. [323] investigated frosting and defrosting on various surfaces with different wetting properties, ranging from superhydrophilic to superhydrophobic. The degree of supercooling depended on wettability, because of the higher energy barrier for larger contact angles. Repeated frosting/defrosting cycles were found to destroy the repellent ability of their superhydrophobic surfaces. Further, surfaces made of rigid material remained repellent after the cyclical frosting/defrosting tests, but flexible surfaces degraded. Varanasi et al. [108] studied the growth of frost on micro patterned superhydrophobic surfaces via ESEM observations. Frosting was found to significantly deteriorate icephobic properties of superhydrophobic surfaces. To achieve low ice adhesion under frosting conditions, they proposed to reduce the contact area between ice and the underlying surface; this is possible by preferential ice nucleation on the surface. A review on synthetic superhydrophobic surfaces as anti-ice and anti-frost surfaces is provided in [324].

*Icing by impacting droplets:* Icing caused by supercooled water impact on a solid surface is a phenomenon with important applications for vehicles and has been studied extensively. While earlier icing studies of supercooled water impact on superhydrophobic surfaces were mainly focused on ice adhesion strength [312, 316], Cao et al. [110] concentrated on the formation of ice from supercooled liquid water and freezing rain. Their results showed the superior performance of superhydrophobic surfaces in keeping objects ice-free. They also showed that reducing the size of nanoparticle fillers in polymer nanocomposites reduces the ice nucleation probability of supercooled liquid. Wang et al. [325] reported superior performance of superhydrophobic aluminum wires in ice accretion tests under freezing rain conditions. Mishchenko et al. [20] extensively studied the icing of supercooled water under dynamic conditions on lithographically prepared micro-nano structured surfaces of varying wettability and geometric features under a wide range of water temperatures, surface temperatures and impact angles. Their study showed the potential of micro-nano structured superhydrophobic surfaces with densely packed features as efficient passive anti-icing devices under non humid conditions (RH~5%) down to - 25°C. A comprehensive theoretical model was also proposed to accommodate droplet dynamics, heat transfer and ice nucleation and verified with experimental observations [326]. Highly porous and self-cleaning superhydrophobic surfaces made from functionalized carbon nanotubes were also shown to have superior repellency against supercooled water droplet impact [320]. Lower droplet surface contact time, reduced heat transfer and low surface energy of the functionalized carbon nanotubes may be the reason for their superior performance. Ice adhesion and growth on superhydrophobic surfaces from impacting supercooled water drops revealed that the time delay in ice initiation increases with decreasing contact angle hysteresis and that surfaces with small roughness features are the most robust in ice repellency [327]. A study of the dynamics of supercooled liquid impact on superhydrophobic surfaces revealed that ice nucleation delay on superhydrophobic surfaces is more important at moderate degrees of supercooling, while at much lower temperatures, close to the homogeneous nucleation temperature, bulk and air−water interface nucleation effects become dominant [328].

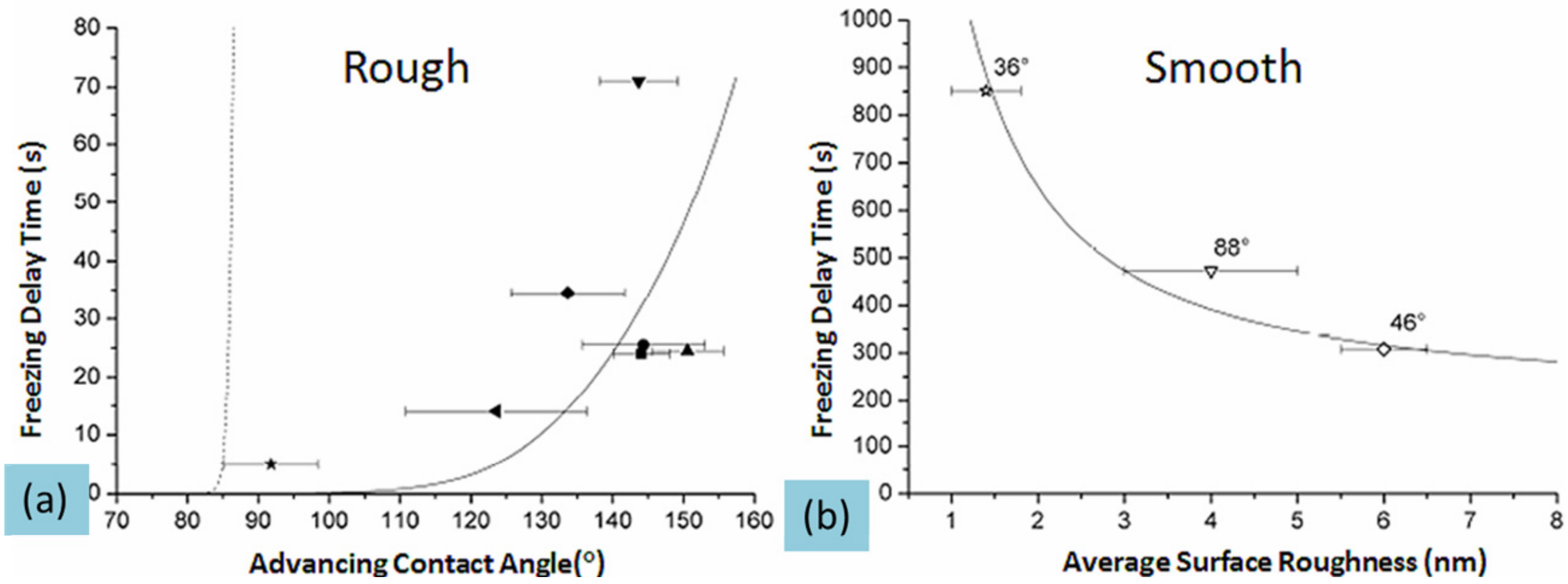

**Figure 20. Influence of surface roughness and wettability on freezing delay time during impact of subcooled droplets. (a) Freezing delay time vs. advancing contact angle for highly corrugated (rough) surfaces (average roughness of the order of microns). Contrary to the classical nucleation theory (dotted curve), the extended model reported in [105] (solid curve) offers a better fit for the experimental data (symbols). (b) Freezing delay time vs. surface roughness for smooth surfaces (average roughness of the order of nm). The value beside each symbol indicates the measured water contact angle on the respective surface. The extended model is indicated by the solid curve. Reprinted with permission from [105]. Copyright 2011, American Chemical Society.**

Due to low residence time and slower ice nucleation rate, superhydrophobic surfaces are being explored as one possible solution for preventing aircraft icing. Antonini et al. [329] have studied anti icing performance of airfoils with and without superhydrophobic coatings under supercooled droplet impingement tests inside a wind tunnel. Superhydrophobic surfaces not only reduced the energy consumption (in keeping the rear side of the aircraft wing ice-free using electrical heaters embedded in the airfoil) by up to 80%, but also changed the morphology of accreted ice, which was easier to remove by deicing systems. A comprehensive predictive model encompassing surface properties and in-flight conditions was used to evaluate the anti-icing performance of different coatings [330]. Freezing of supercooled water on surfaces of different hydrophobicity was studied under shearing air flow and different humidity condition, and the critical shear velocity was measured and correlated with environmental conditions [106].

Jung et al. [105] compared the time to freeze a given flow rate of supercooled liquid droplets impacting on surfaces that ranged from hydrophilic to superhydrophobic. They found that surfaces with nanometer-size roughness and high wettability nucleate later than typical superhydrophobic surfaces with multiscale roughness and low wettability. In that work, observations of repeated droplet collisions on the same location up to the point of ice formation revealed a previously unseen regime for liquid-on-liquid bounce under atmospheric pressure and at subfreezing surface temperatures. This bouncing behavior is beneficial because it can reduce the total amount of water collected on surfaces. Figure 20 shows freezing delay times vs. (a) contact angle $\theta$, and (b) average surface roughness. The data points are separated according to roughness of the corresponding surface; for rough surfaces (average roughness >> critical nucleus size; for water at -20°C, $r_c$ = 2.2nm), shown in Figure 20a, the freezing delay time is independent of roughness (see [105]). The classical heterogeneous nucleation theory is shown by the dotted curve in Figure 20a and deviates significantly from the experimental data (symbols). A similar discrepancy was reported by Gorbunov et al. [331]. An extension to the classical theory was formulated in [105] taking into account the reduction in excess entropy of water in the proximity of a surface, as compared to bulk liquid. Figure 20b shows the freezing delay vs. roughness with measured contact angle for the smooth surfaces (average roughness ≤ critical nucleus size). The adjusted model (solid curve in Figure 20b) reproduces the significant roughness dependency of freezing delay for very small values of roughness. The results of Figure 20 are important in revealing that ultra-smooth non-hydrophobic surfaces resist considerably longer against icing than typical multiscale hydrophobic surfaces. Poulikakos and co-workers [332] elucidated key aspects of droplet impact on surfaces that are severely undercooled, leading to the formation of frost and icing. Multiscale surfaces with small gaps between the asperities (both at micro- and nanoscales) yielded the largest resistance to millimeter drop impalement. The best performing surface impressively showed rebound at −30 °C for drop impact velocity of 2.6 m $s^{-1}$; see Figure 21a.

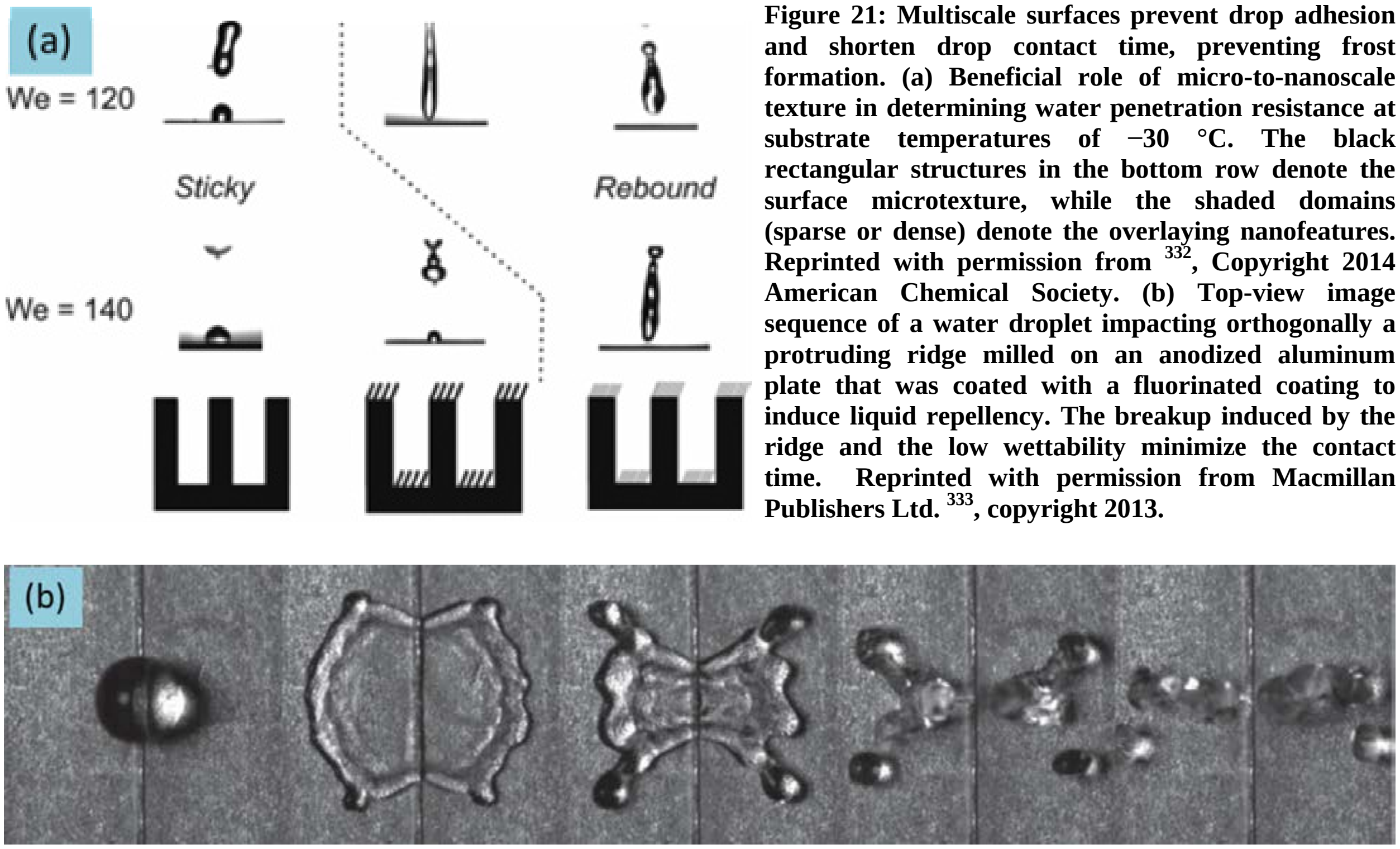

**Figure 21: Multiscale surfaces prevent drop adhesion and shorten drop contact time, preventing frost formation. (a) Beneficial role of micro-to-nanoscale texture in determining water penetration resistance at substrate temperatures of −30 °C. The black rectangular structures in the bottom row denote the surface microtexture, while the shaded domains (sparse or dense) denote the overlaying nanofeatures. Reprinted with permission from [332], Copyright 2014 American Chemical Society. (b) Top-view image sequence of a water droplet impacting orthogonally a protruding ridge milled on an anodized aluminum plate that was coated with a fluorinated coating to induce liquid repellency. The breakup induced by the ridge and the low wettability minimize the contact time. Reprinted with permission from Macmillan Publishers Ltd. [333], copyright 2013.**

In the situation of impacting droplets, it seems reasonable to assume that the risk of icing increases with the contact time between liquid and solid, as contact time controls the extent to which mass and energy are exchanged between the impacting drop and the solid target. Bird et al. [333] therefore defined the problem of designing surfaces with minimum contact time during drop impact. They investigated a range of surfaces with varying degree of hydrophobicity, for which the contact time was comparable to the lowest-order oscillation period for a spherical drop. They reduced the contact time further with multiscale surfaces combining O(100μm) ridges and a superhydrophobic texture. Figure 21**b** depicts an impact sequence of a water droplet striking an anodized aluminum oxide plate with a milled macroscopic texture, pitted microtexture and a fluorinated coating (to induce superhydrophobicity). The droplet separates in half around the protruding ridge. Since the recoil time $\tau \sim (m/\sigma)^{1/2}$, the smaller drops with lower mass $m$ spread and rebound faster from the solid surface, reducing the probability for freezing if the plate is at subfreezing temperatures.

*Frosting*: Compared to icing, frosting directly from the gas phase (desublimation) onto superhydrophobic surfaces is much more challenging and has not been investigated extensively yet. An initial study [20] in this area indicated that superhydrophobic surfaces retard frosting via metastability of the triple point line [334].

*Other anti-icing surfaces*: Besides superhydrophobic surfaces, a new kind of surface, called liquid suffused solid (SLIP) surface, with very low contact angle hysteresis has also been proposed [111] and evaluated to effectively address the issue of icing and frosting. SLIP surfaces also inhibit ice nucleation in supercooled liquid very effectively due to their ultra-smooth texture and high degree of surface homogeneity [335] which act in tandem to lower the temperature for ice nucleation initiation. Though these surfaces have shown promising results, they currently have drawbacks, such as cloaking [336]. The entropy reduction of the crystalline phase near a charged solid surface also influences the heterogeneous nucleation rate of ice at the interface. Recently, zwitter-wettability surfaces (which rapidly absorb molecular water from the environment, but simultaneously appear hydrophobic when probed with water droplets) made from molecular water absorbing polymeric molecules were reported to display frost resisting capability [337].

*Biphilic surfaces*: Frosting/icing studies on biphilic surfaces have been almost entirely lacking, and only recently started to appear in the literature. Mishchenko et al. [194] used functionalized, hydrophilic polymers and particles deposited on the tips of superhydrophobic posts to induce precise geometrical control over water condensation and freezing at the micrometer scale. Control of freezing behavior was demonstrated via deposition of ice-nucleating AgI nanoparticles on the tips of these structures. The hydrophilic polymer with AgI particles on the tips was used to achieve templating of ice nucleation at the micrometer scale. They presented preliminary results indicating that control over ice crystal size, spatial symmetry, and position might be possible with this method.

## 4. Technology sustainability: energy, economic and environmental benefits

Advanced surface engineering for phase change heat transfer will impact, among many others, three important industries: (a) manufacturing, (b) electricity generation, and (c) automotive.

*Energy savings in manufacturing:* Thermal transfer and energy storage processes directly impact chemicals, petroleum refining, as well as the plastic and rubber industries. These industries collectively consume more than 10% of the total energy used in the United States, which represents 97.3 quadrillion Btu in 2011 [338]. For comparison, the total energy consumption per capita in Japan represents half that in the USA. For shell-and-tube vapor generators, a 20-50% improvement of global heat transfer coefficient using engineered surfaces would reduce by a corresponding amount the surface areas (and likewise size and material costs) of heat exchangers needed in the chemical and manufacturing industry, and facilitate process integration [339].

*Energy savings in the thermal generation of electricity:* Figure 22 shows that electricity production represents more than 40% (40 Quad) of the total energy used in the United States. Electricity is overwhelmingly produced via phase change thermal processes involving devices such as condensers, evaporators, steam turbines. The conversion losses, from heat to electricity, account for more than 25% (25 Quad) of the total US energy consumption. Advanced surface engineering for phase change heat transfer would significantly reduce conversion losses by minimizing exergetic losses[339]. Also, heat exchangers with engineered surfaces for dielectric fluids would improve by about 15% the efficiency of low-ΔT Rankine cycles used in refrigeration, solar power, and air conditioning.

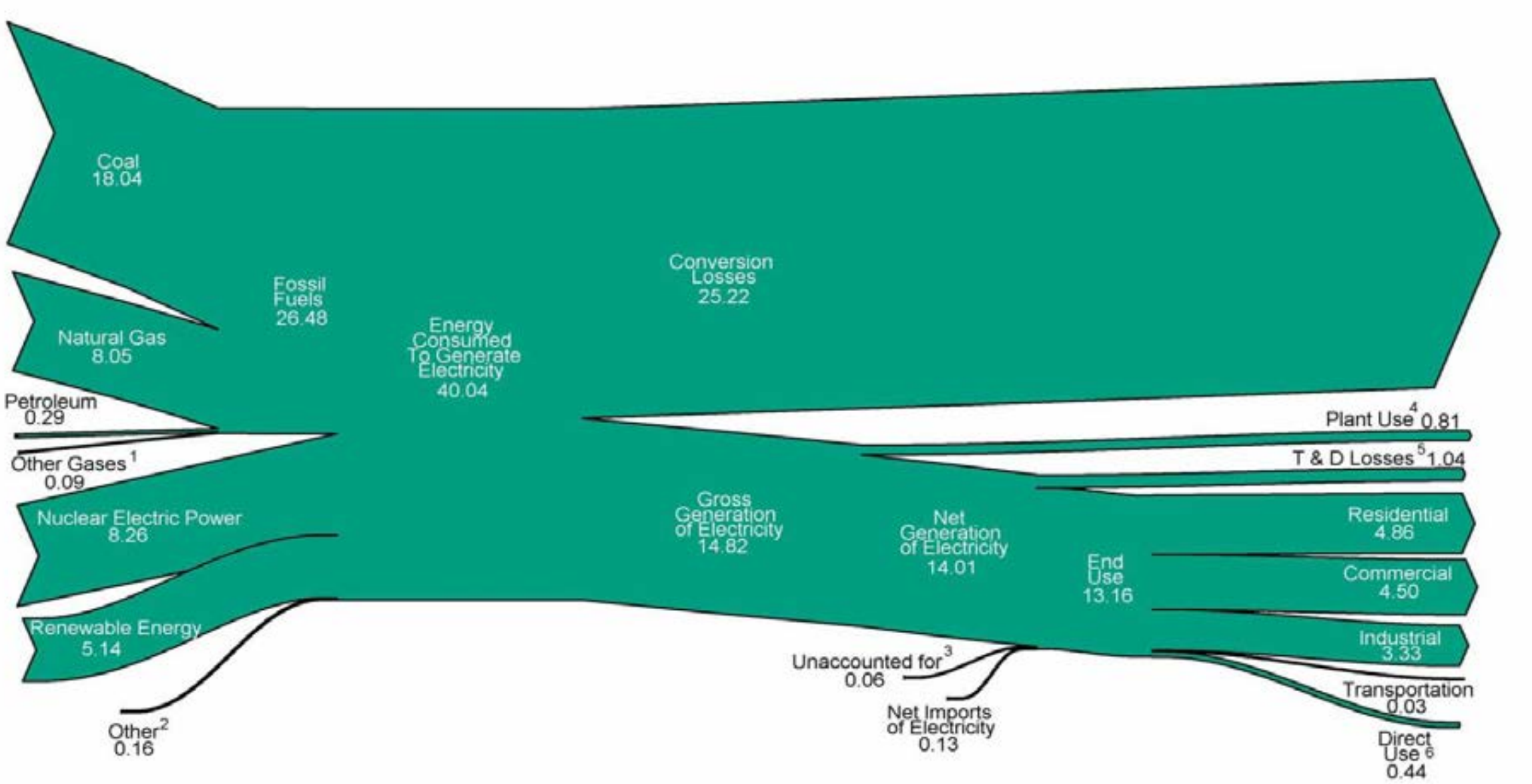

**Figure 22. US electricity flow, in Quad, from production source to consumption sectors, year 2011. See [338] for original image and detailed caption. One Quad (quadrillion BTU) equals $10^{15}$ BTU or approximately $10^{18}$ Joule.**

*Economic and energy impact to the automotive industry:* Electric and hybrid vehicles use energy in a more sustainable manner than vehicles with thermal combustion engines. Engineered surfaces for phase change heat transfer will make hybrid and electric drive vehicles more affordable, thereby increasing their market share. It has been estimated that power electronics constitute 40% of the total cost of the traction drive system in hybrid vehicles [340]. Recent work at the US National Renewable Energy Laboratory (NREL) has demonstrated that two-phase immersion cooling systems can significantly improve thermal performance by reducing the total thermal resistance (junction-to-liquid/coolant) of an automotive power electronics module by more than 50%, compared to the conventional single-phase automotive cooling system [341].

Importantly, micro and nano-engineerd surfaces for phase change heat transfer represent an enhancement process that could be implemented either by retrofitting existing heat exchanger equipment or in manufacturing newer-generation heat exchangers.

## 5. Summary and outlook

The present review illustrates the rationale, techniques and challenges of engineering the texture and chemistry of surfaces for phase change heat transfer. While the idea of engineering surfaces for enhancing phase change heat transfer is not new [208], the degree of sophistication with which surfaces can be engineered today—owing to advances in micro- and nano-fabrication methods over the last two decades—has caused a resurgence of interest in the subject. Examples of the variety of methods available to the "surface engineer" have been presented, and a large choice of modifications of surface energy and textures is possible, from the nanoscale to the centimeter scale.

Based on several examples described in this review, it is the conviction of the authors that optimum surfaces for phase change heat transfer need to address the specificities of phase change heat transfer, in a similar way as *a key matches its lock*. In that sense, surfaces specifically need to be engineered as:

- Possessing multiscale texture, to address the multiple scales of phase change
- Featuring non-homogeneous wettability, to address the biphasic nature of phase change
- Adaptive, to address the transients and various regimes associated with phase change.

Very recently, several surfaces featuring multiscale textures and spatial variations of wettability have shown improved boiling [38, 58], condensation [288] and frost-avoidance performance [333]. Few surfaces currently possess the ability to adapt their features to the various regimes associated with phase change, but such strategies are feasible (e.g. using either passive thermal sensitive liquid crystal coatings [342], or magnetophoretic assembly of nanoparticles into nanofins for on-demand hot spot cooling [343]).

While this review has exposed the related fundamentals and state of the art techniques to fabricate such surfaces, we identify below several remaining challenges important to the design, manufacturing and widespread use of engineered surfaces for phase change heat transfer

There is a need for *design rules.* Phase change heat transfer covers such a wide range of length scales, from the nucleation scales (O(nm)) to the larger length scales associated with flow instabilities, which are O(cm). The time scales also vary over six orders of magnitude, from those relevant to nucleation (O(ms)) to those of frost growth (O(hours)). The way to obtain design rules is not clear at present. Several modeling techniques are available [66, 226, 344, 345], but they do not cover the multiples scales associated with phase change heat transfer on engineered surfaces. Computation time is typically prohibitive. Several theoretical analyses are also available, but typically focus on one or another regime, length scale or timescale. Molecular dynamics (MD) combined with multi-scale approaches will be valuable [346] in bridging the gap between the macroscale (which is of ultimate practical use) and the molecular scale processes, which designate behavior at the micro and macroscale. Several experimental methods exist to characterize phase change heat transfer, but there is a tremendous difficulty in handling the transients and multiple interfaces associated with boiling, the small temperature differences associated with condensation, and the solidification associated with freezing. Recent techniques such as liquid crystal chromatography can capture microscale spatial variations in substrate temperature [257]. Other recent experimental techniques are worthy of considering, such as Micro-Raman thermometry, thermal SEM, environmental SEM.

There is a need for *better physical understanding* of the physical aspects of phase change. In boiling, the evolution of the microlayer thickness and its contribution to the boiling heat transfer is still poorly quantified [62] and a subject of continued debate [347]. While the influence of wettability (static and dynamic) on nucleation has been studied from the field’s inception, its influence on nucleate and critical phase change heat transfer is still an open question. For instance, Dhir et al. [47] state that, "*In spite of all the advances in the computational techniques for solving boiling problems, the one variable that has still not been modeled correctly is the contact angle (static or dynamic).*" Moreover, a recent report redefines the threshold angle for wettability transition (from wettable to non-wettable), and makes a case for its dependence on surface tension (eg. for hexadecane this transition occurs at $125^o$, which is far removed from the $90^o$ value applying for water).[348] There is also a need to study how surface coatings modify the thermal conductivity of surfaces. The physics behind DwC nucleation and its dependence on wetting are poorly understood, as the temperature difference involved in the process is often in the range of the instrumentation error. In frosting, boiling and condensation there is a strong need for experimental and theoretical studies on the poorly understood transition between the nucleation-driven high HTC phase change heat transfer and the single phase heat transport.

Another challenge is extending the knowledge base and techniques mostly developed for water to *technical fluids such as refrigerants*, used in the heating, ventilation and air conditioning industries. Due to the very low surface tension of refrigerants, it is difficult to fabricate surfaces that are not wetted by them. A recent investigation of DwC of low surface tension fluids motivates further studies on development of ultra-low-surface-energy modifiers and lubricant-impregnated surfaces compatible to fluorinated refrigerants to achieve a major breakthrough in this area.[349]

Also, technical applications call for surfaces that are *durable* and stable over time. Nanostructures with very small cross-sections are prone to failure under stress as shown in frost formation studies [319, 350, 351]. The longevity of a surface enhancement feature is important. Indeed, simple surface enhancements like sandblasting can dissipate "within a day" [49], mostly because the additional cavities degas slowly, or hydrophobic spots could delaminate. Chemical stability is also a challenge for several superhydrophilic patterning techniques, as is the case with UV-irradiation of $TiO_2$ containing coatings which revert to their original hydrophobic state when stored [352]. Thermal stability is critical for surfaces for phase change heat transfer. Additionally, micro and nanostructures may exhibit different thermal behavior compared to bulk materials [353]. The resistance to fouling [354] of these surfaces also needs to be investigated. It is difficult to claim *a priori* that enhanced surfaces are more or less prone to fouling than other enhanced surfaces or flat surfaces. Interestingly, low surface energy has also been shown to reduce and prevent fouling [182], and nanostructures with pitches and features significantly smaller than the typical size fouling particles have been shown to mitigate fouling [355-357], for both superhydrophobic and superhydrophilic samples [3].

There are several obstacles that need to be addressed to produce *commercially viable surfaces* for controlling phase change heat transfer. These issues have been reviewed in [3, 49, 355, 358-360].

## Acknowledgements

DA and CMM acknowledge funding from NSF Grants CBET-1066426, 1236030 and 1235867. The authors acknowledge insightful discussions with Gilbert Moreno and Sreekant Narumanchi from NREL, as well as with the participants of the 2013 International Workshop on Micro and Nano Structures for Phase Change Heat Transfer, organized by Yoav Peles (RPI) and Evelyn Wang (MIT). With his academic offspring CJK, the authors also acknowledge the exceptional mentoring of Art Bergles, a selfless leader in the community of enhanced heat transfer, who passed away in March 2014.

## References

1. R. M. Manglik and M. A. Jog, Molecular-to-Large-Scale Heat Transfer With Multiphase Interfaces: Current Status and New Directions, *Journal of Heat Transfer*, 2009, **131**, 121001.
2. F. P. Incropera and D. P. DeWitt, *Fundamentals of Heat and Mass Transfer*, Wiley, 1995.
3. K. Page, M. Wilson, N. J. Mordan, W. Chrzanowski, J. Knowles and I. P. Parkin, Study of the adhesion of Staphylococcus aureus to coated glass substrates, *Journal of Materials Science*, 2011, **46**, 6355-6363.
4. M. J. Moran, H. N. Shapiro, B. R. Munson and D. P. DeWitt, *Introduction to thermal systems engineering*, John Wiley and Sons, Danvers, MA, 2003.
5. J. G. Collier, *Convective Boiling and Condensation*, McGraw-Hill, 1972.
6. A. Bar-Cohen, M. Arik and M. Ohadi, Direct liquid cooling of high flux micro and nano electronic components, *Proceedings of the IEEE*, 2006, **94**, 1549-1570.
7. J. P. McHale and S. V. Garimella, Bubble nucleation characteristics in pool boiling of a wetting liquid on smooth and rough surfaces, *International Journal of Multiphase flow*, 2010, **36**, 249-260.
8. C. Y. Han and P. Griffith, *The Mechanism of Heat Transfer in Nucleate Pool Boiling*, MIT, 1962.
9. F. Demiray and J. Kim, Microscale heat transfer measurements during pool boiling of FC-72: effect of subcooling, *International Journal of Heat and Mass Transfer*, 2004, **47**, 3257-3268.
10. Y. Y. Jiang, H. Osada, M. Inagaki and N. Horinouchi, Dynamic modeling on bubble growth, detachment and heat transfer for hybrid-scheme computations of nucleate boiling, *International Journal of Heat and Mass Transfer*, 2013, **56**, 640-652.
11. I. Golobic, J. Petkovsek and D. B. R. Kenning, Bubble growth and horizontal coalescence in saturated pool boiling on a titanium foil, investigated by high-speed IR thermography, *International Journal of Heat and Mass Transfer*, 2012, **55**, 1385-1402.
12. G. Son and V. K. Dhir, Numerical simulation of nucleate boiling on a horizontal surface at high heat fluxes, *International Journal of Heat and Mass Transfer*, 2008, **51**, 2566-2582.
13. V. K. Dhir, H. S. Abarajith and D. Li, Bubble Dynamics and Heat Transfer during Pool and Flow Boiling, *Heat Transfer Engineering*, 2007, **28**, 608-624.
14. J. Boreyko and C.-H. Chen, Self-Propelled Dropwise Condensate on Superhydrophobic Surfaces, *Physical Review Letters*, 2009, **103**.
15. C.-H. Chen, Q. Cai, C. Tsai, C.-L. Chen, G. Xiong, Y. Yu and Z. Ren, Dropwise condensation on superhydrophobic surfaces with two-tier roughness, *Applied Physics Letters*, 2007, **90**, 173108.
16. S. Anand, A. T. Paxson, R. Dhiman, J. D. Smith and K. K. Varanasi, Enhanced Condensation on Lubricant-impregnated nanotextured surfaces, *Langmuir*, 2012, **6**, 10122-10129.
17. K. Rykaczewski, J. H. J. Scott and A. G. Fedorov, Electron beam heating effects during environmental scanning electron microscopy imaging of water condensation on superhydrophobic surfaces, *Applied Physics Letters*, 2011, **98**, 093106.
18. K.-S. Lee, S. Jhee and D.-K. Yang, Prediction of the frost formation on a cold flat surface, *International Journal of Heat and Mass Transfer*, 2003, **46**, 3789-3796.
19. Y. Hayashi, A. Aoki, S. Adachi and K. Hori, Study of Frost Properties Correlating with Frost Formation Types, *Journal of Heat Transfer*, 1977, **99**, 239-245.
20. L. Mishchenko, B. Hatton, V. Bahadur, J. A. Taylor, T. Krupenkin and J. Aizenberg, Design of ice-free nanostructured surfaces based on repulsion of impacting water droplets, *ACS Nano*, 2010, **4**, 7699-7707.
21. N. A. Patankar, Supernucleating surfaces for nucleate boiling and dropwise condensation heat transfer, *Soft Matter*, 2010, **6**, 1613-1620.
22. H. S. Ahn, C. Lee, H. Kim, H. Jo, S. Kang, J. Kim, J. Shin and M. H. Kim, Pool boiling CHF enhancement by micro/nanoscale modification of zircaloy-4 surface, *Nuclear Engineering and Design*, 2010, **240**, 3350-3360.
23. Y. H. Kim, S. J. Kim, K. Y. Suh, J. L. Rempe, F. B. Cheung and S. B. Kim, Internal vessel cooling feasibility attributed by critical heat flux in inclined rectangular gap, *Nuclear Technology*, 2006, **154**, 13-40.
24. J. C. M. Leung, K. A. Gallivan, R. E. Henry and S. G. Bankoff, Critical heat flux predictions during blowdown transients, *International Journal of Multiphase flow*, 1981, **7**, 677-701.
25. K. Kurokawa, *The Fukushima nuclear accident independent investigation comission*, The National Diet of Japan, Japan, 2012.
26. J. Kim, *U Maryland, Mechanical Engineering, personal communication with D.A. on the use of dimensionless analysis in pool boiling*, May 2013.
27. R. W. Schrage, *A theoretical study of interphase mass transfer* New York, NY, 1953.
28. J. C. Corman and M. H. McLaughlin, Boiling augmentation with structured surfaces, *ASHRAE Transactions*, 1976, **82**, 906-918.
29. O. N. Mankovskij, O. B. Ioffe, L. G. Fibgant and A. R. Tolczinskij, About boiling mechanism on flooded surface with capillary-porous coating, *Ing. Fiz. J*, 1976, **30**, 975-982.
30. H. Ogata and W. Nakayama, Heat transfer to subcritical and supercritical helium in centrifugal acceleration fields 1. Free convection regime and boiling regime, *Cryogenics*, 1977, **17**, 461-470.

31. K. Nishikawa, T. Ito and K. Tanaka, Enhanced heat transfer by nucleate boiling on a sintered metal layer, *Heat Transfer - Japanese Research*, 1979, **8**, 65-81.
32. W. Nakayama, T. Daikoku, H. Kuwahara and T. Nakajima, Dynamic model of enhanced boiling heat transfer on porous surfaces, *Journal of Heat Transfer*, 1980, **102**, 451-456.
33. Y. Takata, S. Hidaka, M. Masuda and T. Ito, Pool boiling on a superhydrophilic surface, *International Journal of Energy Research*, 2003, **27**, 111-119.
34. C. Li and G. P. Peterson, Parametric study of pool boiling on horizontal highly conductive microporous coated surfaces, *Journal of Heat Transfer-Transactions of the ASME*, 2007, **129**, 1465-1475.
35. C. Li, Z. Wang, P. I. Wang, Y. Peles, N. Koratkar and G. P. Peterson, Nanostructured copper interfaces for enhanced boiling, *Small*, 2008, **4**, 1084-1088.
36. J. A. Weibel, S. V. Garimella and M. T. North, Characterization of evaporation and boiling from sintered powder wicks fed by capillary action, *International Journal of Heat and Mass Transfer*, 2010, **53**, 4204-4215.
37. D. Cooke and S. G. Kandlikar, Effect of open microchannel geometry on pool boiling enhancement, *International Journal of Heat and Mass Transfer*, 2012, **55**, 1004-1013.
38. S. G. Kandlikar, Controlling bubble motion over heated surface through evaporation momentum force to enhance pool boiling heat transfer, *Applied Physics Letters*, 2013, **102**, 051611.
39. S. Nukiyama, Maximum and minimum values of heat transmitted from metal to boiling water under atmospheric pressure, *Jpn. Soc. Mech. Eng.*, 1934, **37**, 367-374.
40. B. B. Mikic and W. M. Rohsenow, A new correlation of pool boiling data including effect of heating surface characteristics, *Journal of Heat Transfer*, 1969, **91**, 245-250.
41. K.-Y. Hsu, On the size range of active nucleation cavities on a heating surface, *ASME Journal of Heat Transfer*, 1962, **84**, 207-216.
42. C. H. Wang and V. K. Dhir, Effect of surface wettability on active nucleation site density during pool boiling of water on a vertical surface, *Journal of Heat Transfer-Transactions of the ASME*, 1993, **115**, 659-669.
43. A. R. Betz, J. Xu, H. Qiu and D. Attinger, Do surfaces with mixed hydrophilic and hydrophobic areas enhance pool boiling?, *Applied Physics Letters*, 2010, **97**, 141909.
44. H. Jo, H. S. Ahn, S. Kang and M. H. Kim, A study of nucleate boiling heat transfer on hydrophilic, hydrophobic and heterogeneous wetting surfaces, *International Journal of Heat and Mass Transfer*, 2011, **54**, 5643-5652.
45. G. S. Hwang and M. Kaviany, Critical heat flux in thin, uniform particle coatings, *International Journal of Heat and Mass Transfer*, 2006, **49**, 844-849.
46. N. Zuber, *Hydrodynamic Aspects of Boiling Heat Transfer, AEC report AECU-4439 and PhD Thesis (UCLA)*, 1959.
47. V. K. Dhir, H. S. Abarajith and G. R. Warrier, From Nano to Micro to Macro Scales in Boiling, Fundamentals and Applications, (Edits) S. Kakac, L.L. Vasiliev, Y.Bayazitoglu, and Y.Yener, Proceedings of NATO-ASI Meeting, NATO Science Series II: Mathematics, Physics and Chemistry, Vol. 193, Kulwer Academic Publishers, The Netherlands., 2005.
48. T. G. Theofanous, T. N. Dinh, J. P. Tu and A. T. Dinh, The boiling crisis phenomenon - Part II: dryout dynamics and burnout, *Experimental Thermal and Fluid Science*, 2002, **26**, 793-810.
49. R. L. Webb, Odyssey of the enhanced boiling surface, *Journal of Heat Transfer*, 2004, **126**, 1051-1059.
50. P. J. Berenson, *Transition Boiling Heat Transfer from a Horizontal Surface*, M.I.T, 1960.
51. *R. L. Hummel,Means for increasing the heat transfer coefficient between a wall and boiling liquid,United States Pat.*, 3207209, 1965.
52. V. P. Carey, *Liquid-Vapor Phase-Change Phenomena*, Taylor & Francis Group, New York, NY, 2008.
53. J. Frenkel, A General Theory of Heterophase Fluctuations and Pretransition Phenomena, *J Chem Phys*, 1939, **7**, 538.
54. N. Basu, G. R. Warrier and V. K. Dhir, Onset of Nucleate Boiling and Active Nucleation Site Density During Subcooled Flow Boiling, *Journal of Heat Transfer*, 2002, **124**, 717.
55. R. T. Knapp, Cavitation and Nuclei, *Trans. ASME*, 1958, **80**, 1321.
56. S. G. Bankoff, The Prediction of Surface Temperature at incipient Boiling, *Chem. Eng. Prog. Symp. Ser.*, 1959, **55**, 87.
57. Y. Qi and J. F. Klausner, Comparison of Nucleation Site Density for Pool Boiling and Gas Nucleation, *Journal of Heat Transfer*, 2006, **128**, 13.
58. A. R. Betz, J. Jenkins, C.-J. Kim and D. Attinger, Boiling heat transfer on superhydrophilic, superhydrophobic, and superbiphilic surfaces, *International Journal of Heat and Mass Transfer*, 2013, **57**, 733-741.
59. C. L. Tien, A Hydrodynamic Model for Nucleate Pool boiling, *Int . J. Heat Mass Transfer*, 1962, **5**, 533-540.
60. H. K. Forster and N. Zuber, Dynamics of Vapor Bubbles and Boiling Heat Transfer, *AIChE*, 1955, **1**, 531-535.
61. S. I. Haider and R. L. Webb, A Transient Micro-convection Model of Nucleate Pool Boiling, *International Journal of Heat and Mass Transfer*, 1997, **40**, 3675-3688.
62. Y. Utaka, Y. Kashiwabara and M. Ozaki, Microlayer structure in nucleate boiling of water and ethanol at atmospheric pressure, *International Journal of Heat and Mass Transfer*, 2013, **57**, 222-230.
63. M. G. Cooper and A. J. P. lloyd, The Microlayer in Nucleate Pool Boiling, *International Journal of Heat and Mass Transfer*, 1969, **12**, 895-913.
64. W. Rohsenow, *A method of correlating heat transfer data for surface boiling of liquids*, MIT, 1951.
65. V. K. Dhir, Boiling heat transfer, *Annual Review of Fluid Mechanics*, 1998, **30**, 365-401.

66. H. S. Abarajith and V. K. Dhir, A Numerical Study of the Effect of Contact Angle on the Dynamics of a Single Bubble during Pool Boiling, New Orleans, Louisiana, 2002.
67. S. S. Kutateladze, On the Transition to Film Boiling under Natural Convection, *Kotloturbostroenie*, 1948, **3**, 10.
68. J. H. Lienhard and V. K. Dhir, *Extended Hydrodynamic Theory to the Peak and Minimum Pool Boiling Heat Fluxes*, NASA CR-2270, 1973.
69. Y. Haramura and Y. Katto, A New Hydrodynamic Model of the Critical Heat Flux, Applicable Widely to Both Pool and Forced Convective Boiling on Submerged Bodies in Saturated Liquids, *International Journal of Heat and Mass Transfer*, 1983, **26**, 389-399.
70. T. D. Bui and V. K. Dhir, Transition boiling heat transfer on a vertical surface, *Journal of Heat Transfer-Transactions of the ASME*, 1985, **107**, 756-763.
71. S. G. Kandlikar, A Theoretical Model to Predict Pool Boiling CHF Incorporating Effects of Contact Angle and Orientation, *Journal of Heat Transfer*, 2001, **123**, 1071.
72. V. K. Dhir and S. P. Liaw, Framework for a Unified Model for Nucleate and transition Pool Boiling, *Journal of Heat Transfer*, 1989, **111**, 739-746.
73. S. Kandlikar and S. Garimella, in *Heat transfer and fluid flow in minichannels and microchannels*, Elsevier, 2006, p. 227.
74. J. W. Rose, Dropwise condensation theory and experiment: A review, *Proceedings of the Institution of Mechanical Engineers, Part A: Journal of Power and Energy*, 2002, **216**, 115-128.
75. C. Graham and P. Griffith, Drop Size Distributions and Heat-Transfer in Dropwise Condensation, *International Journal of Heat and Mass Transfer*, 1973, **16**, 337-346.
76. A. Bejan, *Convective Heat Transfer*, Wiley, 2003.
77. E. Schmidt, W. Schurig and W. Sellschopp, Versuche über die Kondensation von Wasserdampf in Film- und Tropfenform, *Tech. Mech. Thermodyn. (Forsch. Ing. Wes.)*, 1930, Section 10.11.14.
78. A. P. M. Glassford, Practical Model for Molecular Contaminant Deposition Kinetics, *Journal of Thermophysics and Heat Transfer*, 1992, **6**, 656-664.
79. L. H. Chen, C. Y. Chen and Y. L. Lee, Nucleation and growth of clusters in the process of vapor deposition, *Surface Science*, 1999, **429**, 150-160.
80. E. J. Le Fevre and J. W. Rose, A theory of heat transfer by dropwise condensation, Chicago, IL, 1966.
81. J. W. Rose, A theory of heat transfer by dropwise condensation, Chicago, IL, 1967.
82. J. W. Rose, Interphase matter transfer, the condensation coefficient and dropwise condensation, 1998.
83. B. B. Mikic, On Mechanism of Dropwise Condensation, *International Journal of Heat and Mass Transfer*, 1969, **12**, 1311-1323.
84. D. Quere, M. J. Azzopardi and L. Delattre, Drops at rest on a tilted plane, *Langmuir*, 1998, **14**, 2213-2216.
85. S. Kim and K. J. Kim, Dropwise Condensation Modeling Suitable for Superhydrophobic Surfaces, *Journal of Heat Transfer*, 2011, **133**, 081502.
86. H. Tanaka, Measurements of Drop-Size Distributions during Transient Dropwise Condensation, *Journal of Heat Transfer-Transactions of the ASME*, 1975, **97**, 341-346.
87. Y. T. Wu, C. X. Yang and X. G. Yuan, Drop distributions and numerical simulation of dropwise condensation heat transfer, *International Journal of Heat and Mass Transfer*, 2001, **44**, 4455-4464.
88. S. Ulrich, S. Stoll and E. Pefferkorn, Computer simulations of homogeneous deposition of liquid droplets, *Langmuir*, 2004, **20**, 1763-1771.
89. H. Wenzel, Versuche über Tropfenkondensation, *Allg. Wärmetech*, 1957, **8**, 839-845.
90. R. W. Bonner, Correlation for dropwise condensation heat transfer: Water, organic fluids, and inclination, *International Journal of Heat and Mass Transfer*, 2013, **61**, 245-253.
91. X. H. Ma, X. D. Zhou, Z. Lan, Y. M. Li and Y. Zhang, Condensation heat transfer enhancement in the presence of non-condensable gas using the interfacial effect of dropwise condensation, *International Journal of Heat and Mass Transfer*, 2008, **51**, 1728-1737.
92. M. H. M. Grooten and C. W. M. van der Geld, Dropwise condensation from flowing air-steam mixtures: Diffusion resistance assessed by controlled drainage, *International Journal of Heat and Mass Transfer*, 2011, **54**, 4507-4517.
93. W. J. Minkowycz and E. M. Sparrow, Condensation heat transfer in the presence of non-condensables, interfacial resistance, super heating variable properties and diffusion, *International Journal of Heat and Mass Transfer*, 1966, **9**, 1125 - 1144.
94. Y. Utaka and T. Nishikawa, Measurement of condensate film thickness for solutal marangoni condensation applying laser extinction method, *Journal of Enhanced Heat Transfer*, 2003, **10**, 119-129.
95. Y. Utaka and T. Kamiyama, Condensate drop movement in Marangoni condensation by applying bulk temperature gradient on heat transfer surface, *Heat Transfer—Asian Research*, 2008, **37**, 387-397.
96. I. Tanasawa, Advances in Condensation Heat Transfer, New York, 1991.
97. W. Nusselt, Die Oberflachen Kondensation des Wasserdampfes, *Zeitschrift, Verein Deutscher Ingenieure*, 1916, **60**, 541-546.
98. W. M. Rohsenow, Heat Transfer and Temperature Distribution in Laminar Film Condensation, *Trans. ASME J. Fluids Eng.*, 1956, **78**, 1645.

99. P. L. Thibaut Brian, R. C. Reid and Y. T. Shah, Frost deposition on cold surfaces, *Industrial and Engineering Chemistry Fundamentals*, 1970, **9**, 375-380.
100. G. Fortin, J.-L. Laforte and A. Ilinca, Heat and mass transfer during ice accretion on aircraft wings with an improved roughness model, *International Journal of Heat and Mass Transfer*, 2006, **45**, 595-606.
101. J. Iragorry, Y. X. Tao and S. Jia, A critical review of properties and models for frost formation analysis, *HVAC and R Research*, 2004, **10**, 393-420.
102. R. O. Piucco, C. J. L. Hermes, C. Melo and J. R. Barbosa Jr, A study of frost nucleation on flat surfaces, *Experimental Thermal and Fluid Science*, 2008, **32**, 1710-1715.
103. C. C. Ryerson, Ice protection of offshore platforms, *Cold Regions Science and Technology*, 2011, **65**, 97-110.
104. N. H. Fletcher, *The Chemical Physics of Ice*, Cambridge University Press, London, 1970.
105. S. Jung, M. Dorrestijn, D. Raps, A. Das, C. M. Megaridis and D. Poulikakos, Are superhydrophobic surfaces best for icephobicity?, *Langmuir*, 2011, **27**, 3059-3066.
106. S. Jung, M. K. Tiwari, N. V. Doan and D. Poulikakos, Mechanism of supercooled droplet freezing on surfaces, *Nature Communications*, 2012, **3**.
107. B. Na and R. L. Webb, A fundamental understanding of factors affecting frost nucleation, *International Journal of Heat and Mass Transfer*, 2003, **46**, 3797-3808.
108. K. K. Varanasi, T. Deng, J. D. Smith, M. Hsu and N. Bhate, Frost formation and ice adhesion on superhydrophobic surfaces, *Applied Physics Letters*, 2010, **97**, 234102.
109. H. Lee, J. Shin, S. Ha, B. Choi and J. Lee, Frost formation on a plate with different surface hydrophilicity, *International Journal of Heat and Mass Transfer*, 2004, **47**, 4881-4893.
110. L. L. Cao, A. K. Jones, V. K. Sikka, J. Z. Wu and D. Gao, Anti-Icing Superhydrophobic Coatings, *Langmuir*, 2009, **25**, 12444-12448.
111. P. Kim, T. S. Wong, J. Alvarenga, M. J. Kreder, W. E. Adorno-Martinez and J. Aizenberg, Liquid-infused nanostructured surfaces with extreme anti-ice and anti-frost performance, *ACS Nano*, 2012, **6**, 6569-6577.
112. S. Jung, M. K. Tiwari and D. Poulikakos, Frost halos from supercooled water droplets, *Proceedings of the National Academy of Sciences of the United States of America*, 2012, **109**, 16073-16078.
113. B. Na and R. L. Webb, Mass transfer on and within a frost layer, *International Journal of Heat and Mass Transfer*, 2004, **47**, 899-911.
114. Z. Liu, X. Zhang, H. Wang, S. Meng and S. Cheng, Influences of surface hydrophilicity on frost formation on a vertical cold plate under natural convection conditions, *Experimental Thermal and Fluid Science*, 2007, **31**, 789-794.
115. R. Le Gall, J. M. Grillot and C. Jallut, Modelling of frost growth and densification, *International Journal of Heat and Mass Transfer*, 1997, **40**, 3177-3187.
116. R. L. Webb, The evolution of enhanced surface geometries for nucleate boiling, *Heat Transfer Engineering*, 1981, **2**, 46-69.
117. P. J. Berenson, Experiments on pool-boiling heat transfer, *International Journal of Heat and Mass Transfer*, 1962, **5**, 985-999.
118. *R. L. Webb,Heat transfer surface having a high boiling heat transfer coefficient,United States Pat.*, 3696861A, 1972.
119. F. Zhou, A. Izgorodin, R. Hocking, L. Spiccia and D. MacFarlane, Electrodeposited MnOx Films from Ionic Liquid for Electrocatalytic Water Oxidation, *Advanced Energy Materials*, 2012, **2**, 1013-1021.
120. Z. Jiang, Y. Tang, Q. Tay, Y. Zhang, O. I. Malyi, D. Wang, J. Deng, Y. Lai, H. Zhou, X. Chen, Z. Dong and Z. Chen, Understanding the Role of Nanostructures for Efficient Hydrogen Generation on Immobilized Photocatalysts, *Advanced Energy Materials*, 2013, **3**, 1368-1380.
121. *M. M. Dahl and L. E. Erb,Liquid heat Exchanger Interface and Method,United States Pat.*, 3990862, 1976.
122. W. Jiang and A. P. Malshe, A novel cBN composite coating design and machine testing: A case study in turning, *Surface and Coatings Technology*, 2011, **206**, 273-279.
123. C.-J. Kim and A. E. Bergles, in *Particulate Phenomena and Multiphase Transport*, Hemisphere, Washington DC, 1988, vol. 2, pp. 3-18.
124. *S. M. You and J. P. O'Connor,Boiling enhancement coating,United States Pat.*, 5814392, 1998.
125. Y. Xia and G. M. Whitesides, Soft litography, *Annual Review of Materials Research*, 1998, **28**, 153-184.
126. C. Lu and R. H. Lipson, Interference litography: a powerful tool for fabricating periodic structures, *Laser and Photonics Reviews*, 2009, **4**, 568-580.
127. Plymouth Grating Laboratory, Scanning-Beam Interference Lithography http://www.plymouthgrating.com/Technology/Technology%20Page.htm.
128. G. Sun, J. I. Hur, X. Zhao and C.-J. Kim, Fabrication of Very-High-Aspect-Ratio Micro Metal Posts and Gratings by Photoelectrochemical Etching and Electroplating, *J. MEMS*, 2011, **20**, 876-884.
129. C. Lee and C. J. Kim, Influence of surface hierarchy of superhydrophobic surfaces on liquid slip, *Langmuir*, 2011, **27**, 4243-4248.
130. J. A. Weibel, S. S. Kim, T. S. Fisher and S. V. Garimella, Carbon Nanotube Coatings for Enhanced Capillary-Fed Boiling from Porous Microstructures, *Nanoscale and Microscale Thermophysical Engineering*, 2012, **16**, 1-17.
131. Y.-W. Lu and S. G. Kandlikar, Nanoscale Surface Modification Techniques for Pool Boiling Enhancement: A Critical Review and Future Directions, *Heat Transfer Engineering*, 2011, **32**, 827-842.

132. K. Gerasopoulos, M. McCarthy, P. Banerjee, X. Fan, J. N. Culver and R. Ghodssi, Biofabrication methods for the patterned assembly and synthesis of viral nanotemplates, *Nanotechnology*, 2010, **21**.
133. K.-H. Chu, R. Enright and E. N. Wang, Structured surfaces for enhanced pool boiling heat transfer, *Applied Physics Letters*, 2012, **100**, 241603.
134. C.-H. Choi and C. J. Kim, Fabrication of dense array of tall nanostructures over a very large sample area with sidewall profile and tip sharpness control, *Nanotechnology*, 2006, **17**, 5326-5333.
135. K. Du, W. I., W. Mao, W. Xu and C. H. Choi, Large-area pattern transfer of metallic nanostructures on glass substrates via interference lithography, *Nanotechnology*, 2011, **22**, 285306.
136. T. Morimoto, Y. Sanada and H. Tomonaga, Wet chemical functional coatings for automotive glasses and cathode ray tubes, *Thin Solid Films*, 2001, **392**, 214-222.
137. L. Carrino, G. Moroni and W. Polini, Cold plasma treatment of polypropylene surface: a study on wettability and adhesion, *Journal of Materials Processing Technology*, 2002, **121**, 373-382.
138. K. Bobzin, N. Bagcivan, N. Goebbels, K. Yilmaz, B. R. Hoehn, K. Michaelis and M. Hochmann, Lubricated PVD CrAlN and WC/C coatings for automotive applications, *Surface & Coatings Technology*, 2009, **204**, 1097-1101.
139. J. Genzer and K. Efimenko, Recent developments in superhydrophobic surfaces and their relevance to marine fouling: a review, *Biofouling*, 2006, **22**, 339-360.
140. X. Wang, L. Zhi and K. Mullen, Transparent, Conductive Graphene Electrodes for Dye-Sensitized Solar Cells, *Nano Letters*, 2007, **8**, 323-327.
141. W. Barthlott and C. Neinhuis, Purity of the sacred lotus, or escape from contamination in biological surfaces, *Planta*, 1997, **202**, 1-8.
142. L. Feng, S. Li, Y. Li, H. Li, L. Zhang, J. Zhai, Y. Song, B. Liu, L. Jiang and D. Zhu, Super-hydrophobic surfaces: from natural to artificial, *Advanced Materials*, 2002, **14**, 1857-1860.
143. X. Feng, L. Feng, M. Jin, J. Zhai, L. Jiang and D. Zhu, Reversible super-hydrophobicity to super-hydrophilicity transition of aligned ZnO nanorod films, *Journal of the American Chemical Society*, 2004, **126**, 62-63.
144. G. B. Sigal, M. Mrksich and G. M. Whitesides, Effect of Surface Wettability on the Adsorption of Proteins and Detergents, *Journal of the American Chemical Society*, 1998, **120**, 3464-3473.
145. P. G. de Gennes, Wetting: Statics and Dynamics, *Reviews of Modern Physics*, 1985, **57**, 827-863.
146. P. G. de Gennes, F. Brochard-Wyart and D. Quéré, *Capillarity and wetting phenomena: drops, bubbles, pearls, waves*, 2004.
147. A. Marmur, Hydro- hygro- oleo- omni-phobic? Terminology of wettability classification, *Soft Matter*, 2012, **8**, 6867.
148. T. Young, An Essay on the Cohesion of Fluids, *Philosophical Transactions of the Royal Society*, 1804, **95**, 65-87.
149. A. Dupré and P. Dupré, *Théorie mécanique de la chaleur*, Paris, 1869.
150. R. N. Wenzel, Resistance of solid surface to wetting by water, *Industrial & Engineering Chemistry*, 1936, **28**, 988-994.
151. A. B. D. Cassie and S. Baxter, Wettability of porous surfaces, *Transactions of the Faraday Society*, 1944, **40**, 546-551.
152. X. J. Feng and L. Jiang, Design and Creation of Superwetting/Antiwetting Surfaces, *Advanced Materials*, 2006, **18**, 3063-3078.
153. L. Feng, Y. Zhang, J. Xi, Y. Zhu, N. Wang, F. Xia and L. Jiang, Petal effect: A superhydrophobic state with high adhesive force, *Langmuir*, 2008, **24**, 4114-4119.
154. C. Dorrer and J. Rühe, Some thoughts on superhydrophobic wetting, *Soft Matter*, 2009, **5**, 51.
155. M. Nosonovsky and B. Bhushan, Biomimetic Superhydrophobic Surfaces: Multiscale Approach, *Nano Letters*, 2007, **7**, 2633-2637.
156. C. Dorrer and J. Ruehe, Condensation and wetting transitions on microstructured ultrahydrophobic surfaces, *Langmuir*, 2007, **23**, 3820-3824.
157. I. U. Vakarelski, N. A. Patankar, J. O. Marston, D. Y. Chan and S. T. Thoroddsen, Stabilization of Leidenfrost vapour layer by textured superhydrophobic surfaces, *Nature*, 2012, **489**, 274-277.
158. R. E. Johnson and R. H. Dettre, Contact angle, Wettability and Adhesion, *Advances in Chemistry series*, 1964, **43**.
159. D. Oner and T. J. McCarthy, Ultrahydrophobic surfaces. Effect ot topography length scales on wettability, *Langmuir*, 2000, **16**, 7777-7782.
160. D. Richard and D. Quere, Viscous drops rolling on a tilted non-wettable solid, *Europhysics Letters*, 1999, **48**, 286-291.
161. V. K. Dhir, Nucleate and transition boiling heat transfer under pool and external flow conditions, *International J Heat Fluid Flow*, 1991, **12**, 290-314.
162. Y. Takata, S. Hidaka, J. M. Cao, T. Nakamura, H. Yamamoto, M. Masuda and T. Ito, Effect of surface wettability on boiling and evaporation, *Energy*, 2005, **30**, 209-220.
163. R. Wang, K. Hashimoto, A. Fujishima, M. Chikuni, E. Kojima, A. Kitamura, M. Shimohigoshi and T. Watanabe, Light-induced amphiphilic surfaces, *Nature*, 1997, **388**, 431-432.
164. W. A. Zisman, in *Contact Angle, Wettability, and Adhesion*, 1964, ch. 2, pp. 1-51.
165. H. T. Phan, N. Caney, P. Marty, S. Colasson and J. Gavillet, How does surface wettability influence nucleate boiling?, *Comptes Rendus Mécanique*, 2009, **337**, 251-259.
166. H. T. Phan, N. Caney, P. Marty, S. Colasson and J. Gavillet, Surface wettability control by nanocoating: The effects on pool boiling heat transfer and nucleation mechanism, *International Journal of Heat and Mass Transfer*, 2009, **52**, 5459-5471.

167. S. P. Liaw and V. K. Dhir, Effect of surface wettability on transition boiling heat transfer from a vertical surface., San Francisco, CA, 1986.
168. X. H. Ma, J. W. Rose, D. Q. Xu, J. F. Lin and B. X. Wang, Advances in dropwise condensation heat transfer: Chinese research, *Chemical Engineering Journal*, 2000, **78**, 87-93.
169. Q. Zhao and B. M. Burnside, Dropwise condensation of steam on ion-implanted condenser surfaces, *Heat Recovery Systems & Chp*, 1994, **14**, 525-534.
170. G. Azimi, R. Dhiman, H.-M. Kwon, A. T. Paxson and K. K. Varanasi, Hydrophobicity of rare-earth oxide ceramics, *Nat Mater*, 2013, **12**, 315-320.
171. R. I. Vachon, G. H. Nix, G. E. Tanger and R. O. Cobb, Pool boiling heat transfer from Teflon-coated stainless steel, *Journal of Heat Transfer*, 1969, **91**, 364-369.
172. C. D. Bain, E. B. Troughton, Y. T. Tao, J. Evall, G. M. Whitesides and R. G. Nuzzo, Formation of monolayer films by the spontaneous assembly of organic thiols from solution onto gold, *Journal of the American Chemical Society*, 1989, **111**, 321-335.
173. K. M. Balss, C. T. Avedisian, R. E. Cavicchi and M. J. Tarlov, Nanosecond imaging of microboiling behavior on pulsed-heated Au films modified with hydrophilic and hydrophobic self-assembled monolayers, *Langmuir*, 2005, **21**, 10459-10467.
174. B. Bourdon, R. Rioboo, M. Marengo, E. Gosselin and J. De Coninck, Influence of the Wettability on the Boiling Onset, *Langmuir*, 2012, **28**, 1618-1624.
175. O. C. Thomas, R. E. Cavicchi and M. J. Tarlov, Effect of surface Wettability on fast transient microboiling behavior, *Langmuir*, 2003, **19**, 6168-6177.
176. L. C. F. Blackman, M. J. S. Dewar and H. Hampson, An investigation of compounds promoting the dropwise condensation of steam, *Journal of Applied Chemistry*, 1957, **7**, 160-171.
177. D. W. Tanner, D. Pope, C. J. Potter and D. West, Heat transfer in dropwise condensation-Part II Surface chemistry, *International Journal of Heat and Mass Transfer*, 1965, **8**, 427-436.
178. E. F. Hare, E. G. Shafrin and W. A. Zisman, Properties Of Films Of Adsorbed Fluorinated Acids, *The Journal of Physical Chemistry*, 1954, **58**, 236-239.
179. Q. Zhao, D. C. Zhang, J. F. Lin and G. M. Wang, Dropwise condensation on L-B film surface, *Chemical Engineering and Processing*, 1996, **35**, 473-477.
180. E. Forrest, E. Williamson, J. Buongiorno, L.-W. Hu, M. Rubner and R. Cohen, Augmentation of nucleate boiling heat transfer and critical heat flux using nanoparticle thin-film coatings, *International Journal of Heat and Mass Transfer*, 2010, **53**, 58-67.
181. C.-C. Hsu and P.-H. Chen, Surface wettability effects on critical heat flux of boiling heat transfer using nanoparticle coatings, *International Journal of Heat and Mass Transfer*, 2012, **55**, 3713-3719.
182. J. D. Smith, A. J. Meuler, H. L. Bralower, R. Venkatesan, S. Subramanian, R. E. Cohen, G. H. McKinley and K. K. Varanasi, Hydrate-phobic surfaces: fundamental studies in clathrate hydrate adhesion reduction, *Physical Chemistry Chemical Physics*, 2012, **14**, 6013-6020.
183. A. T. Paxson , J. L. Yagüe, K. K. Gleason and K. K. Varanasi, Stable Dropwise Condensation for Enhancing Heat Transfer via the Initiated Chemical Vapor Deposition (iCVD) of Grafted Polymer Films, *Adv. Mater.*, 2013, **26**, 418-423.
184. D. S. Wen and B. X. Wang, Effects of surface wettability on nucleate pool boiling heat transfer for surfactant solutions, *International Journal of Heat and Mass Transfer*, 2002, **45**, 1739-1747.
185. S. Morgenthaler, C. Zink and N. D. Spencer, Surface-chemical and -morphological gradients, *Soft Matter*, 2008, **4**, 419-434.
186. L. Zhai, M. C. Berg, F. C. Cebeci, Y. Kim, J. M. Milwid, M. F. Rubner and R. E. Cohen, Patterned superhydrophobic surfaces: Toward a synthetic mimic of the Namib Desert beetle, *Nano Letters*, 2006, **6**, 1213-1217.
187. A. R. Parker and C. R. Lawrence, Water capture by a desert beetle, *Nature*, 2001, **414**, 33-34.
188. M. K. Chaudhury and G. M. Whitesides, How to make water run uphill, *Science*, 1992, **256**, 1539-1541.
189. *R. F. Gaertner,Method and Means for Increasing the Heat Transfer Coefficient between a Wall and Boiling Liquid,United States Pat.*, 3301314, 1967.
190. G. P. Lopez, H. A. Biebuyck, C. D. Frisbie and G. M. Whitesides, Imaging of Features on Surfaces by Condensation Figures, *Science*, 1993, **260**, 647-649.
191. N. L. Abbott, J. P. Folkers and G. M. Whitesides, Manipulation of the wettability of surfaces on the 0.1-micrometer to 1-micrometer scale through micromachining and molecular self-assembly, *Science*, 1992, **257**, 1380-1382.
192. S. C. Thickett, C. Neto and A. T. Harris, Biomimetic Surface Coatings for Atmospheric Water Capture Prepared by Dewetting of Polymer Films, *Advanced Materials*, 2011, **23**, 3718-3722.
193. K. K. Varanasi, M. Hsu, N. Bhate, W. Yang and T. Deng, Spatial control in the heterogeneous nucleation of water, *Applied Physics Letters*, 2009, **95**.
194. L. Mishchenko, J. Aizenberg and B. D. Hatton, Spatial Control of Condensation and Freezing on Superhydrophobic Surfaces with Hydrophilic Patches, *Advanced functional materials*, 2013, **40**, 546-551.
195. V. Jokinen, L. Sainiemi and S. Franssila, Complex droplets on chemically modified silicon nanograss, *Advanced Materials*, 2008, **20**, 3453-3456.

196. A. Lee, M.-W. Moon, H. Lim, W.-D. Kim and H.-Y. Kim, Water harvest via dewing, *Langmuir*, 2012, **28**, 10183-10191.
197. K. Tadanaga, J. Morinaga, A. Matsuda and T. Minami, Superhydrophobic-superhydrophilic micropatterning on flowerlike alumina coating film by the sol-gel method, *Chemistry of Materials*, 2000, **12**, 590-592.
198. *E. D. Branson, P. B. Shah, S. Singh and C. J. Brinker,Preparation of hydrophobic coatings,US Pat.*, 7,485,343, 2009.
199. R. P. Garrod, L. G. Harris, W. C. E. Schofield, J. McGettrick, L. J. Ward, D. O. H. Teare and J. P. S. Badyal, Mimicking a stenocara beetle's back for microcondensation using plasmachemical patterned superhydrophobic-superhydrophilic surfaces, *Langmuir*, 2007, **23**, 689-693.
200. S. J. Pastine, D. Okawa, B. Kessler, M. Rolandi, M. Llorente, A. Zettl and J. M. J. Frechet, A facile and patternable method for the surface modification of carbon nanotube forests using perfluoroarylazides, *Journal of the American Chemical Society*, 2008, **130**, 4238-4239.
201. E. K. Her, T. J. Ko, K. R. Lee, K. H. Oh and M. W. Moon, Bioinspired steel surfaces with extreme wettability contrast, *Nanoscale*, 2012, **4**, 2900-2905.
202. S. P. R. Kobaku, A. K. Kota, D. H. Lee, J. M. Mabry and A. Tuteja, Patterned Superomniphobic-Superomniphilic Surfaces: Templates for Site-Selective Self-Assembly, *Angewandte Chemie-International Edition*, 2012, **51**, 10109-10113.
203. T. M. Schutzius, I. S. Bayer, G. M. Jursich, A. Das and C. M. Megaridis, Superhydrophobic-superhydrophilic binary micropatterns by localized thermal treatment of polyhedral oligomeric silsesquioxane (POSS)-silica films, *Nanoscale*, 2012, **4**, 5378-5385.
204. E. Ueda and P. A. Levkin, Emerging Applications of Superhydrophilic-Superhydrophobic Micropatterns, *Advanced Materials*, 2013, **25**, 1234-1247.
205. M. S. Sarwar, Y. H. Jeong and S. H. Chang, Subcooled flow boiling CHF enhancement with porous surface coatings, *International Journal of Heat and Mass Transfer*, 2007, **50**, 3649-3657.
206. X. Zhou and K. Bier, Pool boiling heat transfer from a horizontal tube coated with oxide ceramics, *Int. J. Refrig.*, 1997, **20**, 552-560.
207. J. Zimmermann, M. Rabe, G. R. J. Artus and S. Seeger, Patterned superfunctional surfaces based on a silicone nanofilament coating, *Soft Matter*, 2008, **4**, 450-452.
208. M. Jakob and W. Fritz, Versuche über den Verdampfungsvorgang, *Forschung im Ingenieurwesen*, 1931, **2**, 435-447.
209. C. Corty and A. S. Foust, Surface variables in nucleate boiling, *Chemical Engineering Progress Symposium Series*, 1955, **51**, 1-12.
210. H. M. Kurihara and J. E. Myers, The effects of superheat and surface roughness on boiling coefficients, *AIChe Journal*, 1960, **6**, 83-91.
211. A. E. Bergles and R. M. Manglik, Current progress and new developments in enhanced heat and mass transfer, *Journal of Enhanced Heat Transfer*, 2013, **20**, 1-15.
212. S. G. Bankoff, Entrapment of Gas in the spreading of a liquid over a rough surface, *A.I.Ch.E. Journal*, 1958, 24-26.
213. S. G. Bankoff, Ebullition from solid surfaces in the presence of pre-existing gaseous phase, *Trans. ASME*, 1957, **79**, 735.
214. J. J. Lorenz, B. B. Mikic and W. M. Rohsenow, The effects of Surface Conditions on Boiling Characteristics, *M.I.T. report*, 1972.
215. B. J. Zhang, K. J. Kim and H. Yoon, Enhanced heat transfer performance of alumina sponge-like nano-porous structures through surface wettability control in nucleate pool boiling, *International Journal of Heat and Mass Transfer*, 2012, **55**, 7487-7498.
216. C.-J. Kim, Structured Surfaces for Enhanced Nucleate Boiling, M.S. Thesis, 1985.
217. H. B. Clark, P. S. Strenge and J. Westwater, Active Sites for Nucleate Boiling, *Chem. Eng. Prog. Symp. Ser.*, 1959, **55**, 103-110.
218. S. R. Yang and R. H. Kim, A Mathematical Model of the Nucleation Site Density in Terms of the Surface Characteristics, *International Journal of Heat and Mass Transfer*, 1988, **31**, 1127-1135.
219. P. Griffith and J. D. Wallis, *The Role of Surface Conditions in Nucleate Boiling*, MIT, 1958.
220. M. Shoji, Studies of boiling chaos: a review, *International Journal of Heat and Mass Transfer*, 2004, **47**, 1105-1128.
221. P. J. Marto and W. Rohsenow, Effects of Surface Conditions on Nucleate Pool Boiling of Sodium, *Journal of Heat Transfer*, 1966, **88**, 196-203.
222. *R. M. Milton,Heat Exchange System,United States Pat.*, 3384154, 1968.
223. *R. M. Milton,Heat Exchange System,United States Pat.*, 3523577, 1970.
224. *R. M. Milton,Heat Exchange System With Porous Boiling Layer,United States Pat.*, 3587730, 1971.
225. L. H. Chien and R. L. Webb, Visualization of pool boiling on enhanced surfaces, *Experimental Thermal and Fluid Science*, 1998, **16**, 332-341.
226. L.-H. Chien and R. L. Webb, A nucleate boiling model for structured enhanced surfaces, *International Journal of Heat and Mass Transfer*, 1998, **41**, 2183-2195.
227. S. Ujereh, T. S. Fisher and I. Mudawar, Effect of carbon nanotube arrays on nucleate pool boiling, *International Journal of Heat and Mass Transfer*, 2007, **50**, 4023-4038.
228. R. F. Gaertner, *Effect of Heater Surface Chemistry on the Level of Burnout Heat heat flux in pool boiling*, General Electric, 1963.

229. B. Bourdon, P. Di Marco, R. Rioboo, M. Marengo and J. De Coninck, Enhancing the onset of pool boiling by wettability modification on nanometrically smooth surfaces, *International Communications in Heat and Mass Transfer*, 2013, **45**, 11-15.
230. Y. Takata, S. Hidaka and T. Uraguchi, Boiling Feature on a Super Water-Repellent Surface, *Heat Transfer Engineering*, 2006, **27**, 25-30.
231. M.-C. Lu, R. Chen, V. Srinivasan, V. P. Carey and A. Majumdar, Critical heat flux of pool boiling on Si nanowire array-coated surfaces, *International Journal of Heat and Mass Transfer*, 2011, **54**, 5359-5367.
232. R. Chen, M. C. Lu, V. Srinivasan, Z. Wang, H. H. Cho and A. Majumdar, Nanowires for enhanced boiling heat transfer, *Nano Letters*, 2009, **9**, 548-553.
233. Z. Yao, Y. W. Lu and S. G. Kandlikar, Effects of nanowire height on pool boiling performance of water on silicon chips, *International Journal of Thermal Sciences*, 2011, **50**, 2084-2090.
234. Z. Yao, Y.-W. Lu and S. G. Kandlikar, Direct growth of copper nanowires on a substrate for boiling applications, *Micro & Nano Letters*, 2011, **6**, 563-566.
235. X. Dai, X. Huang, F. Yang, X. Li, J. Sightler, Y. Yang and C. Li, Enhanced nucleate boiling on horizontal hydrophobic-hydrophilic carbon nanotube coatings, *Applied Physics Letters*, 2013, **102**.
236. T. J. Hendricks, S. Krishnan, C. Choi, C.-H. Chang and B. Paul, Enhancement of pool-boiling heat transfer using nanostructured surfaces on aluminum and copper, *International Journal of Heat and Mass Transfer*, 2010, **53**, 3357-3365.
237. S. Li, R. Furberg, M. S. Toprak, B. Palm and M. Muhammed, Nature-inspired boiling enhancement by novel nanostructured macroporous surfaces, *Advanced functional materials*, 2008, **18**, 2215-2220.
238. R. Furberg, B. Palm, S. Li, M. Toprak and M. Muhammed, The Use of a Nano- and Microporous Surface Layer to Enhance Boiling in a Plate Heat Exchanger, *Journal of Heat Transfer-Transactions of the ASME*, 2009, **131**.
239. H. S. Ahn, H. J. Jo, S. H. Kang and M. H. Kim, Effect of liquid spreading due to nano/microstructures on the critical heat flux during pool boiling, *Applied Physics Letters*, 2011, **98**.
240. J. Shen, C. Graber, J. Liburdy, D. Pence and V. Narayanan, Simultaneous droplet impingement dynamics and heat transfer on nano-structured surfaces, *Experimental Thermal and Fluid Science*, 2010, **34**, 496-503.
241. C. Y. Lee, M. M. H. Bhuiya and K. J. Kim, Pool boiling heat transfer with nano-porous surface, *International Journal of Heat and Mass Transfer*, 2010, **53**, 4274-4279.
242. V. Sathyamurthi, H. S. Ahn, D. Banerjee and S. C. Lau, Subcooled Pool Boiling Experiments on Horizontal Heaters Coated With Carbon Nanotubes, *Journal of Heat Transfer-Transactions of the ASME*, 2009, **131**.
243. H. D. Kim and M. H. Kim, Effect of nanoparticle deposition on capillary wicking that influences the critical heat flux in nanofluids, *Applied Physics Letters*, 2007, **91**, 014104.
244. J. Y. Chang and S. M. You, Boiling Heat transfer phenomena from micro-porous and porous surfaces in saturated FC-72, *International Journal of heat and Mass transfer* 1997, **40**, 4437-4447.
245. G. Moreno, S. Narumanchi and C. King, Pool Boiling Heat Transfer Characteristics of HFO-1234yf on Plain and Microporous-Enhanced Surfaces, *Journal of Heat Transfer*, 2013, **135**, 111014.
246. B. Feng, K. Weaver and G. P. Peterson, Enhancement of critical heat flux in pool boiling using atomic layer deposition of alumina, *Applied Physics Letters*, 2012, **100**, 053120.
247. S. Launay, A. G. Fedorov, Y. Joshi, A. Cao and P. M. Ajayan, Hybrid micro-nano structured thermal interfaces for pool boiling heat transfer enhancement, *Microelectronics Journal*, 2006, **37**, 1158-1164.
248. R. L. Webb, Odyssey of the Enhanced Boiling Surface, *ASME Conference Proceedings*, 2004, **2004**, 961-969.
249. S. G. Liter and M. Kaviany, Pool-boiling CHF enhancement by modulated porous-layer coating: theory and experiment, *International Journal of Heat and Mass Transfer*, 2001, **44**, 4287-4311.
250. S. Kim, H. D. Kim, H. Kim, H. S. Ahn, H. Jo, J. Kim and M. H. Kim, Effects of nano-fluid and surfaces with nano structure on the increase of CHF, *Experimental Thermal and Fluid Science*, 2010, **34**, 487-495.
251. Y. Nam and Y. S. Ju, Bubble nucleation on hydrophobic islands provides evidence to anomalously high contact angles of nanobubbles, *Applied Physics Letters*, 2008, **93**, 103115.
252. B. J. Suroto, M. Tashiro, S. Hirabayashi, S. Hidaka, M. Kohno and Y. Takata, Effects of hydrophobic-spot periphery and subcooling on nucleate pool boiling from a mixed-wettability surface, *Journal of Thermal Science and Technology*, 2013, **8**, 294-308.
253. X. Wang, Y. Song and H. Wang, An experimental study of bubble formation on a microwire coated with superhydrophobic micropatterns, *Heat Transfer Research*, 2013, **44**, 59-70.
254. A. E. Bergles and H. L. Morton, *Survey and evaluation of techniques to augment convective heat transfer*, Cambridge, Mass. : M.I.T. Dept. of Mechanical Engineering, 1965.
255. A. G. Williams, S. S. Nandapurkar and F. A. Holland, A review of methods for enhancing heat transfer rates in surface condensers, *Transactions of the Institution of Chemical Engineers and the Chemical Engineer*, 1968, **46**, CE367-CE373.
256. R. Gregorig, Film condensation on finely rippled surfaces with consideration of surface tension, *Zeitschrift fur Angewandte Mathematik und Physik*, 1954, **5**, 36-49.
257. G. D. Bansal, S. Khandekar and K. Muralidhar, Measurement of Heat Transfer During Drop-Wise Condensation of Water on Polyethylene, *Nanoscale and Microscale Thermophysical Engineering*, 2009, **13**, 184-201.

258. R. Enright, N. Miljkovic, J. L. Alvarado, K. Kim and J. W. Rose, Dropwise Condensation on Micro- and Nanostructured Surfaces, *Nanoscale and Microscale Thermophysical Engineering*, 2014, **In press**.
259. N. Miljkovic, R. Enright and E. N. Wang, Effect of Droplet Morphology on Growth Dynamics and Heat Transfer during Condensation on Superhydrophobic Nanostructured Surfaces, *ACS Nano*, 2012, **6**, 1776-1785.
260. J. B. Boreyko and C. P. Collier, Dewetting Transitions on Superhydrophobic Surfaces: When are Wenzel Drops Reversible?, *J. Phys. Chem. C*, 2013, DOI: 10.1021/jp4053083.
261. T. Haraguchi, R. Shimada, S. Kumagai and T. Takeyama, The effect of Polyvynilidene Chloride coating thickness on promotion of Dropwise Steam Condensation, *International Journal of Heat and Mass Transfer*, 1991, **34**, 3047-3054.
262. P. J. Marto, D. J. Looney, J. W. Rose and A. S. Wanniarachchi, Evaluation of organic coatings for the promotion of dropwise condensation of steam, *International Journal of Heat and Mass Transfer*, 1986, **29**, 1109-1117.
263. S. Vemuri and K. J. Kim, An experimental and theoretical study on the concept of dropwise condensation, *International Journal of Heat and Mass Transfer*, 2006, **49**, 649-657.
264. S. Vemuri, K. J. Kim, B. D. Wood, S. Govindaraju and T. W. Bell, Long term testing for dropwise condensation using self-assembled monolayer coatings of n-octadecyl mercaptan, *Applied Thermal Engineering*, 2006, **26**, 421-429.
265. G. X. Pang, J. D. Dale and D. Y. Kwok, An integrated study of dropwise condensation heat transfer on self-assembled organic surfaces through Fourier transform infra-red spectroscopy and ellipsometry, *International Journal of Heat and Mass Transfer*, 2005, **48**, 307-316.
266. Q. Yang and A. Gu, Dropwise condensation on SAM and electroless composite coating surfaces, *Journal of Chemical Engineering of Japan*, 2006, **39**, 826-830.
267. L. Yin, Y. Wang, J. Ding, Q. Wang and Q. Chen, Water condensation on superhydrophobic aluminum surfaces with different low-surface-energy coatings, *Applied Surface Science*, 2012, **258**, 4063-4068.
268. B. S. Sikarwar, N. K. Battoo, S. Khandekar and K. Muralidhar, Dropwise Condensation Underneath Chemically Textured Surfaces: Simulation and Experiments, *Journal of Heat Transfer-Transactions of the ASME*, 2011, **133**.
269. M. Izumi, S. Kumagai, R. Shimada and N. Yamakawa, Heat transfer enhancement of dropwise condensation on a vertical surface with round shaped grooves, *Experimental Thermal and Fluid Science*, 2004, **28**, 243-248.
270. R. D. Narhe and D. A. Beysens, Water condensation on a super-hydrophobic spike surface, *Europhysics Letters*, 2006, **75**, 98-104.
271. Y. C. Jung and B. Bhushan, Wetting behaviour during evaporation and condensation of water microdroplets on superhydrophobic patterned surfaces, *Journal of Microscopy-Oxford*, 2008, **229**, 127-140.
272. R. Enright, N. Miljkovic, A. Al-Obeidi, C. V. Thompson and E. N. Wang, Condensation on Superhydrophobic Surfaces: The Role of Local Energy Barriers and Structure Length Scale, *Langmuir*, 2012, **28**, 14424-14432.
273. K. Rykaczewski, W. A. Osborn, J. Chinn, M. L. Walker, J. H. J. Scott, W. Jones, C. L. Hao, S. H. Yao and Z. K. Wang, How nanorough is rough enough to make a surface superhydrophobic during water condensation?, *Soft Matter*, 2012, **8**, 8786-8794.
274. K. A. Wier and T. J. McCarthy, Condensation on ultrahydrophobic surfaces and its effect on droplet mobility: Ultrahydrophobic surfaces are not always water repellant, *Langmuir*, 2006, **22**, 2433-2436.
275. A. Lafuma and D. Quere, Superhydrophobic states, *Nat Mater*, 2003, **2**, 457-460.
276. R. D. Narhe and D. A. Beysens, Nucleation and growth on a superhydrophobic grooved surface, *Physical Review Letters*, 2004, **93**.
277. R. D. Narhe and D. A. Beysens, Growth dynamics of water drops on a square-pattern rough hydrophobic surface, *Langmuir*, 2007, **23**, 6486-6489.
278. Y. T. Cheng, D. E. Rodak, A. Angelopoulos and T. Gacek, Microscopic observations of condensation of water on lotus leaves, *Applied Physics Letters*, 2005, **87**.
279. K. K. S. Lau, J. Bico, K. B. K. Teo, M. Chhowalla, G. A. J. Amaratunga, W. I. Milne, G. H. McKinley and K. K. Gleason, Superhydrophobic carbon nanotube forests, *Nano Letters*, 2003, **3**, 1701-1705.
280. C. Journet, S. Moulinet, C. Ybert, S. T. Purcell and L. Bocquet, Contact angle measurements on superhydrophobic carbon nanotube forests: Effect of fluid pressure, *Europhysics Letters*, 2005, **71**, 104-109.
281. X. H. Ma, S. F. Wang, Z. Lan, B. L. Peng, H. B. Ma and P. Cheng, Wetting Mode Evolution of Steam Dropwise Condensation on Superhydrophobic Surface in the Presence of Noncondensable Gas, *Journal of Heat Transfer-Transactions of the Asme*, 2012, **134**.
282. S. Lee, K. Cheng, V. Palmre, M. H. Bhuiya, K. J. Kim, B. J. Zhang and H. Yoon, Heat transfer measurement during dropwise condensation using micro/nano-scale porous surface, *International Journal of Heat and Mass Transfer*, 2013, **65**, 619-626.
283. T. Tsuruta, H. Tanaka and S. Togashi, Experimental verification of constriction resistance theory in dropwise condensation heat transfer, *International Journal of Heat and Mass Transfer*, 1991, **34**, 2787-2796.
284. T. Tsuruta and H. Tanaka, A theoretical study on the constriction resistance in dropwise condensation, *International Journal of Heat and Mass Transfer*, 1991, **34**, 2779-2786.
285. K. Rykaczewski, Microdroplet Growth Mechanism during Water Condensation on Superhydrophobic Surfaces, *Langmuir*, 2012, **28**, 7720-7729.
286. N. Miljkovic, R. Enright, S. C. Maroo, H. J. Cho and E. N. Wang, Liquid Evaporation on Superhydrophobic and Superhydrophilic Nanostructured Surfaces, *Journal of Heat Transfer-Transactions of the ASME*, 2011, **133**.

287. N. Miljkovic, R. Enright, Y. Nam, K. Lopez, N. Dou, J. Sack and E. N. Wang, Jumping-Droplet-Enhanced Condensation on Scalable Superhydrophobic Nanostructured Surfaces, *Nano Letters*, 2013, **13**, 179-187.
288. X. Chen, J. Wu, R. Ma, M. Hua, N. Koratkar, S. Yao and Z. Wang, Nanograssed Micropyramidal Architectures for Continuous Dropwise Condensation, *Advanced functional materials*, 2011, **21**, 4617-4623.
289. J. Cheng, A. Vandadi and C.-L. Chen, Condensation heat transfer on two-tier superhydrophobic surfaces, *Applied Physics Letters*, 2012, **101**.
290. T. Liu, W. Sun, X. Sun and H. Ai, Thermodynamic Analysis of the Effect of the Hierarchical Architecture of a Superhydrophobic Surface on a Condensed Drop State, *Langmuir*, 2010, **26**, 14835-14841.
291. T. Q. Liu, W. Sun, X. Y. Sun and H. R. Ai, Mechanism study of condensed drops jumping on super-hydrophobic surfaces, *Colloids and Surfaces a-Physicochemical and Engineering Aspects*, 2012, **414**, 366-374.
292. K. Rykaczewski, A. T. Paxson, S. Anand, X. M. Chen, Z. K. Wang and K. K. Varanasit, Multimode Multidrop Serial Coalescence Effects during Condensation on Hierarchical Superhydrophobic Surfaces, *Langmuir*, 2013, **29**, 881-891.
293. A. Kumar and G. M. Whitesides, Patterned Condensation Figures as Optical Diffraction Gratings, *Science*, 1994, **263**, 60-62.
294. S. Daniel, M. K. Chaudhury and J. C. Chen, Fast drop movements resulting from the phase change on a gradient surface, *Science*, 2001, **291**, 633-636.
295. M. M. Derby, A. Chatterjee, A. Peles and M. K. Jensen, Flow condensation heat transfer enhancement in a mini-channel with hydrophobic and hydrophilic patterns, *Int. J. Heat Mass Transfer*, 2014, **68**, 151-160.
296. R. Xiao, N. Miljkovic, R. Enright and E. Wang, Immersion condensation on oil-infused heterogeneous surface for enhanced heat transfer, *Sci. Reports*, 2013, **3**, 1988 (1981-1986).
297. C. W. Yao, T. P. Garvin, J. L. Alvarado, A. M. Jacobi, B. G. Jones and C. P. Marsh, Droplet contact angle behavior on a hybrid surface with hydrophobic and hydrophilic properties, *Applied Physics Letters*, 2012, **101**.
298. V. K. Croutch and R. A. Hartley, Adhesion of ice to coatings and the performance of ice release coatings, *Journal of Coatings Technology*, 1992, **64**, 41-53.
299. B. Somlo and V. Gupta, A hydrophobic self-assembled monolayer with improved adhesion to aluminum for deicing application, *Mechanics of Materials*, 2001, **33**, 471-480.
300. K. Li, S. Xu, W. Shi, M. He, H. Li, S. Li, X. Zhou, J. Wang and Y. Song, Investigating the effects of solid surfaces on ice nucleation, *Langmuir*, 2012, **28**, 10749-10754.
301. H. Saito, K. Takai and G. Yamauchi, Water- and ice-repellent coatings, *JOCCA-Surface Coatings International*, 1997, **80**, 168-171.
302. T. V. Charpentier, A. Neville, P. Millner, R. W. Hewson and A. Morina, Development of anti-icing materials by chemical tailoring of hydrophobic textured metallic surfaces, *J Colloid Interface Sci*, 2013, **394**, 539-544.
303. F. Arianpour, M. Farzaneh and S. A. Kulinich, Hydrophobic and ice-retarding properties of doped silicone rubber coatings, *Applied Surface Science*, 2013, **265**, 546-552.
304. J. B. Boreyko and C. P. Collier, Delayed Frost Growth on Jumping-Drop Superhydrophobic Surfaces, *ACS Nano*, 2013, **7**, 1618-1627.
305. Q. Zhang, M. He, J. Chen, J. Wang, Y. Song and L. Jiang, Anti-icing surfaces based on enhanced self-propelled jumping of condensed water microdroplets, *Chem Commun (Camb)*, 2013, **49**, 4516-4518.
306. M. He, J. Wang, H. Li and Y. Song, Super-hydrophobic surfaces to condensed micro-droplets at temperatures below the freezing point retard ice/frost formation, *Soft Matter*, 2011, **7**, 3993.
307. Q. Zhang, M. He, X. Zeng, K. Li, D. Cui, J. Chen, J. Wang, Y. Song and L. Jiang, Condensation mode determines the freezing of condensed water on solid surfaces, *Soft Matter*, 2012, **8**, 8285.
308. L. Yin, Q. Xia, J. Xue, S. Yang, Q. Wang and Q. Chen, In situ investigation of ice formation on surfaces with representative wettability, *Applied Surface Science*, 2010, **256**, 6764-6769.
309. P. Guo, Y. Zheng, M. Wen, C. Song, Y. Lin and L. Jiang, Icephobic/anti-icing properties of micro/nanostructured surfaces, *Adv Mater*, 2012, **24**, 2642-2648.
310. L. B. Boinovich, S. N. Zhevnenko, A. M. Emel'yanenko, R. V. Gol'dshtein and V. P. Epifanov, Adhesive strength of the contact of ice with a superhydrophobic coating, *Doklady Chemistry*, 2013, **448**, 71-75.
311. R. Jafari, R. Menini and M. Farzaneh, Superhydrophobic and icephobic surfaces prepared by RF-sputtered polytetrafluoroethylene coatings, *Applied Surface Science*, 2010, **257**, 1540-1543.
312. S. A. Kulinich and M. Farzaneh, Ice adhesion on super-hydrophobic surfaces, *Applied Surface Science*, 2009, **255**, 8153-8157.
313. R. Menini and M. Farzaneh, Elaboration of Al2O3/PTFE icephobic coatings for protecting aluminum surfaces, *Surface and Coatings Technology*, 2009, **203**, 1941-1946.
314. D. K. Sarkar and M. Farzaneh, Superhydrophobic Coatings with Reduced Ice Adhesion, *Journal of Adhesion Science and Technology*, 2009, **23**, 1215-1237.
315. N. Saleema, M. Farzaneh, R. W. Paynter and D. K. Sarkar, Prevention of Ice Accretion on Aluminum Surfaces by Enhancing Their Hydrophobic Properties, *Journal of Adhesion Science and Technology*, 2011, **25**, 27-40.
316. S. A. Kulinich and M. Farzaneh, How wetting hysteresis influences ice adhesion strength on superhydrophobic surfaces, *Langmuir*, 2009, **25**, 8854-8856.
317. M. Nosonovsky and V. Hejazi, Why Superhydrophobic Surfaces Are Not Always Icephobic, *ACS Nano*, 2012, **6**, 8488-8491.

318. A. J. Meuler, J. D. Smith, K. K. Varanasi, J. M. Mabry, G. H. McKinley and R. E. Cohen, Relationships between water wettability and ice adhesion, *ACS Appl Mater Interfaces*, 2010, **2**, 3100-3110.
319. S. A. Kulinich, S. Farhadi, K. Nose and X. W. Du, Superhydrophobic Surfaces: Are They Really Ice-Repellent?, *Langmuir*, 2011, **27**, 25-29.
320. S. Yang, Q. Xia, L. Zhu, J. Xue, Q. Wang and Q.-m. Chen, Research on the icephobic properties of fluoropolymer-based materials, *Applied Surface Science*, 2011, **257**, 4956-4962.
321. N. J. Shirtcliffe, G. McHale and M. I. Newton, The superhydrophobicity of polymer surfaces: Recent developments, *Journal of Polymer Science Part B: Polymer Physics*, 2011, **49**, 1203-1217.
322. C. Peng, S. Xing, Z. Yuan, J. Xiao, C. Wang and J. Zeng, Preparation and anti-icing of superhydrophobic PVDF coating on a wind turbine blade, *Applied Surface Science*, 2012, **259**, 764-768.
323. T. Jing, Y. Kim, S. Lee, D. Kim, J. Kim and W. Hwang, Frosting and defrosting on rigid superhydrohobic surface, *Applied Surface Science*, 2013, **276**, 37-42.
324. A. J. Meuler, G. H. McKinley and R. E. Cohen, Exploiting Topographical Texture To Impart Icephobicity, *ACS Nano*, 2010, **4**, 7048-7052.
325. F. Wang, C. Li, Y. Lv, F. Lv and Y. Du, Ice accretion on superhydrophobic aluminum surfaces under low-temperature conditions, *Cold Regions Science and Technology*, 2010, **62**, 29-33.
326. V. Bahadur, L. Mishchenko, B. Hatton, J. A. Taylor, J. Aizenberg and T. Krupenkin, Predictive model for ice formation on superhydrophobic surfaces, *Langmuir*, 2011, **27**, 14143-14150.
327. M. A. Sarshar, C. Swarctz, S. Hunter, J. Simpson and C.-H. Choi, Effects of contact angle hysteresis on ice adhesion and growth on superhydrophobic surfaces under dynamic flow conditions, *Colloid and Polymer Science*, 2012, **291**, 427-435.
328. A. Alizadeh, M. Yamada, R. Li, W. Shang, S. Otta, S. Zhong, L. Ge, A. Dhinojwala, K. R. Conway, V. Bahadur, A. J. Vinciquerra, B. Stephens and M. L. Blohm, Dynamics of ice nucleation on water repellent surfaces, *Langmuir*, 2012, **28**, 3180-3186.
329. C. Antonini, M. Innocenti, T. Horn, M. Marengo and A. Amirfazli, Understanding the effect of superhydrophobic coatings on energy reduction in anti-icing systems, *Cold Regions Science and Technology*, 2011, **67**, 58-67.
330. J. Xiao and S. Chaudhuri, Design of anti-icing coatings using supercooled droplets as nano-to-microscale probes, *Langmuir*, 2012, **28**, 4434-4446.
331. B. Gorbunov, A. Baklanov, N. Kakutkina, H. L. Windsor and R. Toumi, Ice nucleation on soot particles, *Journal of Aerosol Science*, 2001, **32**, 199-215.
332. T. Maitra, M. K. Tiwari, C. Antonini, P. Schoch, S. Jung, P. Eberle and D. Poulikakos, On the Nanoengineering of Superhydrophobic and Impalement Resistant Surface Textures below the Freezing Temperature, *Nano Lett*, 2014, **14**, 172-182.
333. J. C. Bird, R. Dhiman, H. M. Kwon and K. K. Varanasi, Reducing the contact time of a bouncing drop, *Nature*, 2013, **503**, 385-388.
334. Y. Zhang, X. Yu, H. Wu and J. Wu, Facile fabrication of superhydrophobic nanostructures on aluminum foils with controlled-condensation and delayed-icing effects, *Applied Surface Science*, 2012, **258**, 8253-8257.
335. P. W. Wilson, W. Lu, H. Xu, P. Kim, M. J. Kreder, J. Alvarenga and J. Aizenberg, Inhibition of ice nucleation by slippery liquid-infused porous surfaces (SLIPS), *Phys Chem Chem Phys*, 2013, **15**, 581-585.
336. K. Rykaczewski, S. Anand, S. B. Subramanyam and K. K. Varanasi, Mechanism of Frost Formation on Lubricant-Impregnated Surfaces, *Langmuir*, 2013, **29**, 5230-5238.
337. H. Lee, M. L. Alcaraz, M. F. Rubner and R. E. Cohen, Zwitter-Wettability and Antifogging Coatings with Frost-Resisting Capabilities, *ACS Nano*, 2013, **7**, 2172-2185.
338. *Annual Energy Review 2011*, U.S. Energy Information Administration, 2012.
339. B. Linnhoff, *A user guide on process integration for the efficient use of energy*, Institution of Chemical Engineers (Great Britain), 1994.
340. *Technology and Cost of the MY2007 Toyota Camry HEV - Final Report*, Oak Ridge National Laboratory, 2007.
341. G. Moreno, *Section 5.7 "Two-Phase Cooling Technology for Power Electronics with Novel Coolants", in Advanced Power Electronics and Electric Motors Annual Progress Report, FY 2011*, U.S. Department of Energy Office of Vehicle Technologies, 2012.
342. R. Thevenin, Z. Wu, P. Keller, R. Cohen, C. Clanet and D. Quere, New thermal-sensitive superhydrophobic material, Pittsburgh, PA, USA, 2013.
343. P. Yi, K. Khoshmanesh, A. F. Chrimes, J. L. Campbell, K. Ghorbani, S. Nahavandi, G. Rosengarten and K. Kalantar-zadeh, Dynamic Nanofin Heat Sinks, *Advanced Energy Materials*, 2014, **4**, n/a-n/a.
344. G. Agbaglah, S. Delaux, D. Fuster, J. Hoepffner, C. Josserand, S. Popinet, P. Ray, R. Scardovelli and S. Zaleski, Parallel simulation of multiphase flows using octree adaptivity and the volume-of-fluid method, *Comptes Rendus Mécanique*, 2011, **339**, 194-207.
345. R. Raj, C. Kunkelmann, P. Stephan, J. Plawsky and J. Kim, Contact line behavior for a highly wetting fluid under superheated conditions, *International Journal of Heat and Mass Transfer*, 2012, **55**, 2664-2675.
346. P. Koumoutsakos, Multiscale flow simulations using particles, *Annual Review of Fluid Mechanics*, 2005, **37**, 457-487.
347. J. Kim, Review of nucleate pool boiling bubble heat transfer mechanisms, *International Journal of Multiphase Flow*, 2009, **35**, 1067-1076.

348. K.-Y. Law, Definitions for hydrophilicity, hydrophobicity, and superhydrophobicity: getting the basics right, *J. Phys. Chem. Lett.* , 2014, **5**, 686-688.

349. K. Rykaczewski, A. T. Paxson, M. Staymates, M. L. Walker, X. Sun, S. Anand, S. Srinivasan, G. H. McKinley, J. Chinn, J. H. J. Scott and K. K. Varanasi, Dropwise condensation of low surface tension fluids on omniphobic surfaces, *Scientific Reports*, 2014, **4**, 4158 (4151-4158).

350. S. Farhadi, M. Farzaneh and S. A. Kulinich, Anti-icing performance of superhydrophobic surfaces, *Applied Surface Science*, 2011, **257**, 6264-6269.

351. Y. Wang, J. Xue, Q. Wang, Q. Chen and J. Ding, Verification of icephobic/anti-icing properties of a superhydrophobic surface, *ACS Appl Mater Interfaces*, 2013, **5**, 3370-3381.

352. X. Zhang, H. Kono, Z. Liu, S. Nishimoto, D. A. Tryk, T. Murakami, H. Sakai, M. Abe and A. Fujishima, A transparent and photo-patternable superhydrophobic film, *Chemical Communications*, 2007, 4949-4951.

353. M. Zhang, M. Y. Efremov, F. Schiettekatte, E. A. Olson, A. T. Kwan, S. L. Lai, T. Wisleder, J. E. Greene and L. H. Allen, Size-dependent melting point depression of nanostructures:Nanocalorimetric measurements, *Physical Review B*, 2000, **62**, 10548-10557.

354. T. R. Bott, *Fouling of Heat Exchangers*, Elsevier, New York, 1995.

355. T. Humplik, J. Lee, S. C. O'Hern, B. A. Fellman, M. A. Baig, S. F. Hassan, M. A. Atieh, F. Rahman, T. Laoui, R. Karnik and E. N. Wang, Nanostructured materials for water desalination, *Nanotechnology*, 2011, **22**.

356. C.-H. Choi and C.-J. Kim, in *Green Tribology – Biomimetics, Energy Conservation, and Sustainability*, eds. M. Nosonovsky and B. Bhushan, Springer, 2012, pp. 79-104.

357. S. Y. Heo, J. K. Koh, G. Kang, S. H. Ahn, W. S. Chi, K. Kim and J. H. Kim, Bifunctional Moth-Eye Nanopatterned Dye-Sensitized Solar Cells: Light-Harvesting and Self-Cleaning Effects, *Advanced Energy Materials*, 2014, **4**, n/a-n/a.

358. J. R. Thome, *Enhanced Boiling Heat Transfer*, Hemisphere Publishing Corporation, New York, 1989.

359. J. R. Thome, Enhanced boiling of mixtures, *Chemical Engineering Science*, 1987, **42**, 1909-1917.

360. A. J. Scardino and R. de Nys, Mini review: Biomimetic models and bioinspired surfaces for fouling control, *Biofouling*, 2011, **27**, 73-86.